\documentclass{emulateapj}

\def\gtsima{$\; \buildrel > \over \sim \;$}
\def\ltsima{$\; \buildrel < \over \sim \;$}
\def\prosima{$\; \buildrel \propto \over \sim \;$}
\def\gsim{\lower.7ex\hbox{\gtsima}}
\def\lsim{\lower.7ex\hbox{\ltsima}}
\def\simgt{\lower.7ex\hbox{\gtsima}}
\def\simlt{\lower.7ex\hbox{\ltsima}}
\def\simpr{\lower.7ex\hbox{\prosima}}

\newcommand{\Mpc}{$h^{-1}$\thinspace Mpc}
\newcommand{\hmpc}{$h$\thinspace Mpc$^{-1}$}
\newcommand{\etal}{{\rm et al.~}}

\newcommand{\be}{\begin{equation}}
\newcommand{\ee}{\end{equation}}

\def\apjl{ApJL}
\def\apjs{ApJS}

\edef\aap{A\&A}

\begin{document}  
 
\shorttitle{Rich superclusters}
\shortauthors{Einasto et al.}

\title{Toward understanding rich superclusters}

\author{ M. Einasto\altaffilmark{1}, 
E. Saar\altaffilmark{1}, 
V.J. Mart\'{\i}nez\altaffilmark{2},
J. Einasto\altaffilmark{1}, 
L. J. Liivam\"agi\altaffilmark{1}, 
E. Tago\altaffilmark{1}, 
J.-L. Starck\altaffilmark{3}, 
V. M\"uller\altaffilmark{4},  
P. Hein\"am\"aki\altaffilmark{5},   
P. Nurmi\altaffilmark{5}, 
S. Paredes\altaffilmark{6},
M. Gramann\altaffilmark{1},  
G. H\"utsi\altaffilmark{1}
}

\altaffiltext{1}{Tartu Observatory, EE-61602 T\~oravere, Estonia}
\altaffiltext{2}{Observatori Astron\`omic, Universitat de Val\`encia, Apartat
de Correus 22085, E-46071 Val\`encia, Spain} 
\altaffiltext{3}{CEA-Saclay, DAPNIA/SEDI-SAP, Service d'Astrophysique, F-91191
  Gif 
sur Yvette, France} 
\altaffiltext{4}{Astrophysical Institute Potsdam, An der Sternwarte 16,
D-14482 Potsdam, Germany}
\altaffiltext{5}{Turku University,
Tuorla Observatory, V\"ais\"al\"antie 20, Piikki\"o, Finland} 
\altaffiltext{6}{Departamento de Matem\'atica Aplicada y Estad\'{\i}stica, 
Universidad Polit\'ecnica de Cartagena, 
30203 Cartagena, Spain}

\begin{abstract} 

  We present a morphological study of the two richest superclusters from the
  2dF Galaxy Redshift Survey (SCL126, the Sloan Great Wall, and SCL9, the
  Sculptor supercluster). We use Minkowski functionals, shapefinders, and
  galaxy group information to study the substructure of these superclusters as
  formed by different populations of galaxies. We compare the properties of
  grouped and isolated galaxies in the core region and in the outskirts of
  superclusters.

  The fourth Minkowski functional $V_3$ and the morphological signature 
  $K_1$- $K_2$ show a crossover from low-density morphology (outskirts 
  of supercluster) to high-density morphology (core of supercluster) at 
  mass fraction $m_f \approx 0.7$. The galaxy content 
  and the morphology of the galaxy populations 
  in supercluster   cores and outskirts is different.
  The core regions contain a larger fraction of early type, red galaxies,
  and richer groups than the outskirts of superclusters.
    In the core and outskirt regions   the fine structure of the two prominent 
  superclusters as delineated by galaxies from different populations
   also differs.
  The values of the fourth Minkowski functional $V_3$ show that   
  in the supercluster SCL126 the population of early type, red galaxies is
  more clumpy than the population of late type, blue galaxies, especially in 
  the outskirts of the supercluster. In the contrary, in the supercluster
  SCL9, the clumpiness of the spatial distribution of galaxies of different 
  type and color is quite similar in the outskirts of the supercluster, 
  while in the core region the clumpiness of the late type, blue galaxy
  population is larger than the clumpiness of the early type, red galaxy
  population.
  
   Our results suggest that both local 
  (group/cluster) and global (supercluster) environments are important 
  in forming galaxy morphologies and colors (and determining the star formation 
  activity). 
  
 The differences 
  between the superclusters indicate that these superclusters have 
  different evolutional histories.
  
\end{abstract}

\keywords{cosmology: large-scale structure of the Universe -- clusters
of galaxies; cosmology: large-scale structure of the Universe --
Galaxies; clusters: general}

\section{Introduction}
\label{sec:intro}

Huge superclusters which may contain several tens of 
rich (Abell) clusters are the largest coherent systems in the Universe with
characteristic dimensions of up to 100~\Mpc. 
\footnote{$h$ is the Hubble constant in units of 100~km~s$^{-1}$~Mpc$^{-1}$.} 
As they are very large, and dynamical evolution takes place at a slower rate
for larger scales, superclusters have retained memory of the initial conditions
of their formation, and of the early evolution of structure
\citep{kofman87}.

While we might be able to explain the structure and properties of most
(average) superclusters, explaining rich superclusters is still a
challenge. To start with, even their existence is not well explained by the
main contemporary structure modeling tool, numerical simulations. There is a
number of rich superclusters in our close cosmological neighbourhood -- we
shall list them in a moment -- but the comparison of the
luminosity functions of observed and simulated  (Millennium)
superclusters shows that the fraction of 
really rich superclusters in simulations is much lower than in 
observed smaples 
 \citep{e06}. The extreme cases of observed objects usually
provide the most stringent tests for theories; this motivates the need for a
detailed understanding of the richest superclusters.

The richest relatively close superclusters are the Shapley Supercluster
\citep[][and references therein]{proust06} and the Horologium--Reticulum
Supercluster \citep{rose02,fleenor05,e2003}.  The two superclusters studied in
this paper, the Sloan Great Wall and the Sculptor supercluster, belong also to
this category of very rich superclusters.

The formation of rich superclusters had to begin earlier than smaller 
structures; they are sites of early star and galaxy formation 
\citep[e.g.][]{mob05}, and the first places where systems of galaxies form 
\citep[e.g.][]{ven,ouch}. Thus future deep surveys like the ALHAMBRA Deep survey 
\citep{moles05} are likely to detect (core) regions of rich superclusters at 
very high redshifts. The supercluster environment affects the properties of 
groups and clusters located there \citep{e2003,pl04}. Rich superclusters contain 
high density cores which are absent in poor superclusters \citep[][hereafter 
Paper III]{e2007a}. The fraction of X-ray clusters in rich superclusters is 
larger than in poor superclusters \citep[][hereafter E01]{e2001}, and the core 
regions of the richest superclusters may contain merging X-ray clusters 
\citep{rose02,bar00}. The richest superclusters are more filamentary, less 
compact and more asymmetrical than poor superclusters \citep[][hereafter Paper 
II]{e07b}. Moreover, as we noted above, the fraction of very rich superclusters 
in observed catalogues is larger than models predict \citep{e06}.

\begin{figure*}[ht]
\centering
\resizebox{0.9\textwidth}{!}{\includegraphics*{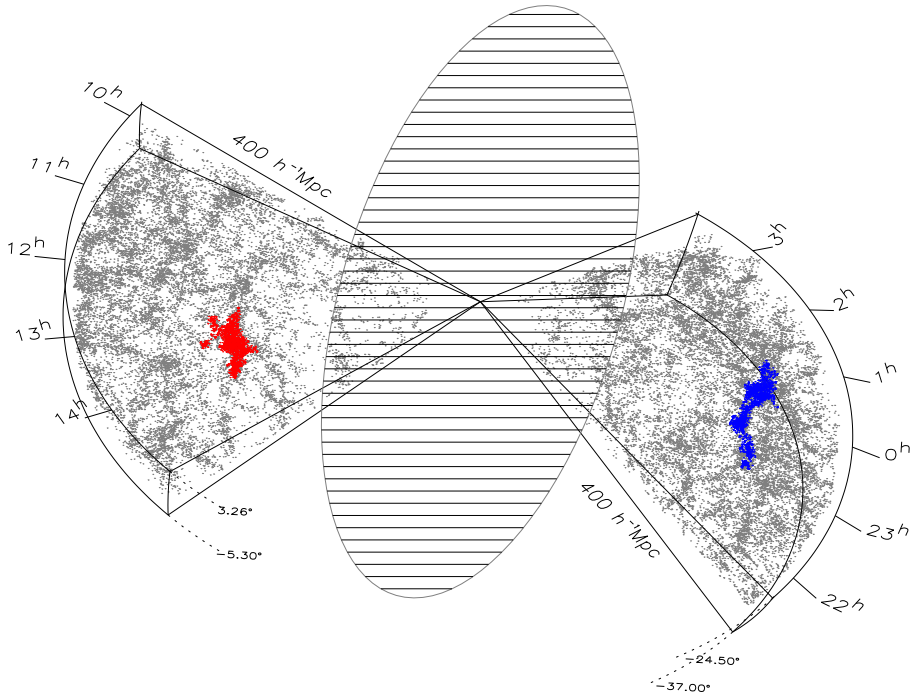}}
\caption{The nearest, almost complete part of the 2dF GRS survey (up to
400 \Mpc). The two richest superclusters in this survey are shown.
The disk showing the orientation of the galactic plane helps to allocate
the 2dF survey in space; 
SCL 126 is located towards the Galactic North pole (left, red color),
SCL 9 is at the right side of the figure (blue color). 
}
\label{fig:silvestre}
\end{figure*}

In the present paper we continue the study of superclusters selected on the
basis of the 2dF Galaxy Redshift Survey. Our paper is devoted to a detailed
study of the two richest superclusters in the 2dF Galaxy Redshift Survey. We
chose them from the catalogue of superclusters of the 2dFGRS by
\citet[][hereafter Paper I]{e07c}.  These are: the supercluster SCL126 in the
Northern sky, and the supercluster SCL9 (the Sculptor supercluster) in the
Southern sky (see Fig.~\ref{fig:silvestre}).
  
The supercluster SCL126 is the most prominent supercluster defined by Abell
clusters in the Northern 2dFGRS  (SCL126 in the catalog of superclusters by
E01, N152 in Paper I) in the direction of the Virgo constellation. This
supercluster has also been called the Sloan Great Wall
\citep{hoyle02,vogeley04,gott05,nichol06}.  The presence of this supercluster
affects the measurements of the correlation function \citep{croton04}, and of
the genus and Minkowski functionals of the SDSS and 2dF redshift surveys
\citep{park05,saar06}. The ``meatball'' shift in the measurements of the
topology in the SDSS data is partly due to this supercluster \citep{gott06}.

The Sculptor supercluster, the most prominent supercluster in the Southern
strip of the 2dFGRS, is among the three richest superclusters in E01; it
contains 25 Abell clusters, six of these are also X-ray
clusters. \citet{zap04} found evidence about the presence of warm-hot diffuse
gas there, which is associated with the inter-cluster galaxy distribution in
this supercluster.

In a recent paper \citep[][hereafter RI]{e07d} we studied the morphology of rich 
superclusters (their shape and internal structure), using Minkowski functionals 
and shapefinders. Our calculations in RI showed that the morphology of the 
richest superclusters from the 2dFGRS is different from each other: the 
supercluster SCL126 resembles a very rich filament (wall) with a high density 
core region, while the supercluster SCL9 can be described as a collection of 
spiders (a multi-spider), consisting of a large number of cores connected by 
relatively thin filaments.

The main aim of the present paper is to understand whether the differences in 
the overall morphology of these two rich superclusters under study are also 
reflected by their fine structure as determined by the distribution of galaxies 
of different luminosity, color and spectral type in core region and in 
outskirts of superclusters. 

The first study to show that in a supercluster early and late type galaxies 
trace the structure of the supercluster in a different manner was performed by 
\citet{gio86} who demonstrated that in the Perseus supercluster, elliptical 
galaxies are mainly located along the central body of the supercluster, while 
spiral galaxies are distributed in the outer regions of the supercluster. The 
presence of a large-scale segregation of galaxies of different type in nearby 
superclusters was shown also by \citet{e87}. In Paper III we showed that rich 
superclusters have a larger fraction of  red, non-star-forming galaxies 
than poor superclusters. Recently, galaxy populations have been studied in core 
regions of some very rich superclusters (\citet{hai06} -- in the Shapley 
supercluster, \citet{pr05} -- in the Pisces-Cetus supercluster). These studies 
showed that rich clusters in the core regions of superclusters contain a large 
fraction of passive galaxies, while actively star forming galaxies are located 
between the clusters.

Thus studies of rich superclusters which contain a large variety of 
environments and possibly a variety of evolutionary phases give us a 
possibility to study the properties of cosmic structures and the 
properties of galaxies therein in a consistent way, helping us to 
understand the role of environment in galaxy evolution.

One method to quantify the  structure of superclusters is to use the Minkowski 
functionals. This type of study has been called morphometry by \citet{hik03}. 
Minkowski functionals, genus and shapefinders, defined on the basis of Minkowski 
functionals have been used earlier to study the 3D topology of the large scale 
structure from 2dF and SDSS surveys \citep{saar06,hik03,park05,gott06,jam06} and 
to characterize the morphology of superclusters \citep{sah98,sheth03, 
sss04,bpr01,kbp02,bas03,bas06} from observations and simulations. These studies 
concern only the ``outer'' shapes of superclusters and do not treat their 
substructure. We expanded this approach in Paper RI by using the Minkowski 
functionals and shapefinders to analyze the full density distribution in 
superclusters, at all density levels.

The fourth Minkowski functional $V_3$ (the Euler characteristic) gives us the 
number of isolated clumps (or voids) in the region \citep{saar06}, meaning that 
we can use it to study the clumpiness of the galaxy distribution inside 
superclusters -- the fine structure of superclusters. We calculate the fourth 
Minkowski functional $V_3$ for galaxies of different populations in 
superclusters for a range of threshold densities, starting with the lowest 
density used to determine superclusters, up to the peak density in the 
supercluster core.  This analysis shows how the fourth Minkowski functional can 
be used in studies of the fine structure of superclusters as delineated by 
different galaxy populations. This study is of an exploratory nature since this 
is the first time when this method are used for studies of the fine structure of 
galaxy populations of individual superclusters.  Employing Minkowski 
functionals, we can see in detail how the morphology of superclusters is traced 
by galaxies of different type.

In our analysis we use also the shapefinders calculated on the basis of the 
Minkowski functionals. In addition, we compare the galaxy content of groups of 
different richness in core regions and in outskirts of superclusters.

The paper is composed as follows. In Section 2 we describe the galaxy data,
the supercluster catalogue and the data on the richest superclusters. In
Sect. 3 we compare the overall galaxy content of the superclusters. In Sect. 4
we describe the use of the fourth Minkowski functional (the Euler
characteristic) to study the fine structure of the superclusters as delineated
by different galaxy populations. In Sect. 5 we study the galaxy content in the
core regions and in the outskirts of the superclusters, and compare galaxy
populations in groups of various richness. In the last sections (6 and 7) we
discuss our results and give the conclusions. In the Appendix we give a
definition of the Minkowski functionals and shapefinders, and describe
different kernels used to calculate the density fields of superclusters.

\section{Data}

\subsection{Rich supercluster data}

We used the 2dFGRS final release \citep{col01,col03}, and the catalogue of
superclusters of galaxies from the 2dF survey (Paper I), applying a redshift
limit $z\leq 0.2$.  When calculating (comoving) distances we used a flat
cosmological model with the standard parameters, the matter density $\Omega_m
= 0.3$, and the dark energy density $\Omega_{\Lambda} = 0.7$ (both in units of
the critical cosmological density).  Galaxies were included in the 2dF GRS, if
their corrected apparent magnitude $b_j$ lied in the interval from $b_1 =
13.5$ to $b_2 = 19.45$.  We used weighted luminosities to calculate the
luminosity density field on a grid of the cell size of 1~\Mpc\, smoothed with
an Epanechnikov kernel of the radius 8~\Mpc; this density field was used to
find superclusters of galaxies. We defined superclusters as connected
non-percolating systems with densities above a certain threshold density; the
actual threshold density used was 4.6 in units of the mean luminosity
density. A detailed description of the supercluster finding algorithm can be
found in Paper I.

In our analysis we also used the data about groups of galaxies from the 2dFGRS
\citep[][hereafter T06]{tago06}.  Groups in this catalogue were determined
using the Friend-of-Friend (FoF) algorithm in which galaxies are linked
together into a system if they have at least one neighbor at a distance less
than the linking length. For details about the group finding algorithm and the
analysis of the selection effects see T06. The catalogues of groups and
isolated galaxies can be found at
\texttt{http://www.aai.ee/$\sim$maret/2dfgr.html}, the catalogues of observed
and model superclusters -- at
\texttt{http://www.aai.ee/$\sim$maret/2dfscl.html}. 
We present additional information about
the morphology of the richest superclusters at 
\texttt{http://www.aai.ee/$\sim$maret/richscl.html} and
\texttt{http://www.aai.ee/$\sim$maret/rich2.html}.

\begin{figure*}[ht]
\centering
\resizebox{0.45\textwidth}{!}{\includegraphics*{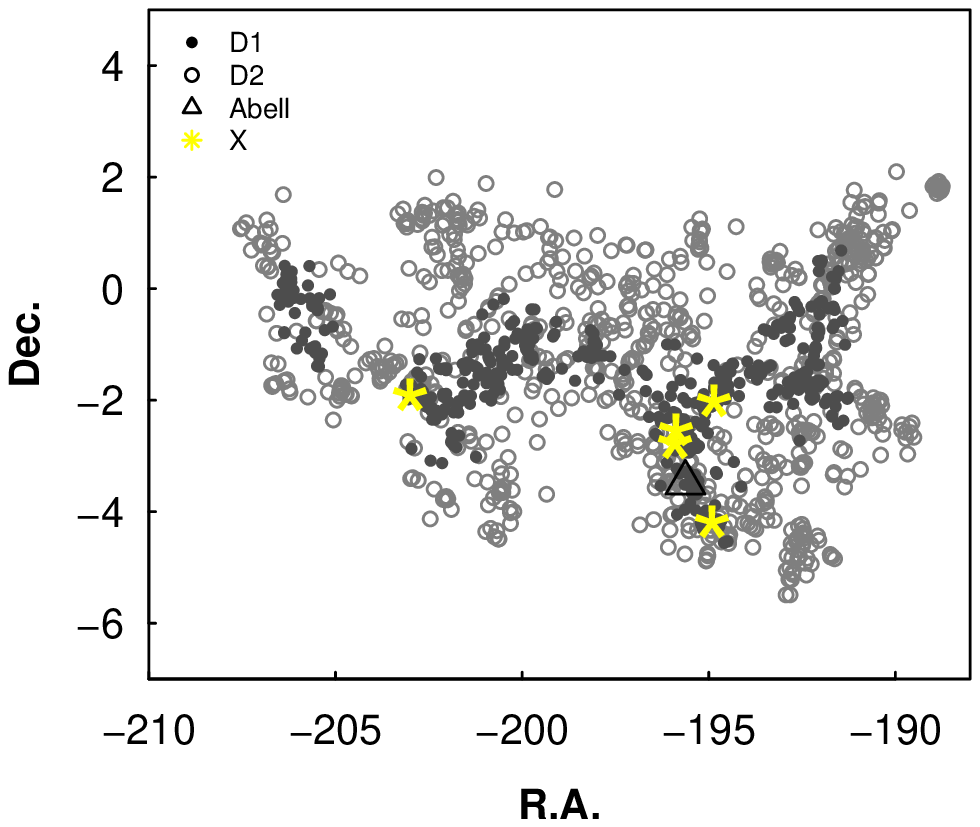}}
\resizebox{0.45\textwidth}{!}{\includegraphics*{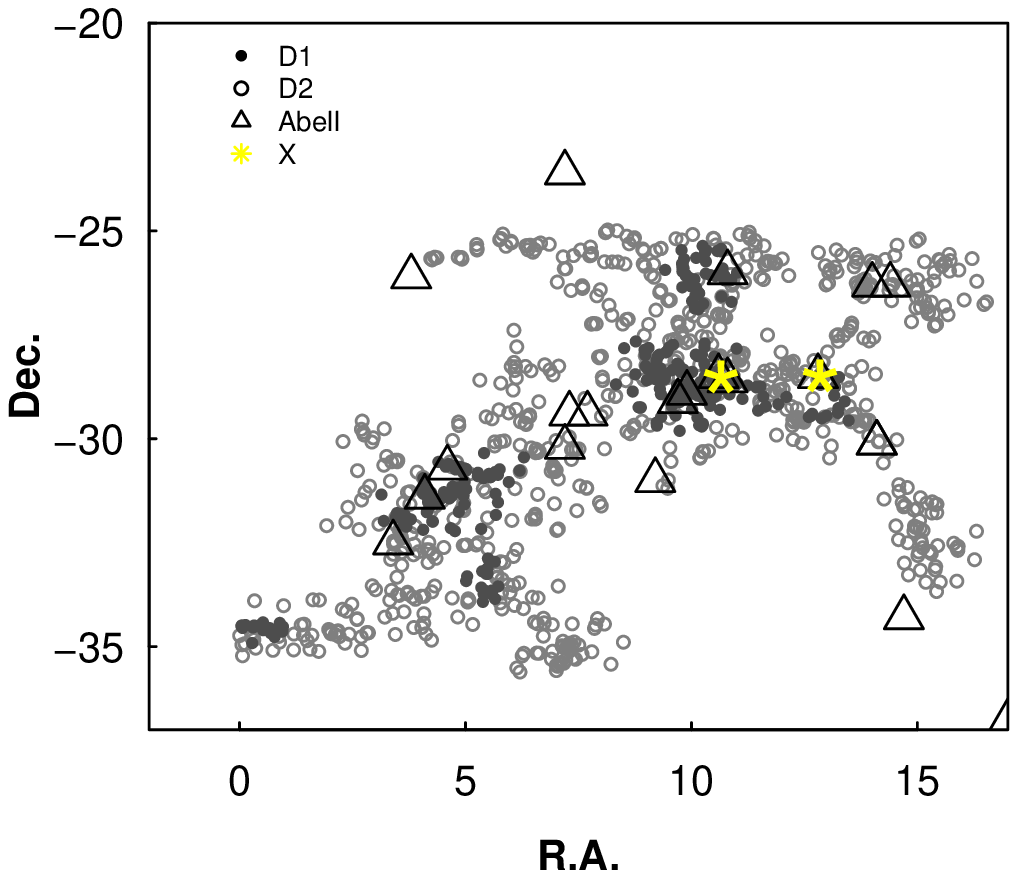}}
\caption{The sky distribution of galaxies, and Abell and X-ray clusters 
in the superclusters SCL126 (left) and SCL9 (right). 
Filled gray circles: galaxies from the core region ($D1$, see sect. 5), 
empty gray circles: galaxies from the outskirt region ($D2$),
triangles: Abell clusters, yellow stars: X-ray clusters.
}
\label{fig:radecx}
\end{figure*}

For the present analysis we select the richest superclusters from the
catalogue of the 2dF superclusters. The data on these superclusters are given in
Table~\ref{tab:1}.  In this Table we give the central coordinates, 
redshifts
and distances of the
superclusters, the numbers of galaxies, groups and Abell and X-ray clusters in
the superclusters, the mean values of the luminosity density within
superclusters and their total luminosities (from Paper II).  In our
morphological analysis we use volume-limited samples of galaxies from these
superclusters. The luminosity limits for each supercluster sample are also
given in Table~\ref{tab:1}. We plot the sky distribution of galaxies, and
Abell and X-ray clusters in Fig.~\ref{fig:radecx}; the location of these
superclusters in space can be seen in Fig.~\ref{fig:silvestre}.

{\scriptsize
\begin{table*}[ht]
\caption{Data on rich superclusters }
\scriptsize
\begin{tabular}{lrrrrrrrrrrrcc} 
\hline 
 ID & R.A. & Dec & Dist & $z$ & $N_{gal}$ &$M_{lim}$ & $N_{vol}$ & $N_{cl}$
 &$N_{gr}$ &$N_{ACO}$&$N_X$ & $\delta_m$ &   $L_{tot}$ \\ 
     & deg & deg & \Mpc  &       &   &&  &  &     &   &              \\
\hline 
SCL126 (N152)& 194.71 & -1.74& 251.2 & 0.085 &  3591  & -19.25&1308 &18 &  40,2 &  9
&  4       & 7.7 &  0.378E+14 \\ 
SCL9 (N34)   &   9.85& -28.94& 326.3 & 0.113 &  3175  & -19.50&1176 &24 &  26,9 &  12
(25) &  2(6) & 8.1 &   0.497E+14\\ 
\label{tab:1}     
\end{tabular} 

\normalsize
\tablecomments{
The supercluster ID after Einasto et al. (2001) with the ID of Paper I in
parenthesis; equatorial coordinates, the comoving distance $D$ for our 
cosmology; redshift $z$,  
the number of galaxies $N_{gal}$ in the supercluster,
the magnitude limit $M_{lim}$ and the number of galaxies $N_{vol}$ for 
the volume limited supercluster. 
$N_{cl}$ and $N_{gr}$ are the numbers of the density field 
clusters and groups according to paper I.  
$N_{ACO}$ gives the number of Abell clusters in this part of
the supercluster that is covered by the 2dF survey; the number inside  
parenthesis is the total number of Abell clusters in this
supercluster, by Einasto et al. (2001) list.
$N_X$ is the number of X-ray clusters,
$\delta_m$ is the mean value of the luminosity density 
in the supercluster  (in units of the mean survey density),
and $L_{tot}$ is the total luminosity of the supercluster (in Solar units).
}
\end{table*}            
}

The most prominent Abell supercluster in the Northern 2dF survey is the
supercluster SCL126 that lies in the direction towards the Virgo
constellation
at a redshift $z = 0.085$ (Paper I).
This supercluster contains 7 Abell clusters: A1620, A1650,
A1651, A1658, A1663, A1692, A1750. Of these clusters, A1650, A1651, A1663, and
A1750 are X-ray clusters. The supercluster SCL126 is almost completely covered
by the 2dF survey volume; only a small part of it lies outside the survey
\citet{e2003}. All Abell and X-ray clusters from this superclusters (9 and 6,
correspondingly) are located in the region of the survey.

The richest supercluster in the Southern Sky is the Sculptor 
supercluster
at a redshift $z = 0.113$ (Paper I).
This supercluster contains 
also several X-ray clusters, and the largest number of Abell clusters in 
our supercluster sample, 25. However, only 12 of them are located within the 
region covered by the 2dF supercluster S34.  These are: A88, A122, 
A2751, A2759, A2778, A2780, A2794, A2798, A2801, and A2844, and two X-
ray clusters, A2811 and A2829.

If we assume that the mean mass-to-light ratio is about 400 (in solar units), 
then we can estimate the masses of the richest superclusters of our sample,
using the estimates of the total luminosity of the superclusters
(Table~\ref{tab:1}): $M_{SCL9} = 2 \times 10^{16} h^{-1} M_{\sun}$,
$M_{SCL126} = 1.5 \times 10^{16} h^{-1} M_{\sun}$.  These estimates are lower
limits only, since the 2dF survey does not fully cover these
superclusters. These masses are of the same order as the masses of other known
very rich superclusters. For example, \citet{proust06} estimate that the total
mass of the Shapley supercluster is at least $M_{tot} = 5\times 10^{16} h^{-1}
M_{\sun}$.  \citet{fleenor05} give the same estimate for the total mass of the
Horologium--Reticulum supercluster. \citet{pr05} estimated that the total mass
of the Pisces-Cetus supercluster is at least $M_{tot} = 1.5 \times 10^{16}
h^{-1} M_{\sun}$.  Thus our two superclusters have total masses similar to
other rich superclusters.

\subsection{Populations of galaxies used in the present analysis}

We characterize galaxies by 
their luminosity, the spectral parameter $\eta$ and by the color index 
$col$ \citep{ma02} and \citep{ma03a,depr03,cole05} as follows.

We divided galaxies into the populations of bright (B )and faint (F) galaxies, 
using an absolute magnitude  limit $M_{bj} = -20.0$ (here and in the following, 
we give absolute magnitudes for $h=1, \quad H=100$ km/sec/Mpc; for any other 
value of $h$, the absolute magnitude $M=M_{h=1}+5\log h$). We chose this limit 
close to the characteristic luminosity $M^*$ of the Schechter luminosity 
function. The value of $M^\star$ is different for different galaxy populations 
\citet{ma03a,depr03,cr05}, having values from $-19.0$ to $-20.9$. We used an 
absolute magnitude limit $M_{bj} = -20.0$ as a compromise between the different 
values. 

The absolute magnitude limits for the faintest galaxies  in our superclusters 
(volume limited samples, $M_{lim}$) are given in Table~\ref{tab:1}. 
Actually, our faint galaxy population is brighter
than the faint galaxies analysed in several other 
superclusters, e.g. in the Shapley supercluster, 
\citep{mercurio06,hai06}, in the supercluster A2199 (a part
of the Hercules supercluster SCL160 in E01 list) in \citet{h219906},
and in the supercluster A901/902 \citep{gray04}. 

We used the spectral parameter $\eta$ to divide galaxies into the 
populations of early (E) and late (S) type. We used for this purpose the 
spectral parameter limit $\eta = -1.4$: $\eta \leq -1.4$ for early type 
galaxies, and  $\eta > - 1.4$ for late type galaxies, and excluded 
galaxies with uncertain determination of $\eta$. More detailed 
morphological types were defined as follows: Type 1 \citep{ken92}: $\eta 
< -1.4$; Type 2: $-1.4 \leq \eta < 1.1$; and Type 3: $1.1 \leq \eta$. 

Moreover, the spectral parameter $\eta$ is correlated with the equivalent width 
of the $H_{\alpha}$ emission line, thus being an indicator of the star formation 
rate in galaxies. 
\citet{ma03b} calls galaxies with $\eta > -1.4$ (late type galaxies) as "generally 
star-forming".  Thus the spectral parameter gives us 
information on the types and star formation activity of galaxies.

{\scriptsize
\begin{table}[ht]
\caption{
The superclusters SCL126 and SCL9:
the number of galaxies in different galaxy populations,
for the whole supercluster (volume limited samples), and 
for the core region ($D1$) and for the outskirts of the superclusters
(D2)
}
 
\begin{tabular}{rllllllll} 
\hline 
(1)&(2)&(3)&(4)&    &(5)&(6)&(7)& (8) \\      
\hline 
           &          \multispan3  SCL126  & &            & \multispan3  SCL9  \\        
Region     & All& $M_{bj} \leq -19.50$ & $D1$ &  $D2$   &              &    All     &   $D1$ &  $D2$  \\        
\\
N$_{gal}$  &1308 & 932& 488    &  820  &                   &1176& 342    &  834  \\                          
N$gr_{10}$ & 405 & 308& 227    &  181  &                   & 247& 137    &  108  \\                         
N$gr_2$    & 576 & 410& 172    &  410  &                   & 576& 137    &  442  \\                         
~N$ig$     & 327 & 214&  89    &  229  &                   & 353&  68    &  284  \\                         
\\                                                                                                 
$E$        & 809 & 603& 340    &  469  &                   & 772& 236    &  536  \\                      
$S$        & 490 & 322& 145    &  345  &                   & 393& 102    &  291  \\                      
\\                                                                                              
$r$        & 937 & 685& 377    & 560   &                   & 835& 251    & 584   \\                      
$b$        & 371 & 247& 111    & 260   &                   & 341&  91    & 250   \\                      
\\                                                                                           
$B$        & 400 & 400& 144    & 256   &                   & 556& 168    & 388   \\                      
$F$        & 908 & 532& 344    & 564   &                   & 620& 174    & 446   \\                      
\label{tab:s1269d1d2ngal}                        
\end{tabular}
\tablecomments{
Columns in the Table are as follows:
\noindent column 1: Population ID (Sec. 2.2),
N$_{gal}$ -- the number of galaxies in a given
density region, N$gr_2$ -- the number of galaxies in poor groups
with 2--9 member galaxies, N$gr_{10}$ -- the number of galaxies
in rich groups with at least 10 member galaxies. 
N$ig$ -- the number of galaxies which do not belong to groups. 
\noindent Columns 2--4 and 6--8: the numbers of galaxies
in a given population, in the superclusters SCL126 and SCL9, correspondingly. 
}
\end{table}
}

We used the rest-frame color index, $col = (B - R)_0$, to divide galaxies 
into the populations of red galaxies (r), $col \geq 1.07$, 
and blue galaxies (b), $col < 1.07$ 
\citep{cole05,wild05}. \citet{wild05} suggest that red galaxies 
are mostly passive, and blue galaxies -- actively star forming. 
However, since there exist also red galaxies showing 
signs of star formation \citep{wolf05,haines07} we name our populations as red 
and blue. 

In the next section we will analyse the relationship between
the spectral parameters and color 
indexes of galaxies in our superclusters   in more detail.

In our analysis we use following ratios:
$E/S$ -- the ratio of the numbers of early and late type galaxies,
$r/b$ -- the ratio of the numbers of red and blue galaxies, and
$B/F$ -- the ratio of the numbers of bright 
($M_{bj}\leq -20.0$) and faint ($M_{bj} > -20.0$) galaxies.

\section{Galaxy content of rich superclusters}

\begin{figure*}[ht]
\centering
\resizebox{0.30\textwidth}{!}{\includegraphics*{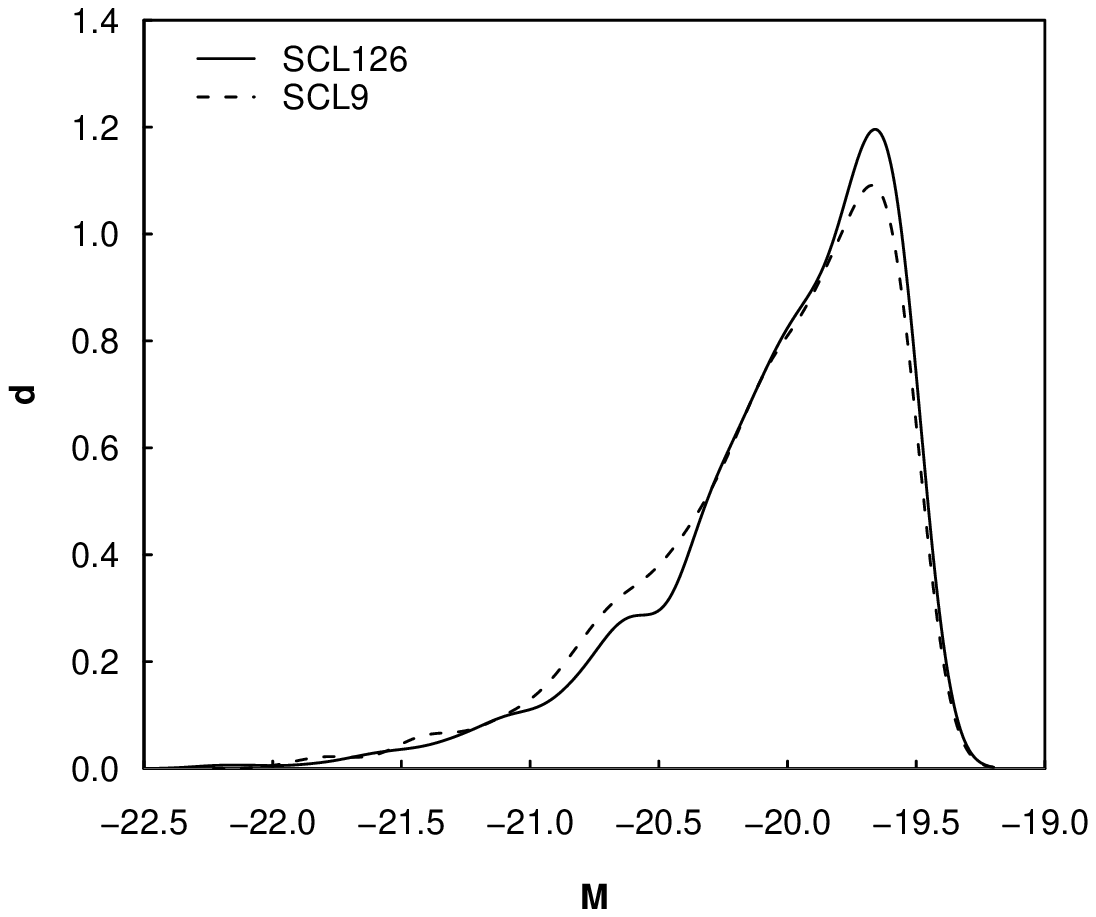}}
\resizebox{0.30\textwidth}{!}{\includegraphics*{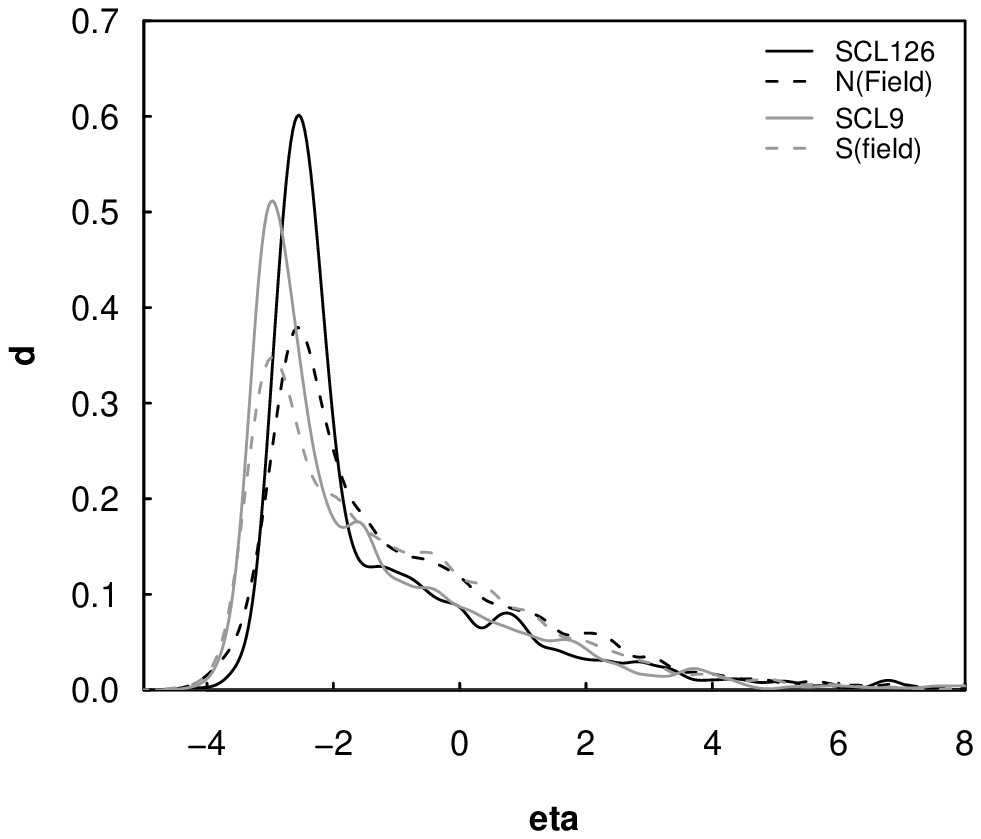}}
\resizebox{0.30\textwidth}{!}{\includegraphics*{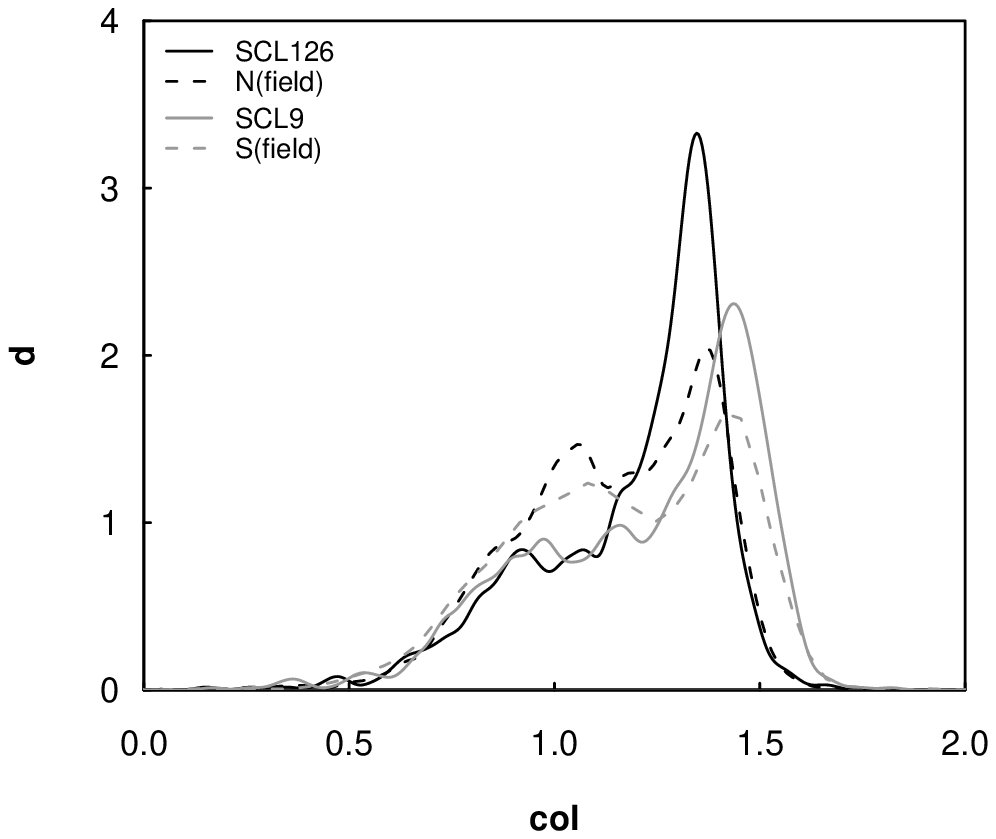}}
\caption{The distribution of the absolute magnitude (left panel),
the spectral parameter $\eta$ (middle panel) and
 the color index $col$ (right panel) of galaxies 
in  the supercluster SCL126 (solid line), and in SCL9 (dashed line), 
both for the absolute magnitude limit $M_{bj}\leq -19.50$). 
}
\label{fig:r2all}
\end{figure*}

First we study an overall galaxy content of the rich superclusters (see Table~\ref{tab:s1269d1d2ngal}) . We plot
the differential luminosity functions, and the distributions of the spectral
parameter $\eta$ and the color index $col$ for galaxies from superclusters
SCL126 and SCL9 in Fig.~\ref{fig:r2all}. 
For comparison we plot also the distribution of the spectral
parameter $\eta$ and the color index $col$ for field galaxies, chosen from the same
redshift interval as our two superclusters, and having the same absolute magnitude
limit ($M_{bj} = -19.50$, see below). There are 1975 galaxies
in the Northern field sample, $N(field)$, and 2927 in the Southern field
sample, $S(field)$. The number of galaxies in field samples is rather small
due to absolute magnitude limit used here, since, in general, galaxies
in the field are fainter than in superclusters (Paper III).
The comparison with field galaxies shows whether the possible differences between 
the distributions of spectral parameters and colors may be due to redshift 
differences between the two superclusters.

All the distributions (probability
densities) shown in this paper have been obtained using the R environment
\citep{ig96}, \texttt{http://www.r-project.org} (the 'stats' package).
This package does not provide the customary error limits; we discuss these
limits in the appendix and show that they are small (about 4--5\%).

In Table~\ref{tab:r2overall} we give the ratios of the numbers of galaxies of 
different type in the superclusters; for SCL126 for two magnitude limits, the 
original (Table ~\ref{tab:1}), and the SCL9 magnitude limit $M_{bj} = -19.50$, 
in order to compare these ratios with those calculated for the supercluster 
SCL9. 

The Kolmogorov-Smirnov test, Fig.~\ref{fig:r2all}, and 
Table~\ref{tab:r2overall} show  differences in the overall 
galaxy content of the two richest 2dFGRS superclusters. 

The Kolmogorov-Smirnov test shows that the probability that
the distributions of luminosities of the two superclusters
are drawn from the same parent sample is 0.135,
having marginal statistical significance. The most important difference
between the luminosities of the galaxies of two superclustes is that
the brightest galaxies in SCL126 are brighter than those in SCL9. 
In Sect. 5 we show that there are also other differences in the distribution
of the brightest galaxies in the two superclusters under study.

{\scriptsize
\begin{table}[ht]
\caption{The galaxy content of the superclusters SCL126 and SCL9. }
\begin{tabular}{lrrr} 
\hline 
Sample     & \multispan2 SCL126 &  SCL9\\      
           &         &          &       \\
\hline 
(1)&(2)&(3)&(4)\\
\hline 
$M_{bj^{lim}}$    & -19.25&  -19.50&    -19.50\\
N$_{gal}$      & 1308  &  932   &      1176\\
F$gr_{10}$     & 0.31  &  0.33  &      0.21\\
F$gr_2$        & 0.44  &  0.44  &      0.49\\
F$ig$          & 0.25  &  0.23  &       0.30\\
~~$E/S$        & 1.65  &  1.87  & 1.96\\
~~$r/b$        & 2.53  &  2.77  & 2.45\\
~~$B/F$        & 0.44  &  0.75  & 0.90\\
\label{tab:r2overall}                        
\end{tabular}
\tablecomments{
The columns in the Table are as follows:
\noindent column 1: Population ID and number ratio type. 
N$_{gal}$ -- the number of galaxies, F$gr_{10}$ -- the fraction of galaxies
in rich groups with at least 10 member galaxies,
F$gr_2$ -- the fraction of galaxies in poor groups with 2--9 member galaxies,  
F$ig$ -- the fraction of galaxies which do not belong to any group. 
\noindent Columns 2--3: absolute magnitude limits and 
the ratio of the numbers of galaxies
in given populations (Sec. 2.2) in the supercluster SCL126. $-19.25$ -- 
galaxies brighter than $M_{bj}\leq -19.25$,
$-19.50$ -- galaxies brighter than $M_{bj}\leq -19.5$.
Column 4: the same ratios 
in similar populations in the supercluster SCL9.
}
\end{table}
}

Next, Table~\ref{tab:r2overall} shows that the ratio of
the numbers of red and blue galaxies ($r/b$)
is slightly larger in the supercluster
SCL126, but the peak value of the color index of galaxies in this
supercluster is smaller than in SCL9. 
The distributions of the spectral parameter $\eta$ and the
color index $col$ both show large differences in the early type (red) galaxy
content of the richest superclusters -- in the supercluster SCL9, galaxies
have larger color index values and smaller spectral parameter values than in
supercluster SCL126.  

According to the Kolmogorov-Smirnov test, the difference between the 
distributions of the spectral parameter $\eta$ and the color index $col$ 
of the two superclusters is highly significant.
However, the comparison with the field galaxies of the same distance interval 
and absolute magnitude limit as the galaxies from superclusters, shows that the 
main reason for the differences of spectral parameters and colors is the 
difference in their distances (Fig.~\ref{fig:r2all}). 
This effect is difficult to quantify exactly, as the color distributions
for supercluster and field galaxies differ in detail, but are close in the
average. We illustrate their similarity by comparing in
Table~\ref{tab:r2red} the quartiles 
for the distributions of the spectral parameter $\eta$ and the color 
index $col$, 
for parameter intervals which correspond to early type and red galaxies ($\eta \le 
-1.4 $ and $col \geq 1.07$, correspondingly), for the absolute magnitude limit 
$M_{bj} = -19.50$. These quartiles are very similar for the supercluster
and field galaxies.

{\scriptsize
\begin{table}[ht]
\caption{The early type and red galaxy content of the superclusters
SCL126 and SCL9. }
\begin{tabular}{lrrrr} 
\hline 
SCL          & $N{gal}$ & lower quartile    &  median   & upper quartile\\      
\hline 
(1)&(2)&(3)&(4)&(5)\\
\hline 
$\eta < -1.4 $&          &  &   &     \\
SCL126       & 603 & -2.72 & -2.49  &  -2.19 \\
N(Field)     & 1053 & -2.80 & -2.47  &  -2.05 \\
\\
SCL9         & 772 & -3.08 & -2.78  &  -2.30 \\
S(Field)     & 1583 & -3.08 & -2.71  &  -2.11 \\
\\
$col > 1.07$ &  &       &        &        \\
SCL126       & 685 &  1.23 &  1.32  &   1.37 \\
N(Field)     & 1235 &  1.19 &  1.30  &   1.37 \\
\\
SCL9         & 835 &  1.27 &  1.39  &   1.47 \\
S(Field)     & 1862 &  1.21 &  1.36  &   1.45 \\

\label{tab:r2red}                        
\end{tabular}
\tablecomments{
The columns in the Table are as follows:
\noindent column 1: Population ID. 
\noindent Column 2: The number of galaxies in a given population. 
\noindent Columns 3--5: The 1st quartile value, the median and upper
quartile values of the spectral parameter for the $\eta$
interval $\eta < -1.4 $ and of the color index $col$ for
the interval $col > 1.07$ in the superclusters SCL126 and SCL9.
}
\end{table}
}

These differences, however, do not affect our calculations of 
the Minkowski functionals for individual superclusters. 
So there is no need to 
correct the colors and spectral parameters for this effect. 

\begin{figure*}[ht]
\centering
\resizebox{0.45\textwidth}{!}{\includegraphics*{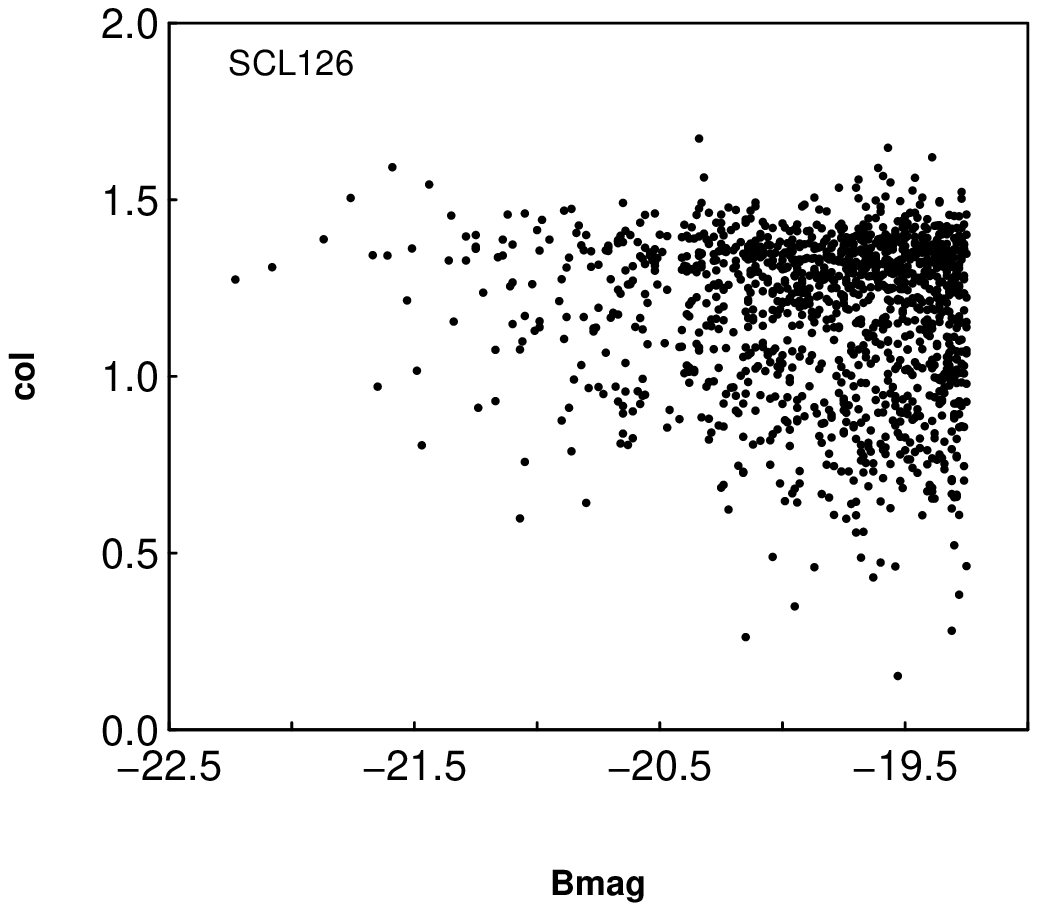}}
\resizebox{0.45\textwidth}{!}{\includegraphics*{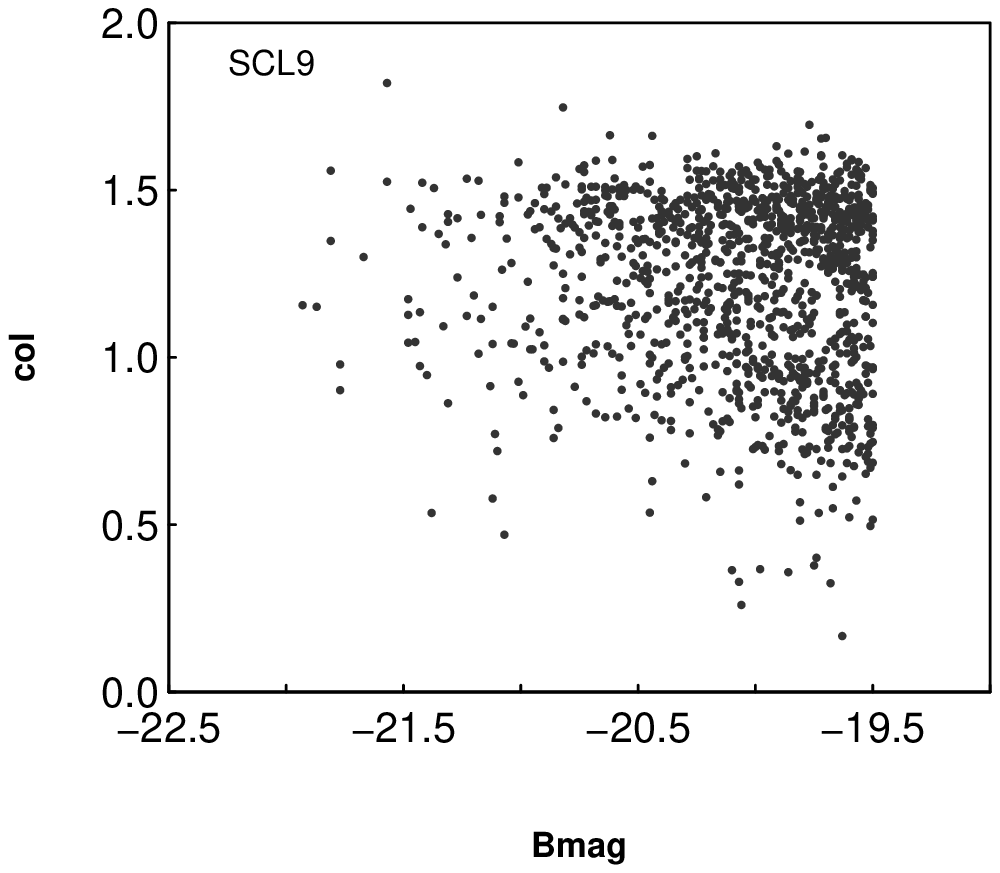}}
\caption{The color-magnitude diagrams for the superclusters
SCL126 (left panel) and SCL9 (right panel). 
}
\label{fig:r2cm}
\end{figure*}

Next we present the color-magnitude diagrams for our two superclusters under 
study (Fig.~\ref{fig:r2cm}). Interestingly, this figure shows that there is no 
correlation between the color and magnitude
among the red galaxies in our superclusters. In this respect 
the color-magnitude diagrams for the superclusters SCL126 and SCL9
are different from those for some other 
superclusters, e.g., for the supercluster A901/902 \citep{gray04} and for the 
core  of the Shapley superclusters \citep{hai06}. 
The reasons for that are probably the use of total magnitudes in our study 
\citep[see also][]{cross01}, in this case the color-magnitude relation is
almost flat \citep{scodeggio01}, and the use of the data about relatively
bright galaxies only.
The studies by \citep{gray04,hai06} include 
much fainter galaxies than those used in our study; for fainter galaxies
the slope of the color-magnitude diagram is larger
than for bright galaxies \citep[see also][]{metcalfe94}
even when using total magnitudes, as mentioned by \citet{scodeggio01}.   
Thus Fig.~\ref{fig:r2cm}
suggests that we can use galaxy colors for classification without further
corrections.

\begin{figure*}[ht]
\centering
\resizebox{0.45\textwidth}{!}{\includegraphics*{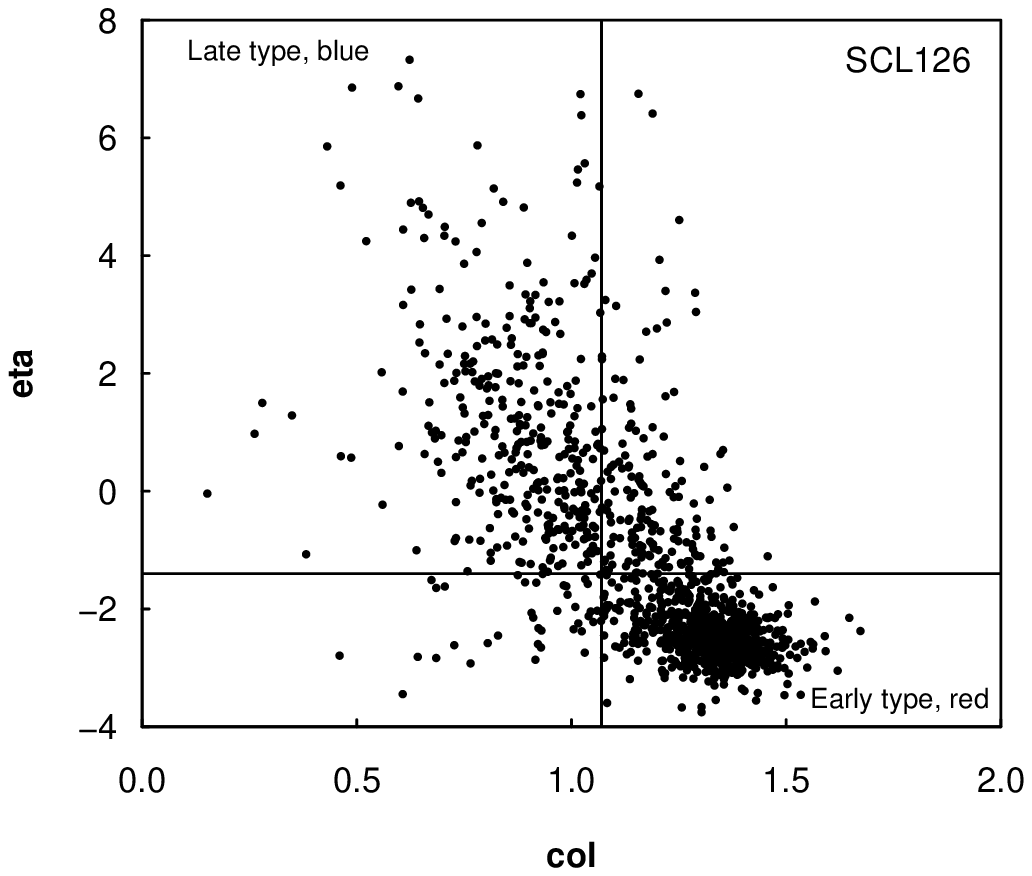}}
\resizebox{0.45\textwidth}{!}{\includegraphics*{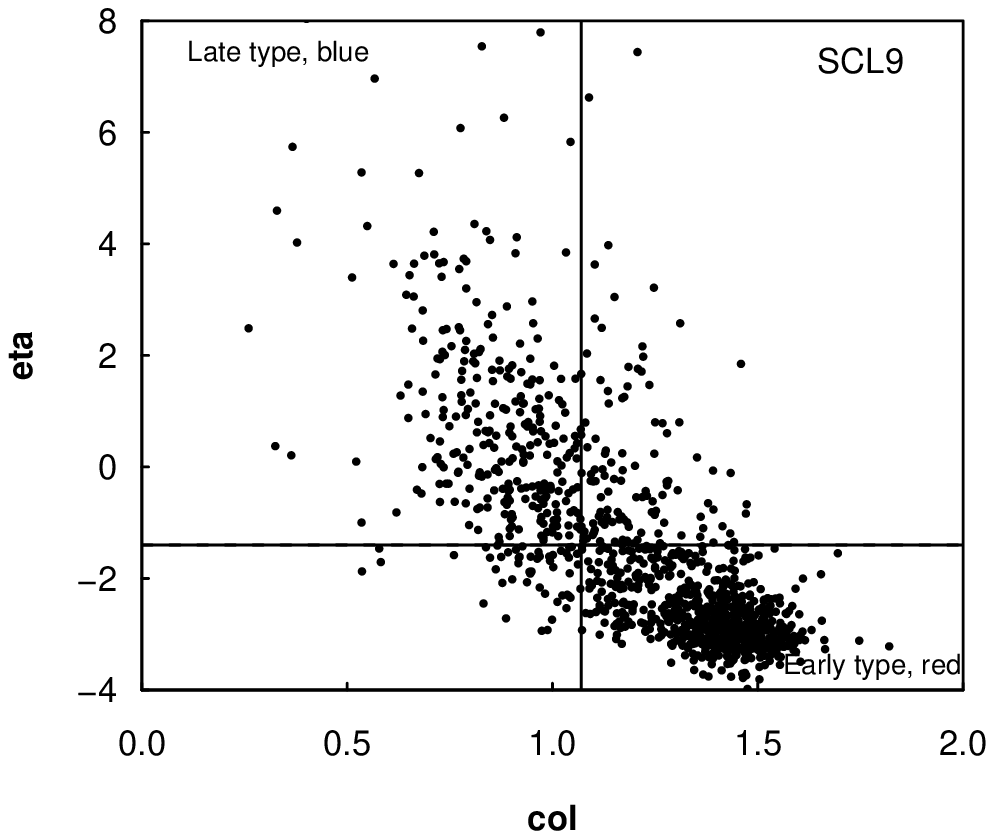}}
\caption{The distribution of 
the spectral parameter $\eta$ and
 the color index $col$ of galaxies 
in  the supercluster SCL126 (left panel), and in SCL9 (right panel). 
The solid lines separate early and late-type galaxies
($\eta = -1.4$), and red and blue galaxies ($col = 1.07$).
}
\label{fig:r2etacol}
\end{figure*}

Finally, we take a closer look to the classification of galaxies by their 
spectral parameters and colors. For that we plot in Fig.~\ref{fig:r2etacol} the 
joint distribution of the spectral parameter $\eta$ and the color index $col$ for 
the superclusters SCL126 and SCL9, in analogy to Fig. 2 in \citet{wild05}. 
We see that most galaxies populate an area at the lower right quadrant, 
where the galaxies  have spectra characteristic to early type galaxies 
(with no star 
formation) and have red colors. The upper left quadrant of this Figure 
is populated by late type, blue galaxies. But there are
also galaxies, which have spectra characteristic to late type ("generally 
star forming") galaxies, and in the same time have red colors. The number of 
these galaxies is small -- 161 in the supercluster SCL126, and 114 in the 
supercluster SCL9.
There are also galaxies with blue colors and spectra characteristic to 
early type galaxies (41 in SCL126 and 62 in SCL9). 
A detailed analysis of the properties of these galaxies is beyond the scope 
of the present paper, but below we discuss where 
these galaxies are located in superclusters and whether the differences between 
the populations of early type and red galaxies (and late type and blue galaxies)
influence the calculation of the Minkowski functionals.

\section{Minkowski Functionals } 
\subsection{Method} 

The supercluster geometry (morphology) is given by their outer
(limiting) isodensity surface, and its enclosed volume. When
increasing the density level over the outer threshold overdensity
$\delta=4.6$ (sect. 2.1), the isodensity surfaces move into the
central parts of the supercluster.  The morphology and topology of the
isodensity contours is (in the sense of global geometry) completely
characterized by the four Minkowski functionals \mbox{$V_0$--$V_3$}.

For a given surface the four Minkowski functionals (from the first to the
fourth) are proportional to: the enclosed volume $V$, the area of the surface
$S$, the integrated mean curvature $C$, and the integrated Gaussian curvature
(or Euler characteristic) $\chi$. The Euler characteristic describes the
topology of the surface; at high densities it gives the number of isolated
clumps (balls) in the region, at low densities -- the number of cavities
(voids) \citep[see, e.g.][]{saar06}. This is the functional we shall use in
this paper (see Appendix for details).
 
We calculate also the shapefinders $H_{1}$ (thickness), $H_{2}$ (width), and
$H_{3}$ (length), which have dimensions of length, and dimensionless
shapefinders $K_1$ (planarity) and $K_2$ (filamentarity) (see Appendix for
definitions). In paper RI we showed that in the $K_1$-$K_2$ shapefinder plane
the morphology of superclusters is described by a curve which is
characteristic of multi-branching filaments; we call it a morphological
signature.
 
In order to estimate the Minkowski functionals, we have to ensure that the
mean density is constant throughout the volume we study. This is the main
reason why we use volume limited samples of supercluster galaxies. This also means
that we have to recalculate the (luminosity) density field; the field used to
select superclusters was calculated on the basis of the full sample.

\begin{figure*}[ht]
\centering
\resizebox{0.45\textwidth}{!}{\includegraphics*{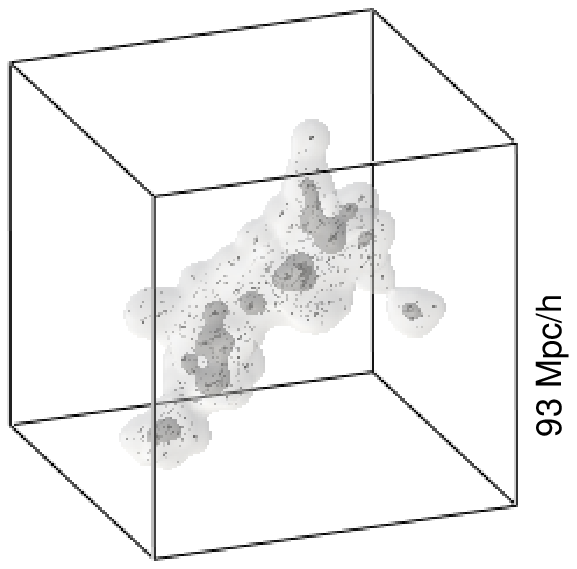}}
\resizebox{0.45\textwidth}{!}{\includegraphics*{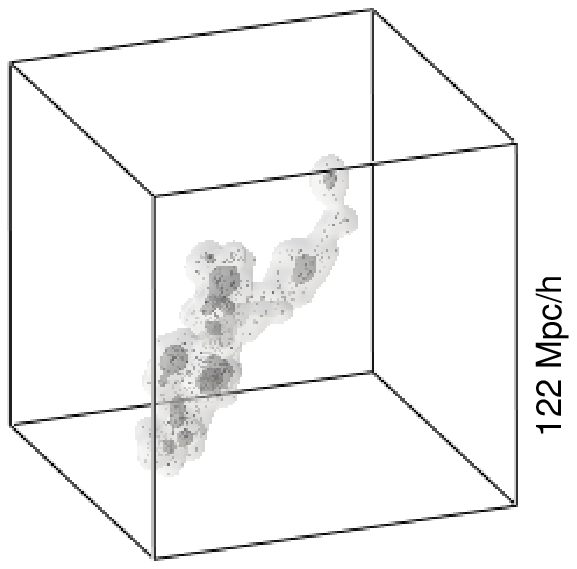}}
\caption{The luminosity density distribution in the two superclusters.
Left panel: SCL126, right panel: SCL9. The isosurfaces show the outer border of
a supercluster and the density level for $m_f=0.7$. Dots show individual
galaxies. 
Darker regions: core (D1), lighter regions - outskirts (D2).}
\label{fig:juhan}
\end{figure*}

To obtain the density field for estimating the Minkowski functionals, we used
a kernel estimator with a $B_3$ box spline as the smoothing kernel, with the
total extent of 16~\Mpc\ 
\citep[for a detailed description see][and Paper RI]{saar06}.
This kernel covers exactly the 16~\Mpc\ extent of the Epanechnikov
kernel, used to obtain the original density field, but it is smoother and
resolves better density field details (its effective width is about 8~\Mpc).
The density field for both superclusters is shown in Fig.~\ref{fig:juhan}.

As the argument labeling the isodensity surfaces, we chose the (excluded) mass
fraction $m_f$ -- the ratio of the mass in regions with density {\em lower}
than the density at the surface, to the total mass of the supercluster. When
this ratio runs from 0 to 1, the iso-surfaces move from the outer limiting
boundary into the center of the supercluster, i.e. the fraction $m_f=0$
corresponds to the whole supercluster, and $m_f=1$ to its highest density
peak. This is the convention adopted in all papers devoted to the morphology
of the large-scale galaxy distribution.

At small mass fractions the isosurface includes the whole supercluster and the 
value of the 4th Minkowski functional $V_3 = 1$.  As we move to higher mass 
fractions, the isosurface includes only the higher density parts of 
superclusters. Individual high density regions in a supercluster, which at low 
mass fractions are joined together into one system, begin to separate from each 
other, and the value of the fourth Minkowski functional ($V_3$) increases. At a 
certain density contrast (mass fraction) $V_3$ has a maximum, showing the 
largest number of isolated clumps in a given supercluster. At still higher 
density contrasts only the high density peaks contribute to the supercluster and 
the value of $V_3$ decreases again.

In Paper RI we showed that according to Minkowski functionals, the
supercluster SCL126 resembles a multibranching filament with a high density
core at a mass fraction $m_f \approx 0.95$. The maximum value of the fourth
Minkowski functional, $V_3$, in this supercluster is 9. The main body of this
filament is seen in Fig.~\ref{fig:radecx} (region $D1$, see below for
definition).
 
The supercluster SCL9 can be described as a multispider rather than a 
rich filament, i.e. this supercluster consists of a large number of 
relatively isolated clumps or cores connected by relatively thin 
filaments, in which the density of galaxies is too low to contribute to 
the higher density parts of this supercluster. The maximum value of the 
fourth Minkowski functional, $V_3$, in this supercluster is 15, i.e.
this supercluster is more clumpy than the supercluster SCL126. The main 
cores of this supercluster are seen in Fig.~\ref{fig:radecx} as region 
$D1$. 

\begin{figure*}[ht]
\centering
\resizebox{0.28\textwidth}{!}{\includegraphics*{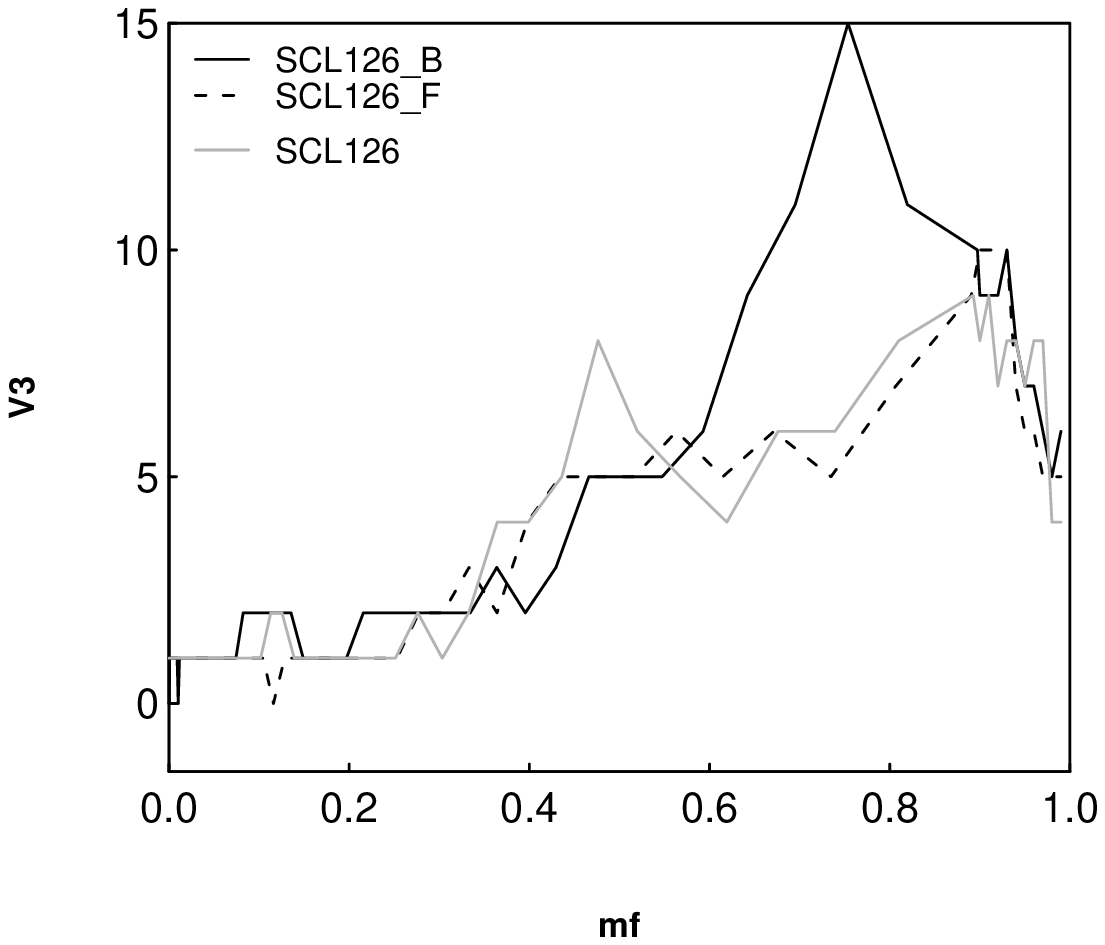}}
\resizebox{0.28\textwidth}{!}{\includegraphics*{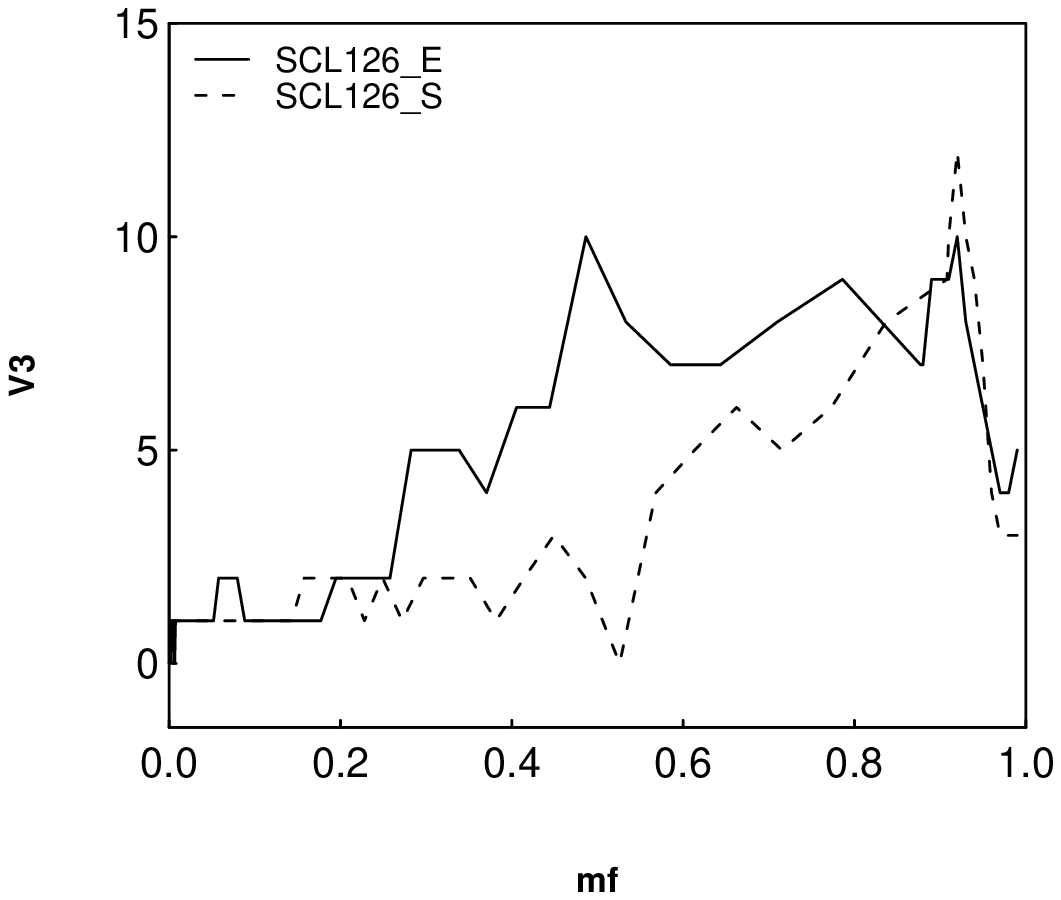}}
\resizebox{0.28\textwidth}{!}{\includegraphics*{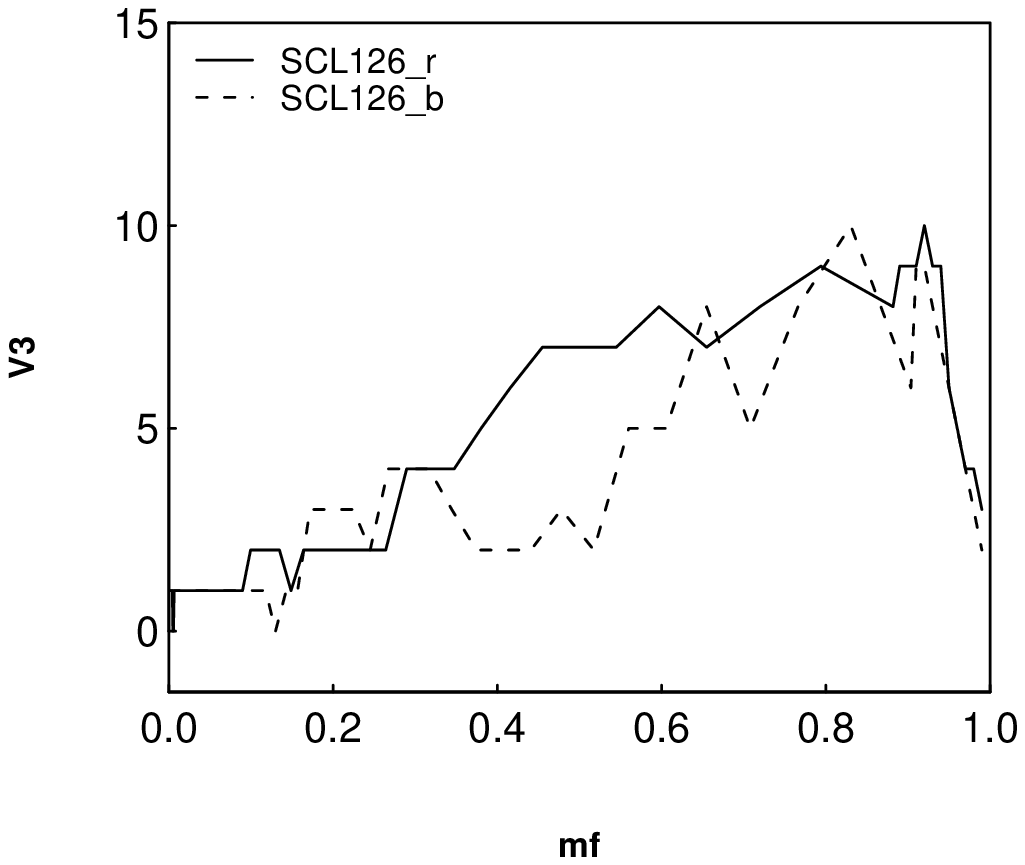}}
\hspace*{2mm}\\
\resizebox{0.28\textwidth}{!}{\includegraphics*{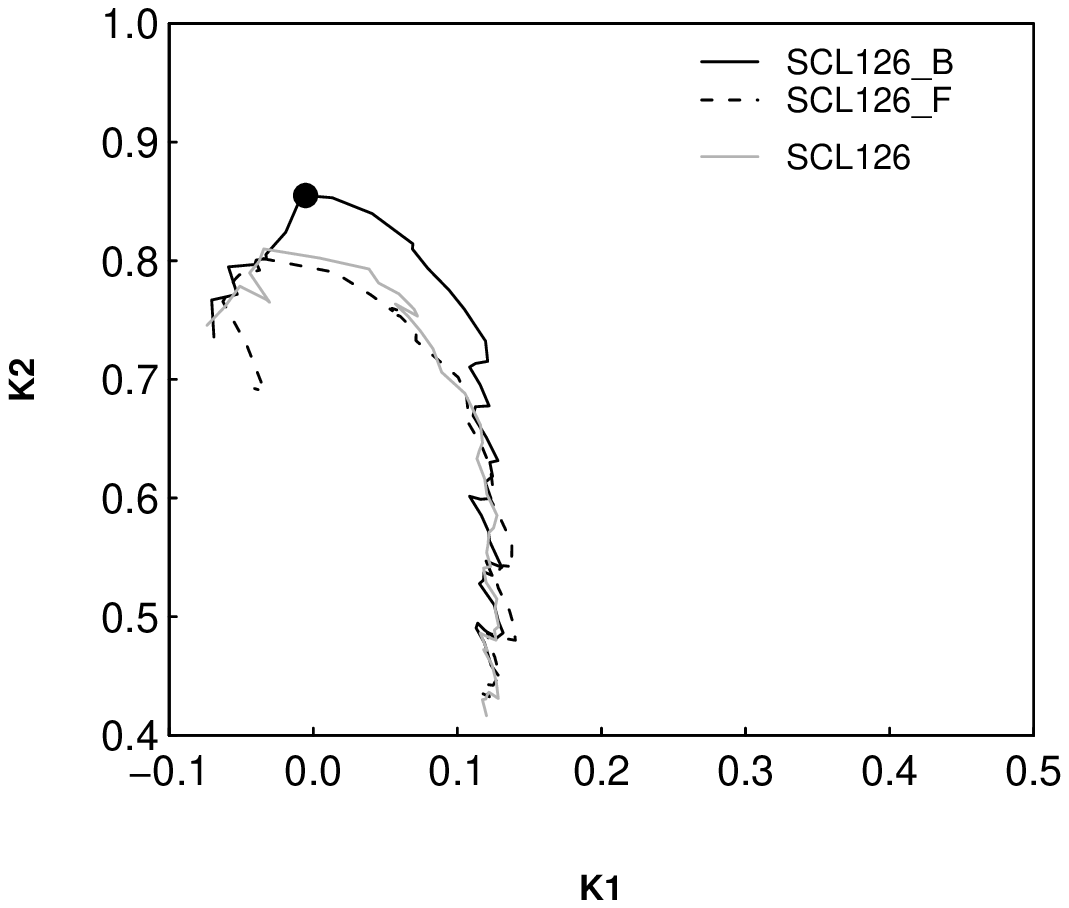}}
\resizebox{0.28\textwidth}{!}{\includegraphics*{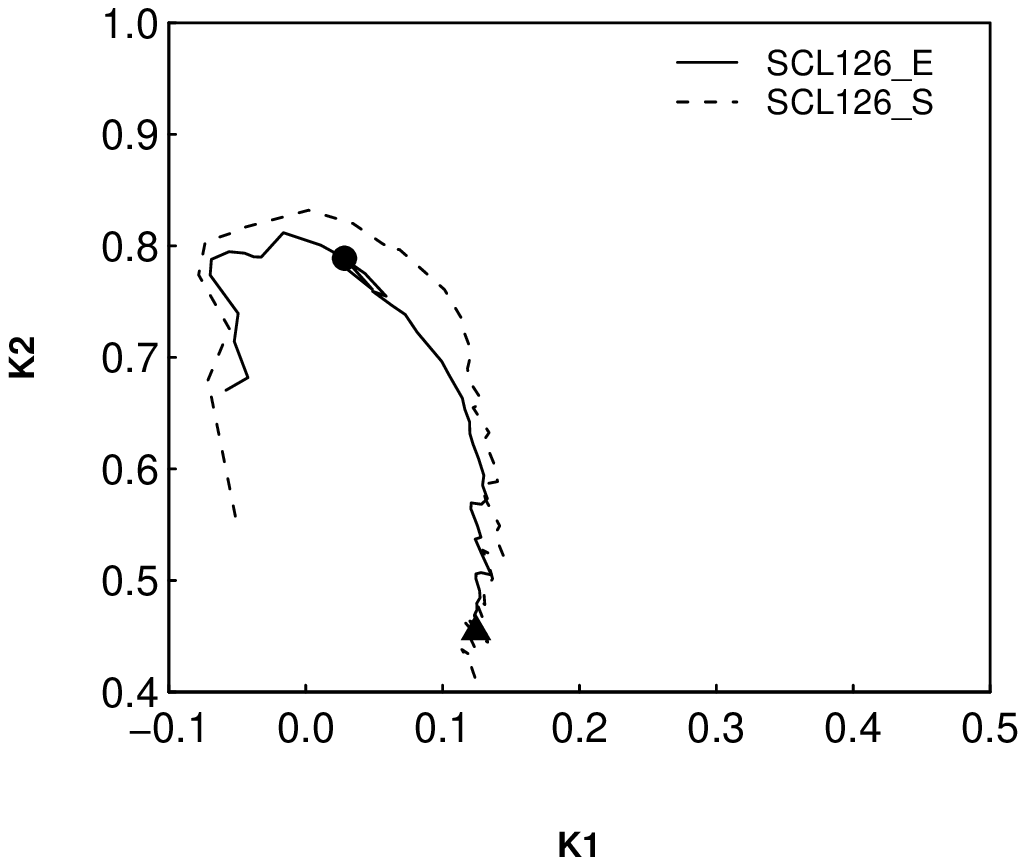}}
\resizebox{0.28\textwidth}{!}{\includegraphics*{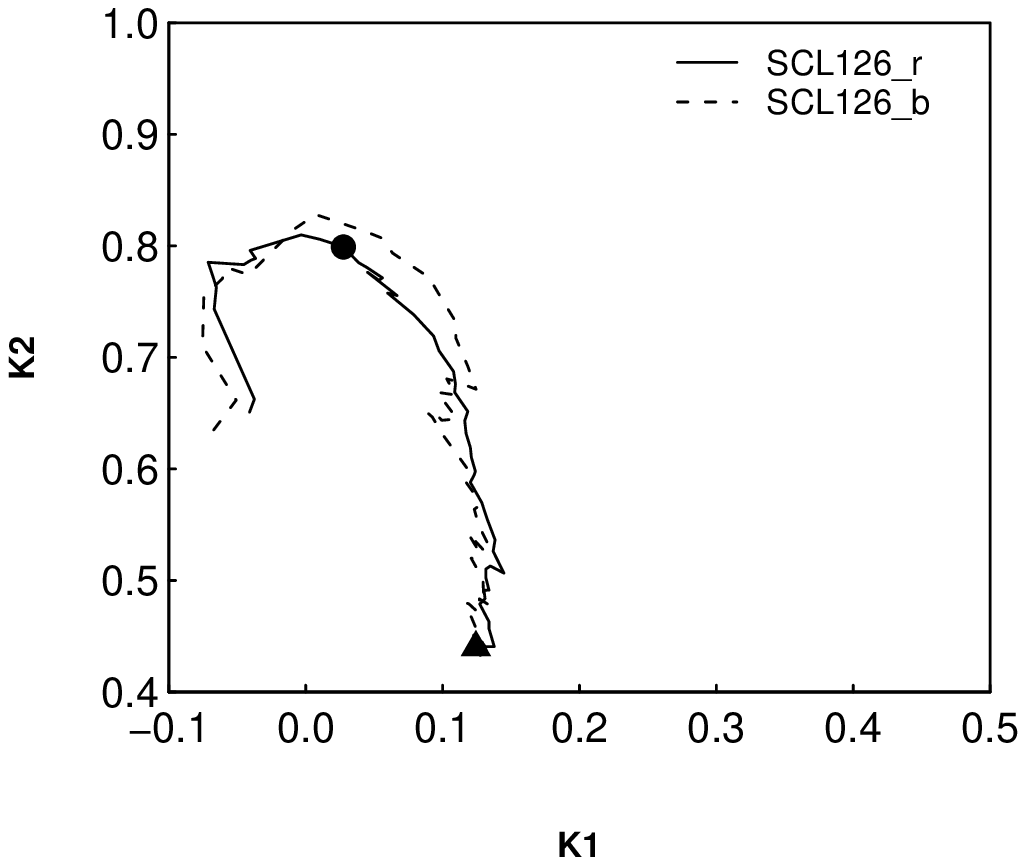}}
\caption{The supercluster SCL126. Upper panels:
the Minkowski functional $V_3$ (the Euler characteristic)
vs the mass fraction $m_f$ for bright (B, $M \leq -20.0$) and 
faint (F, $M > -20.0$) galaxies (left panel), for early and late 
type galaxies (middle panel) and for red and blue galaxies
(right panel). Lower panels: the morphological signatures
$K_{1}$ (planarity) -- $K_{2}$ (filamentarity) for the same populations. 
Triangles show the values of $K_1,~K_2$,
where the mass fraction $m_f = 0.0$ (the whole supercluster), 
and  filled circles -- the values of $K_1,~K_2$, which correspond to
  the $m_f = 0.7$.  
In the left panels we also plot the $V_3$ and  $K_1,~K_2$ curves for the whole
supercluster (light grey line).  
}
\label{fig:v3k12s126}
\end{figure*}

\begin{figure*}[ht]
\centering
\resizebox{0.28\textwidth}{!}{\includegraphics*{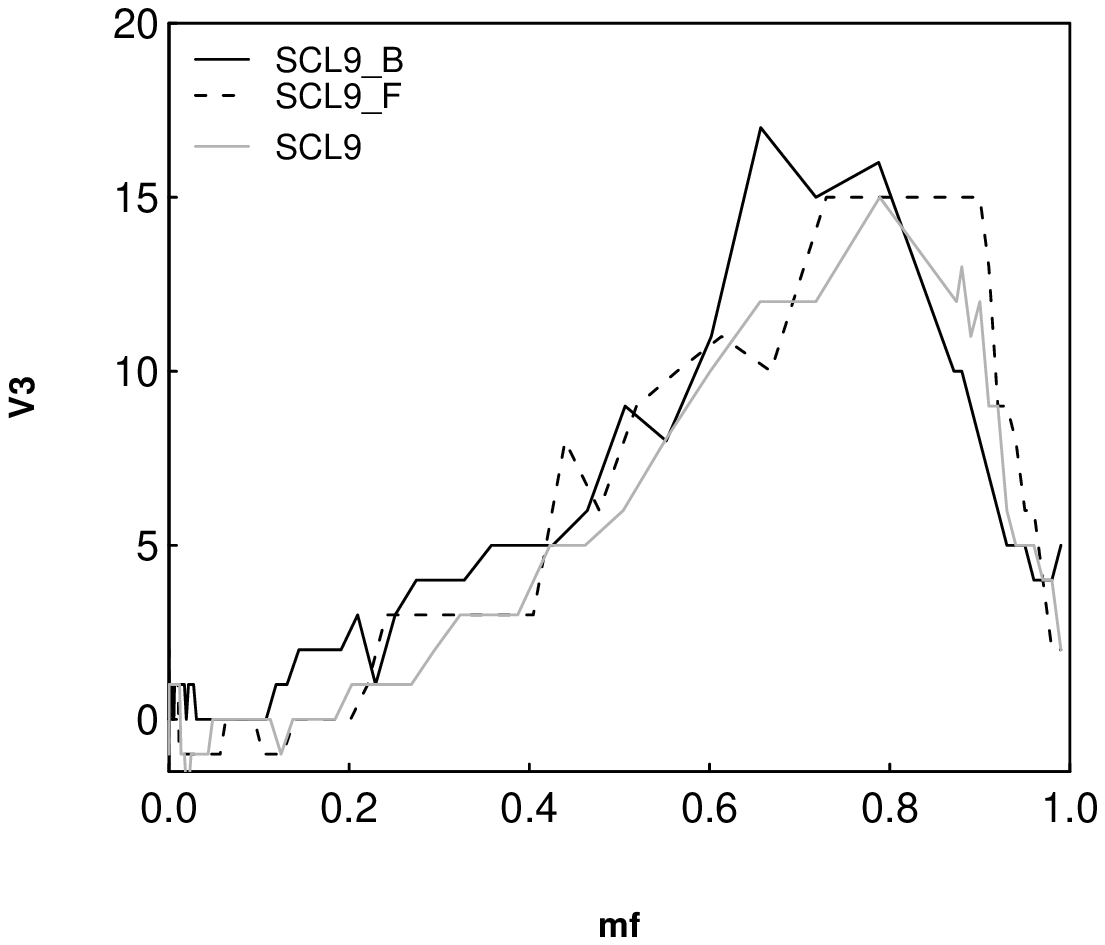}}
\resizebox{0.28\textwidth}{!}{\includegraphics*{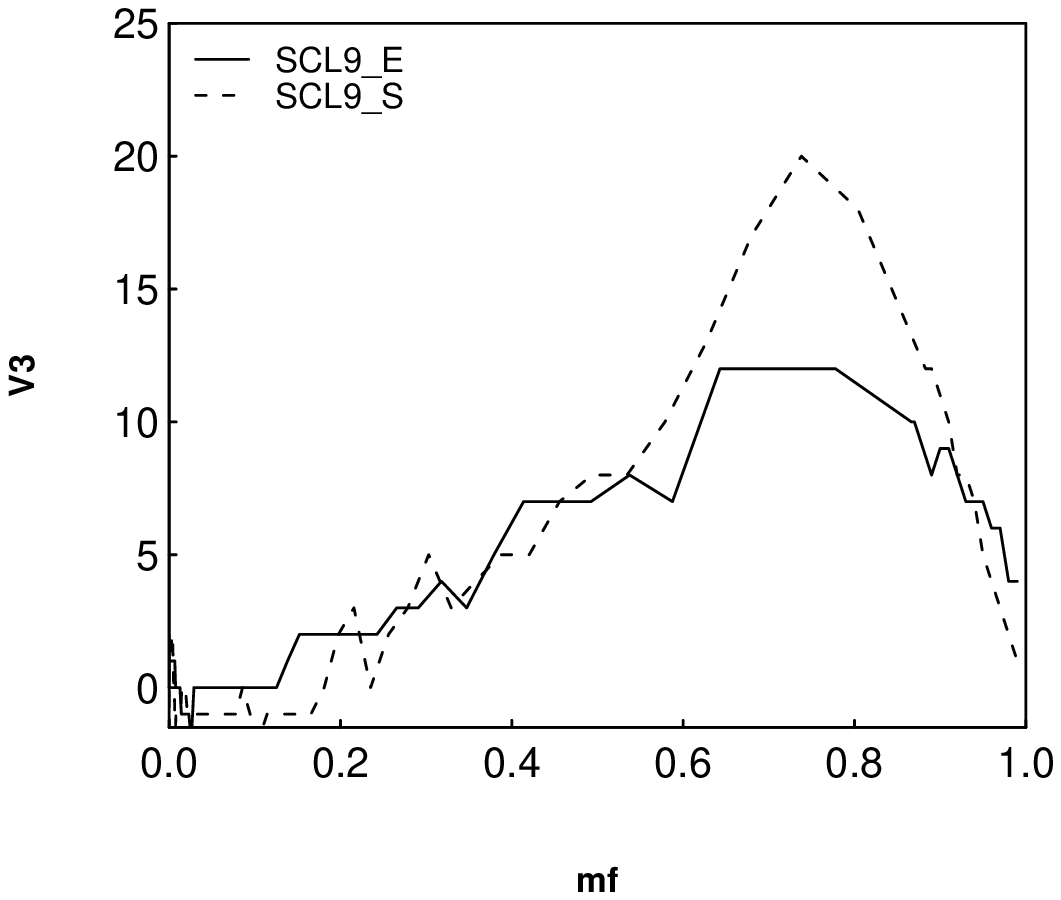}}
\resizebox{0.28\textwidth}{!}{\includegraphics*{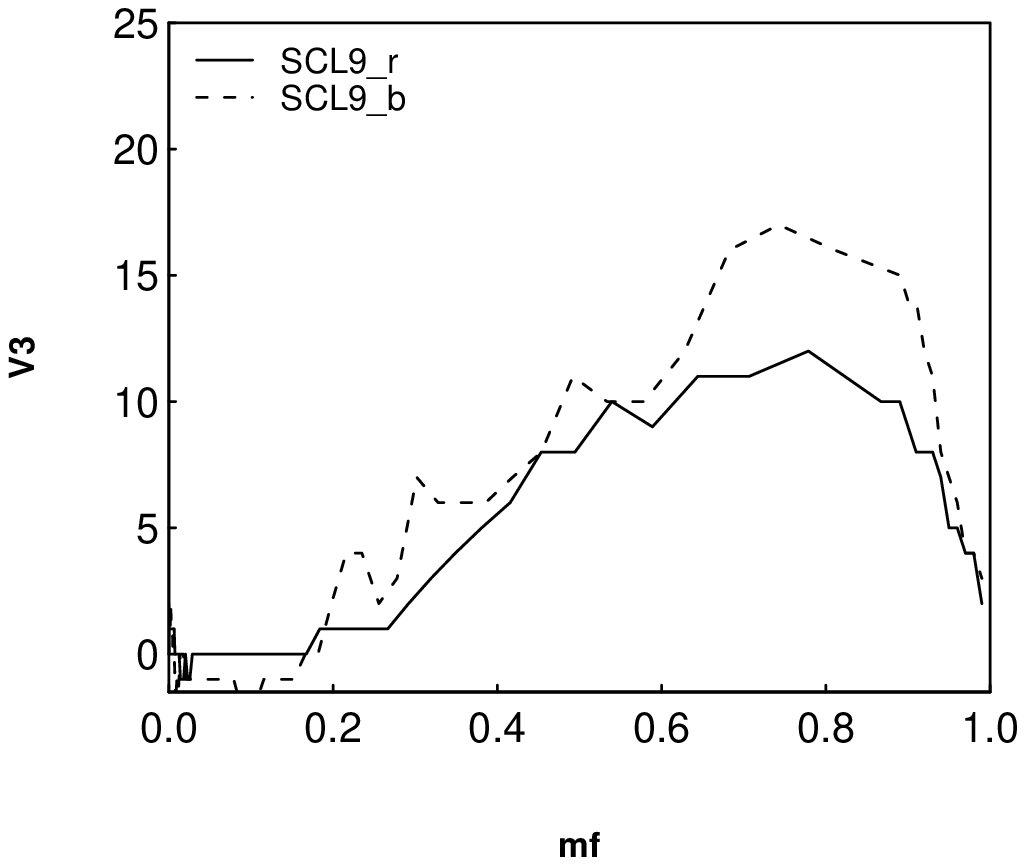}}
\hspace*{2mm}\\
\resizebox{0.28\textwidth}{!}{\includegraphics*{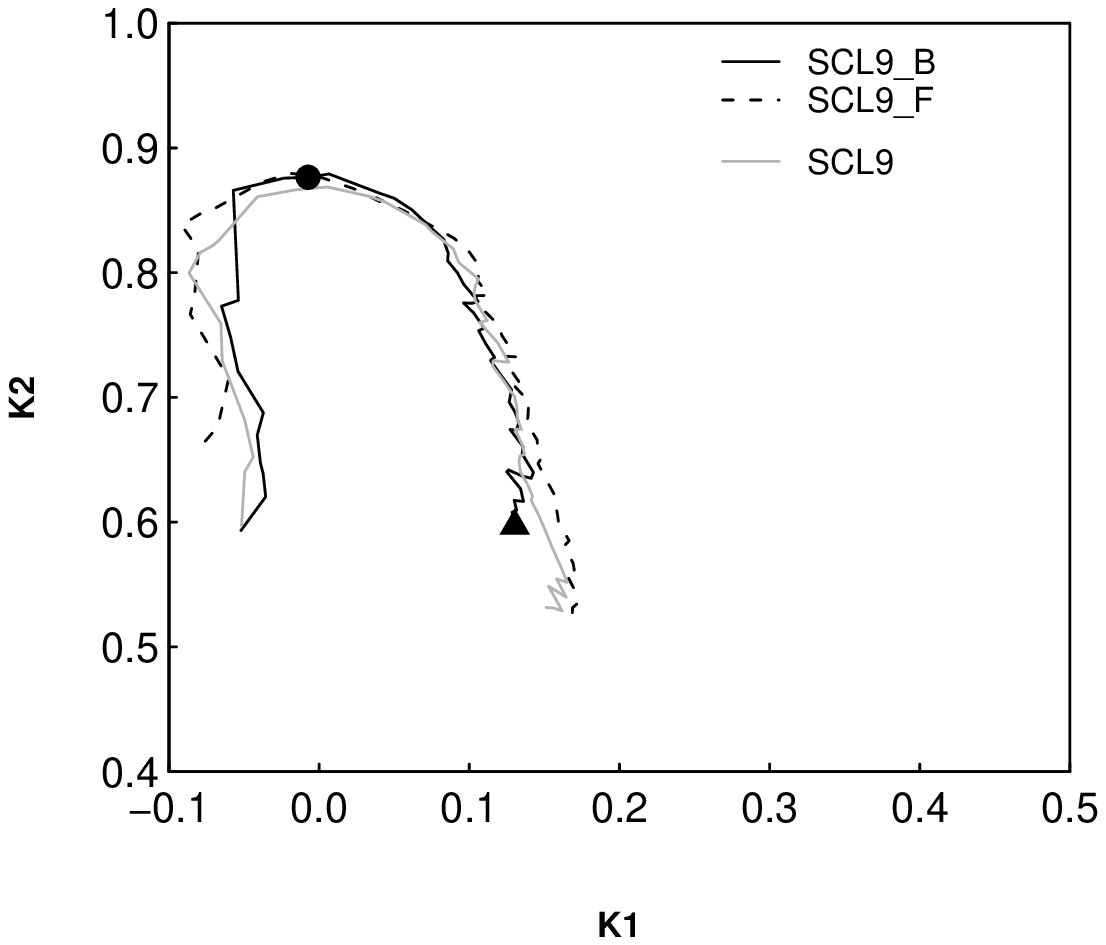}}
\resizebox{0.28\textwidth}{!}{\includegraphics*{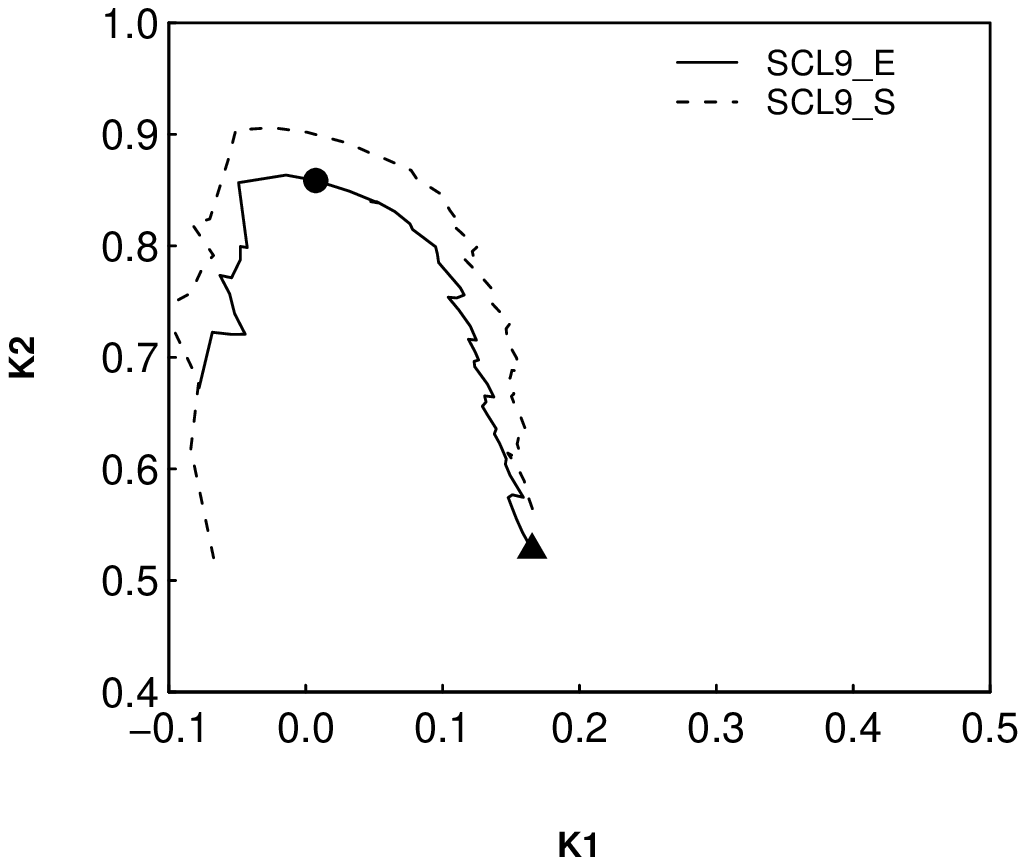}}
\resizebox{0.28\textwidth}{!}{\includegraphics*{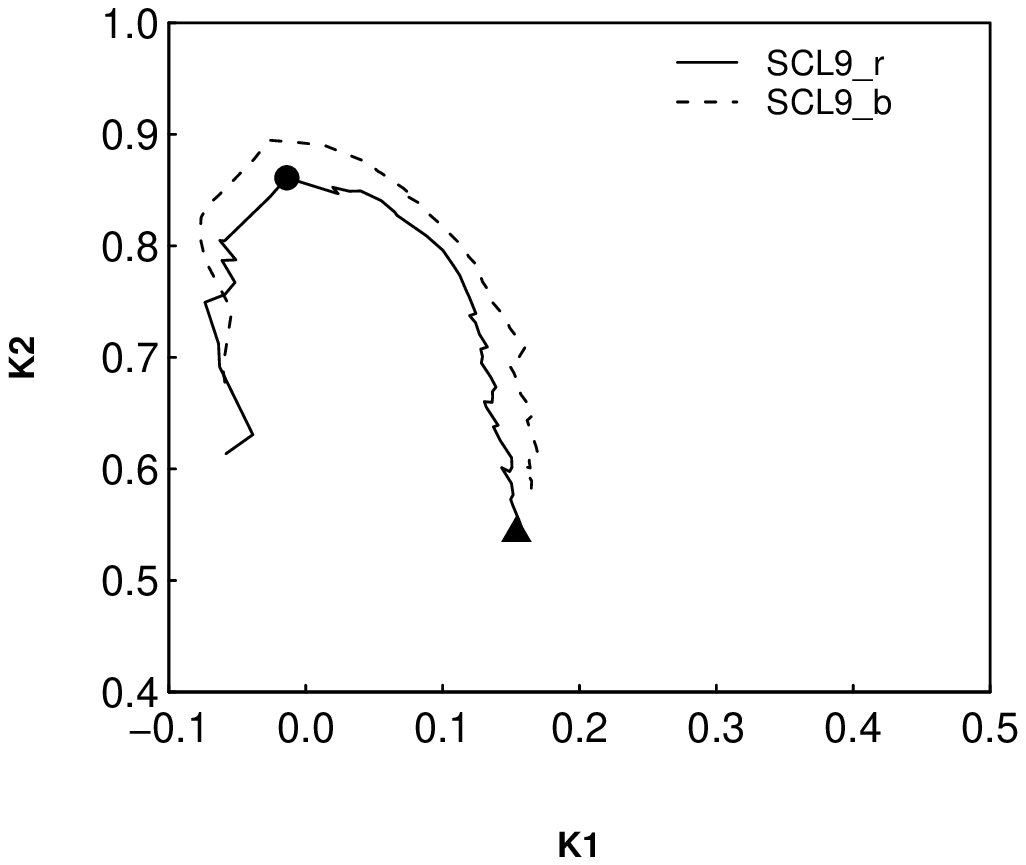}}
\caption{ The supercluster SCL9. Upper panels:
the Minkowski functional $V_3$ (the Euler characteristic)
vs the mass fraction $m_f$ for bright (B, $M \leq -20.0$) and 
faint (F, $M > -20.0$) galaxies (left panel), for early and late 
type galaxies (middle panel) and for red and blue galaxies
(right panel). Lower panels: the morphological signatures
$K_{1}$ (planarity) - $K_{2}$ (filamentarity) for the same populations.   
Symbols as in Fig.~\ref{fig:v3k12s126}.
}
\label{fig:v3k12s9}
\end{figure*}

\begin{figure*}[ht]
\centering
\resizebox{0.28\textwidth}{!}{\includegraphics*{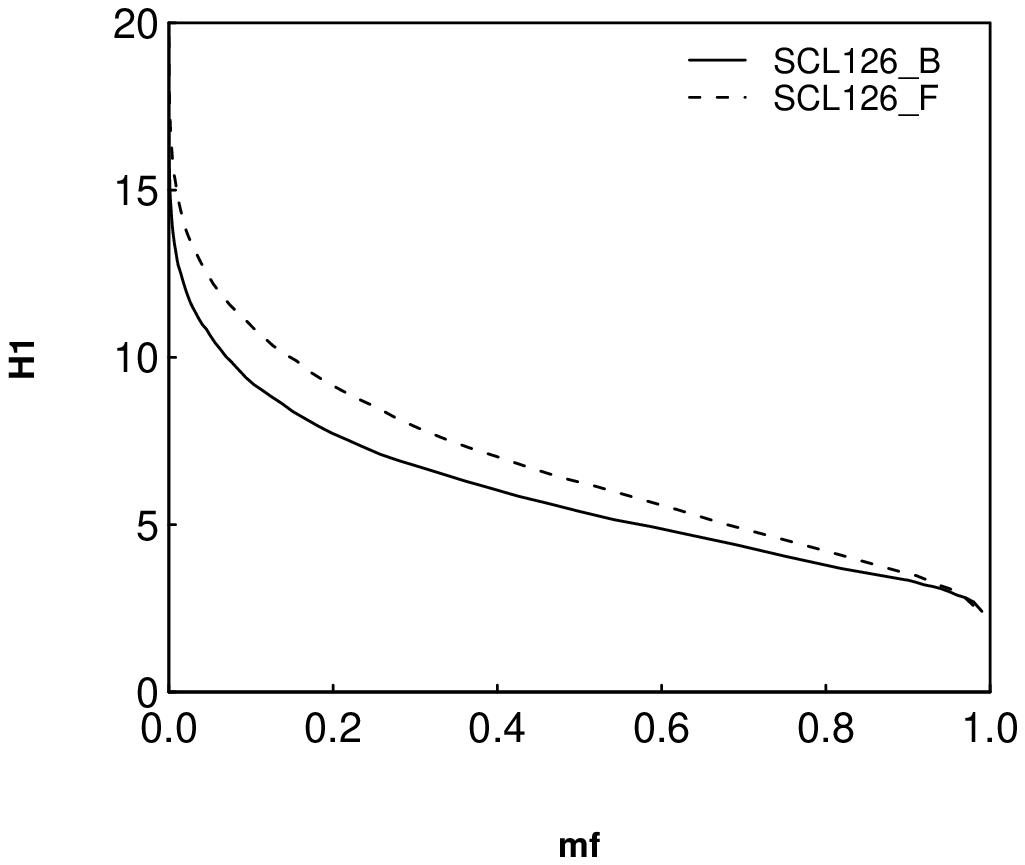}}
\resizebox{0.28\textwidth}{!}{\includegraphics*{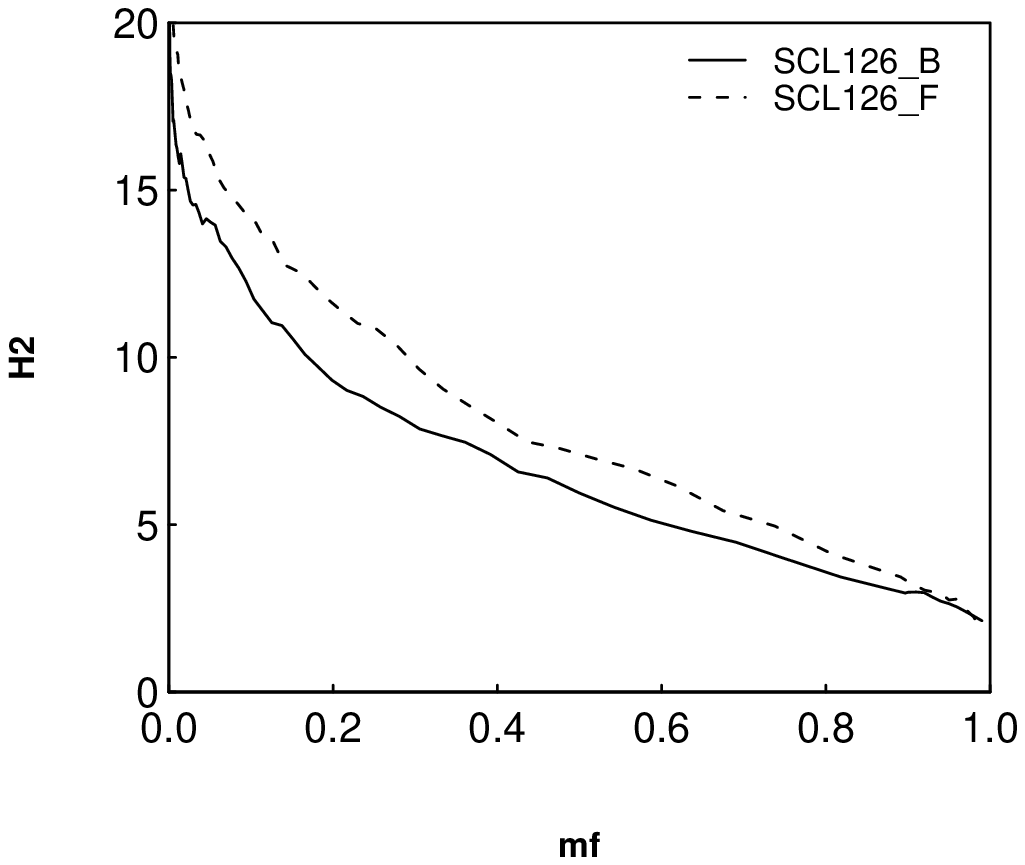}}
\resizebox{0.28\textwidth}{!}{\includegraphics*{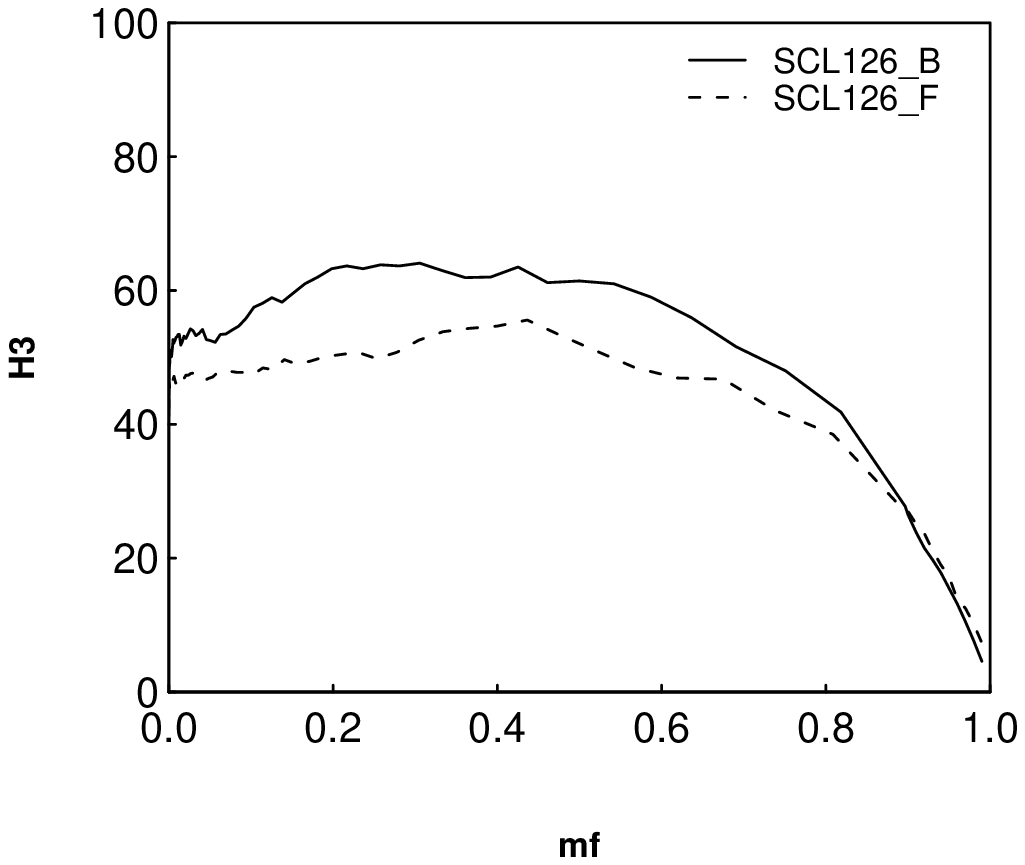}}
\caption{From left to right: the shapefinders 
$H_{1}$ (thickness), $H_{2}$ (width), and  $H_{3}$ (length) (in \Mpc) 
for the supercluster SCL126, for bright (B, $M \leq -20.0$) and 
faint (F, $M > -20.0$) galaxies.
}
\label{fig:h13}
\end{figure*}

Now we shall find the Minkowski functional $V_3$ and morphological signature
$K_1$-$K_2$ for rich superclusters separately for galaxies from different
populations, as marked by their luminosity, the spectral parameter $\eta$ and
the color index $col$ (Figs.~\ref{fig:v3k12s126} and ~\ref{fig:v3k12s9}). To
understand better the morphological signature we show the shapefinders $H_1$,
$H_2$ and $H_3$ (Fig.~\ref{fig:h13}). To not to overcrowd the paper with
Figures we show the shapefinders for the supercluster SCL126 for bright and
faint galaxies only.

\subsection{SCL126} 

Fig.~\ref{fig:v3k12s126} (upper left panel) shows the $V_3$ curves for bright 
and faint galaxies in the supercluster SCL126. At small values of the 
mass fraction $m_f$ the values of $V_3$ are small. At a mass fraction 
$m_f \approx 0.4$ the values of $V_3$ of both bright and faint galaxies 
increase rapidly which shows that at this density level the supercluster 
is splitted into several clumps. For bright galaxies the $V_3$ curve has 
some small plateaus at $m_f \approx 0.6$ and $m_f \approx 0.7$, this 
curve reaches maximum value of about 15 at mass fraction $m_f \approx 
0.9$. 

For faint galaxies the value of $V_3$ increases at mass fraction 
$m_f \approx 0.4$. In the whole 
mass fraction interval the values of $V_3$ for faint galaxies are 
smaller than those for bright galaxies, about 5. This indicates that the 
overall distribution of bright galaxies is clumpy, while the 
distribution of faint galaxies is more homogeneous. The peaks of $V_3$ 
values at very high mass fractions where the maximum value of
$V_3$  for faint galaxies is 10, are due to high - density cores in 
this supercluster.
The $V_3$ curve for the full supercluster is very similar to that of 
$V_3$ curve for faint galaxies showing faint galaxies trace the structure 
of the supercluster rather well.

Fig.~\ref{fig:v3k12s126} (lower left panel) shows the shapefinders $K_1$ 
(planarity) and $K_2$ (filamentarity) for bright and faint galaxies. They are 
calculated on the basis of shapefinders $H_{1}$-$H_{3}$ 
(Fig.~\ref{fig:h13}). Of these, the shapefinder $H_{1}$ is the smallest and 
characterizes the thickness of superclusters. The shapefinder $H_{2}$ is an 
analogy of the width of a supercluster, but it contains information about both 
the area and curvature of an isodensity surface. The shapefinder $H_{3}$ is the 
largest and describes the length of the superclusters. This is not the 
real length of the supercluster, but a measure of the integrated curvature of 
the surface. 

Fig.~\ref{fig:h13} shows that at all mass fractions, the thickness $H_1$ and 
the width $H_2$ of the supercluster as determined using data about bright galaxies 
are smaller than those calculated using data about faint galaxies. Thus, at all 
density levels the distribution of bright galaxies is more compact than the 
distribution of faint galaxies (which is also less clumpy). However, the 
shapefinder $H_{3}$ which is calculated as $H_3=C/4\pi$ (the integrated mean 
curvature C) reflects the clumpiness of a population, and since the distribution 
of bright galaxies is much more clumpy (according to $V_3$), the value of 
$H_{3}$ for bright galaxies is larger than this value for faint galaxies 
(Fig.~\ref{fig:h13}, right panel). This makes also the filamentarity $K_2$ for 
bright galaxies to be larger than the filamentarity of faint galaxies. As a 
result, in the morphological signature (Fig.~\ref{fig:v3k12s126}, lower left 
panel) the values of $K_2$ for bright galaxies are larger. The morphological 
signature of faint galaxies is similar to that of the whole supercluster, this is 
another indication that faint galaxies trace the structure of the supercluster 
well.

In this Figure we marked the value of $K_1$ and $K_2$ at the mass fraction $m_f
\approx 0.7$. Interestingly, we see that at this mass fraction (density level)
the characteristic morphology of supercluster changes, as seen from the change
of the morphological signature.

The values of the fourth Minkowski functional, $V_3$, for galaxies of
different type (Fig.~\ref{fig:v3k12s126}, upper middle panel, 
cassified by the spectral
parameter $\eta$) show that the clumpiness of early type galaxies starts to
increase at low values of the mass fraction, $m_f \approx 0.2 - 0.3$. At the
mass fraction value $0.5$ the value of $V_3$ has a small peak followed by a
minimum; at the mass fraction $m_f \approx 0.7$ the value of $V_3$ increases
again, and has a peak value about 10 at the mass fraction of about 0.9.

The distribution of late-type galaxies in this supercluster is much 
less clumpy, as show the values of $V_3$.  The number of isolated clumps 
of these galaxies grows only at rather high mass fraction values, $m_f > 
0.6$ ($V_3 = 6$), and has a peak at $m_f \approx 0.9$ (here $V_3 = 12$). 
The largest differences in the clumpiness of the spatial distribution of
galaxies of different type are, 
according to the fourth Minkowski functional, at mass fractions $m_f < 
0.6$, which describe the outer (lower density) region of the 
supercluster.

Fig.~\ref{fig:v3k12s126} (lower middle panel) shows the morphological
signature for galaxies of different type. The value of $K_2$ (filamentarity)
that contains information both about the spatial extent and clumpiness of the
data, is larger for late-type galaxies. In this figure an important feature is
that, again, at the mass fraction (density level) $m_f \approx 0.7$ the
characteristic morphology of supercluster changes.

The curves of the fourth Minkowski functional for red galaxies are
rather similar to those of early type galaxies (Fig.~\ref{fig:v3k12s126},
upper right panel, the color index $col$). Here we again see an increase of the
values of $V_3$ at the mass fractions $m_f \approx 0.4$ and $m_f \approx 0.6$, and
a maximum at $m_f \approx 0.9$ ($V_3 = 10$).

This panel shows that blue galaxies form a few isolated
clumps already at relatively low mass fraction values, but those clumps have
low density, and at higher mass fractions some of them do not
contribute to the supercluster any more; the value of $V_3$ decreases (being
smaller than 6) and then increases again at mass fractions $m_f \approx
0.6-0.7$. At this mass fraction the clumpiness of the blue
galaxy distribution becomes comparable to that of red galaxies. 
At very high values
of the mass fraction, $m_f > 0.9$, blue galaxy clumps are not seen, and the
value of $V_3$ decreases rapidly. The value of $V_3$ for red galaxies is
larger than for blue galaxies at almost all density levels (mass fraction
values).

At the mass fraction $m_f \approx 0.7$ (Fig.~\ref{fig:v3k12s126}, lower right
panel) the characteristic morphology of the supercluster, as described by
red and blue galaxies, changes.

Thus Fig.~\ref{fig:v3k12s126} shows that  the differences in clumpiness 
between galaxies of different type and color (and star formation rate) in 
the supercluster SCL126 are the largest, according to the fourth Minkowski 
functional, at mass fractions $m_f < 0.7$, which describe the outer (lower 
density) regions of the supercluster. In outer parts of this supercluster, the 
distribution of late type and blue galaxies is much less clumpy 
(more homogeneous) that the distribution of early type, red galaxies.

The fourth Minkowski functional  $V_3$ for early type galaxies 
(Fig.~\ref{fig:v3k12s126}, middle panel) has a small peak at mass 
fractions of about $0.4 < m_f < 0.6$ where the value of $V_3 = 10$. This 
indicates that at this intermediate density level early type galaxies 
form some isolated clumps which at still higher density level (mass 
fraction) do not contribute to supercluster any more, and the value of 
$V_3$ decreases again. At the same mass fractions $V_3$ for red galaxies 
(right panel) has a value $V_3 = 7$ with no peak. Therefore, these 
additional isolated clumps are due to early type blue galaxies.  The 
galaxies that are classified as late type but have red colors are 
located in intermediate density filaments which connect clumps of early 
type galaxies to the main body of the supercluster.

\subsection{SCL9}

Next let us study the values of $V_3$ for galaxies of different 
populations in the supercluster SCL9 (Fig.~\ref{fig:v3k12s9}, upper 
right panel).  Here we see that the $V_3(m_f)$ curve is rather different 
from that for SCL126: at small mass fractions, the values of $V_3$ for 
bright and faint galaxies (left panel) are small, but at a mass fraction 
value of about 0.4, the values of $V_3$ increase. This increase is more 
rapid for bright galaxies than for faint galaxies. The values of $V_3$ 
for bright galaxies reach a maximum at a mass fraction $m_f 
\approx 0.7$, and the maximum values of $V_3$ are larger than in the 
supercluster SCL126, about 18. For faint galaxies the $V_3$ curve 
reaches the maximum value, 15, at the mass fraction $m_f \approx 0.7$. The 
clumpiness of both the bright and faint galaxy distribution
in the supercluster SCL9 is 
larger than that in the supercluster SCL126. Also in this supercluster 
the $V_3$ curve for the full supercluster is very similar to that of 
the $V_3$ curve for faint galaxies showing that faint galaxies trace the 
structure of the supercluster well.

In the middle right panel of Fig.~\ref{fig:v3k12s9} ($V_3$ for early and late
type galaxies, defined by the spectral parameter $\eta$) we see a continuous
increase of the number of isolated clumps as delineated by galaxies of
different type. Interestingly, up to mass fractions of about $m_f = 0.7$ the
$V_3$ curves for galaxies of different type almost coincide ($V_3 = 12)$, which
is opposite to what we saw in SCL126. Then the number of clumps in the
distribution for late type galaxies increases rapidly and reaches the maximum
value, 20, at mass fractions between 0.7 and 0.8. The distribution of late
type galaxies in the core region of the supercluster SCL9 is even more clumpy
than the distribution of early type galaxies.

The curves for $V_3$ for red  and blue
galaxies (divided using the color index $col$, upper left
panel of Fig.~\ref{fig:v3k12s9}) are
quite similar to those we saw above. Up to the mass fractions $m_f \approx
0.7$ (the outer regions of the supercluster) the $V_3$ curves for galaxies of
different color almost coincide. In the core region of the supercluster the
distribution of blue galaxies is very clumpy, the maximum
number of clumps is 17, while the maximum number of isolated clumps in the
distribution of red galaxies is 12. Larger values of $V_3$ for blue
 galaxies suggest again that in this supercluster, blue
galaxies are located in numerous clumps, while the distribution of red galaxies
is smoother.

This figure shows that at intermediate mass fractions, $0.4 < m_f < 0.6$,
the values of $V_3$ for blue galaxies show a small
peak not seen in the $V_3$ curve for 
the late type galaxies. This peak is generated by those galaxies that
have blue colors, but spectra characteristic to early type galaxies.
This is consistent with the overall large clumpiness of blue galaxies in
this supercluster.

In Fig.~\ref{fig:v3k12s9} (lower panels) we plot the morphological signature for 
galaxies of different populations in SCL9. These figures show that also in 
this supercluster the filamentarity of bright galaxies is larger than the
filamentarity 
of faint galaxies. The morphological signature of faint galaxies is similar to 
that of the whole supercluster. The filamentarity $K_2$ 
for late type, blue galaxies
is larger than that for early type (and red) 
galaxies due 
to their larger clumpiness. However, we see in these figures again that at the
mass fraction $m_f \approx 0.7$ the morphology of the supercluster 
changes.

\subsection{Summary} 

The differences in the distribution of galaxies from different populations in the
superclusters SCL126 and SCL9 are related to different overall morphology of 
these superclusters. In particular, in the supercluster SCL126 at  mass 
fractions $m_f < 0.7$ the differences in the fine structure formed by galaxies of 
different populations are large: 
here the maximum values of $V_3$ are, correspondingly, 8 for red galaxies and 5 
for blue galaxies, showing that the distribution
of blue, late type galaxies is more homogeneous than the distribution
of red, early type galaxies. 
At mass fractions $m_f > 0.7$ the  clumpiness of the distribution of galaxies of 
different type and color in the supercluster SCL126 is similar. In the contrary, 
in the supercluster SCL 9 the clumpiness of galaxy populations
of different type and 
color is quite similar in the outskirts of the supercluster
(e.g., the values of $V_3$ are the same, 7--8 both for the early and late type
galaxies). At mass fractions $m_f > 0.7$ 
in this supercluster, the clumpiness of the late type, blue galaxy population
is larger (the maximum value of $V_3$ is 17)  than the clumpiness of the early 
type, red galaxy population (for those 
galaxies the maximum value is $V_3 = 12$). 
Overall, the clumpiness of all galaxy populations 
in the supercluster SCL126 is smaller than the clumpiness of galaxy populations 
in SCL9.

This shows that in superclusters under study there exists a region of mass 
fractions $m_f \approx 0.7$, where we can observe in the behaviour of both the 
fourth Minkowski functionals for galaxies of different type and of the 
morphological signature a crossover from low- density morphology to high-density 
morphology. Based on that, we choose the mass fraction $m_f=0.7$ to delimit the 
supercluster cores.  In the following analysis we use this density level as 
separating the high density core regions of superclusters ($D1$) from their 
outer regions with lower densities (outskirts, $D2$), see Fig.~\ref{fig:radecx}.
 
The density level $m_f = 0.7$ approximately corresponds to region where the
luminosity density contrast $\delta = 10$ (Papers I and III). In Paper III we showed
that high density cores with $\delta > 10.0$ are characteristic to rich
superclusters. Typical densities in poor superclusters are lower, $\delta <
10.0$, comparable to densities in the outskirts of rich superclusters.

In both superclusters, about 3/4 of late type galaxies with red colors, as well 
as those galaxies showing blue colors and early type spectra, are located 
at intermediate densities in the outskirt regions.
Fig.~\ref{fig:v3k12s126} and Fig.~\ref{fig:v3k12s9} show that although 
due to these galaxies 
the values of the fourth Minkowski functional $V_3(m_f)$ for galaxies of 
different type and color differ in some details, 
the overall shapes of $V_3(m_f)$ curves show no systematic differences.

Fig.~\ref{fig:radecx} shows the main body of the supercluster SCL126 as a rich
filament (region $D1$; paper RI). The main core of the supercluster at
R.A. about 195 deg contains four Abell clusters (three of them are also X- ray
sources).  This region has a diameter of about 10~\Mpc\ \citep{e2003}.  The
X-ray cluster at R.A. about 203 deg is Abell 1750, a merging binary cluster
\citep{don,bel}.

In the supercluster SCL9 the regions of highest density form several separate
concentrations of galaxies (a "multispider", Fig.~\ref{fig:radecx} and RI).
One of them, the main center of SCL9, contains 5 Abell clusters and one X-ray
cluster. There are Abell clusters also in outskirts of this supercluster.  In
this supercluster the total number of Abell clusters is larger than in the
supercluster SCL126, but they do not form such a high concentration of Abell
and X-ray clusters that is observed in the core region of SCL126.

In the following section we study the distribution of galaxies from various
populations in the core region and in outskirts of superclusters, using
additionally the information about the group membership for galaxies from
different populations.

We mention here that the values of $V_3(m_f)$ reflect the distribution of 
galaxies at scales determined by the width of our smoothing kernel, B3, and this 
corresponds to larger scales than the typical scale at which 
the relation between  galaxy type and local
density applies, i.e. our results do not reflect directly features in 
galaxy distribution with characteristic scale less than about 1 - 2 \Mpc.

Thus the differences in clumpiness of 
galaxies of different type at high mass fractions (in supercluster cores)
does not contradict to the presence of the luminosity - density relation at 
small scales. At the same time, the brightest galaxies of both early and late
type galaxies are located in high density regions and this is seen also 
in  $V_3(m_f)$ curves at high mass fraction \citep{h88,e91,park07}.

\section{Substructure of superclusters in core region and in outskirts.
Rich and poor groups}

\subsection{Core and outskirts} 

Groups of galaxies are additional tracers of substructure in superclusters.
The group richness is a local density indicator, if we compare the properties
of galaxies in various local environments (Paper III). We divide groups by
their richness as follows: rich ($N_{gal} \geq 10$) groups and clusters (we
denote this sample as $Gr_{10}$) and poor groups ($N_{gal} < 10$, $Gr_2$). In
Table~\ref{tab:r2overall} we give the fractions of galaxies in these groups
for the whole superclusters.

\begin{figure*}[ht]
\centering
\resizebox{0.35\textwidth}{!}{\includegraphics*{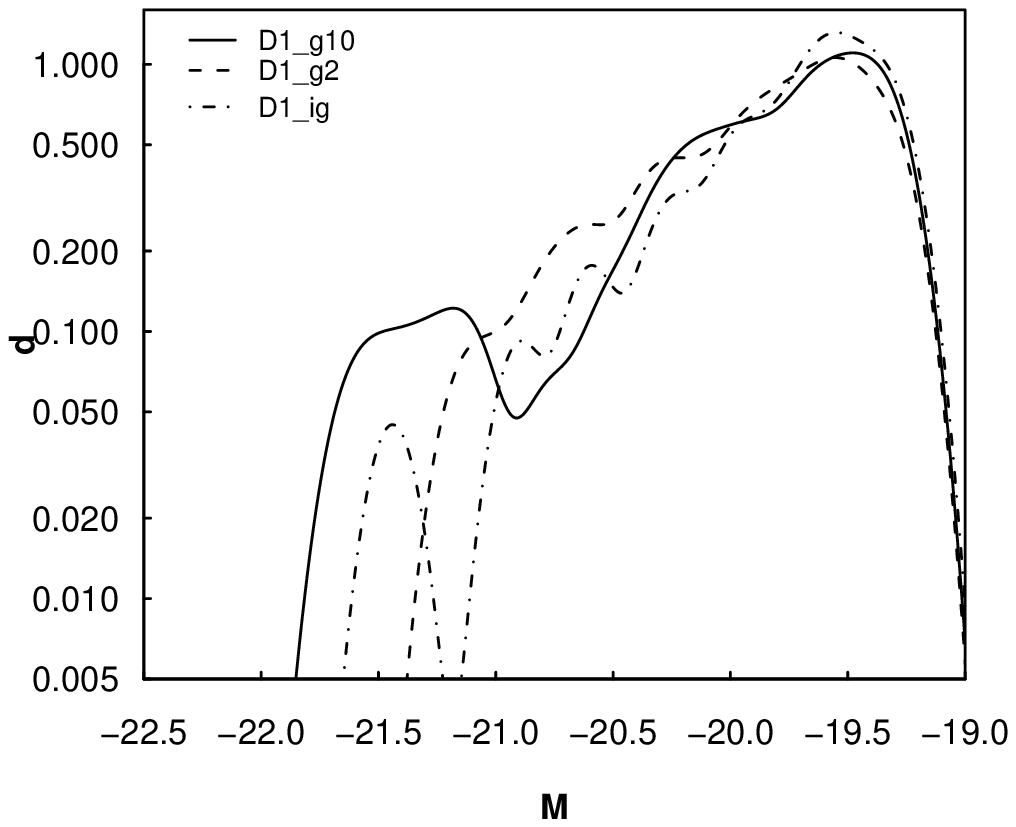}}
\resizebox{0.35\textwidth}{!}{\includegraphics*{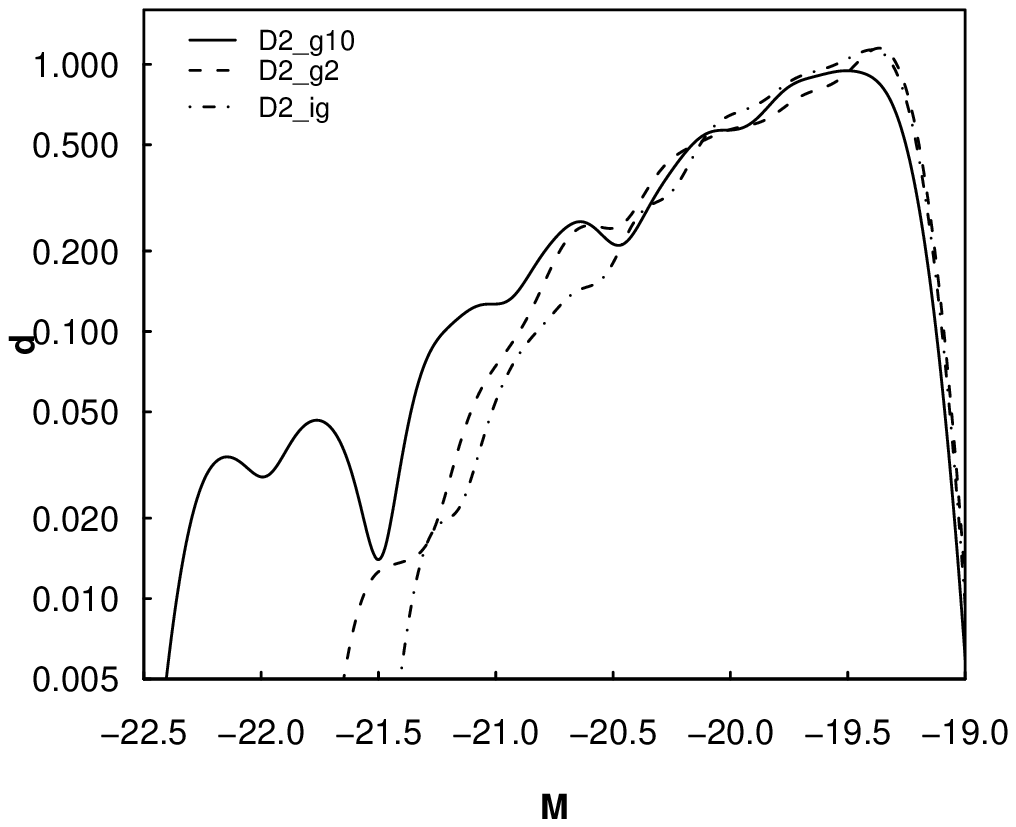}}
\hspace*{2mm}\\
\resizebox{0.35\textwidth}{!}{\includegraphics*{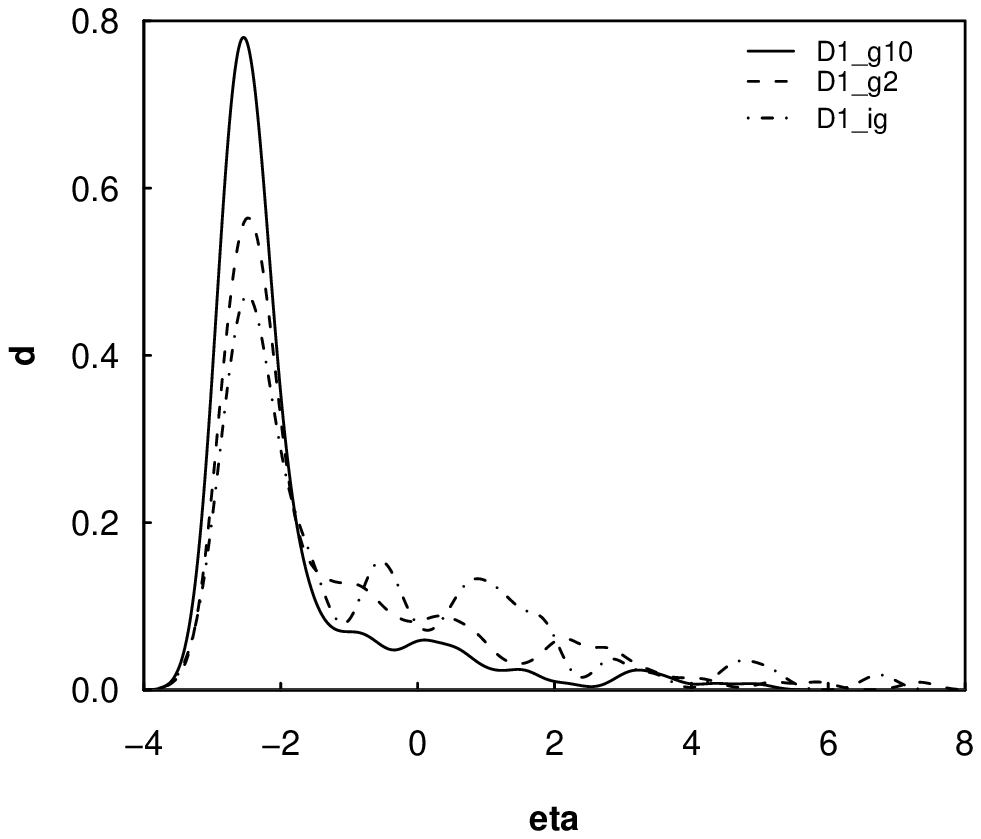}}
\resizebox{0.35\textwidth}{!}{\includegraphics*{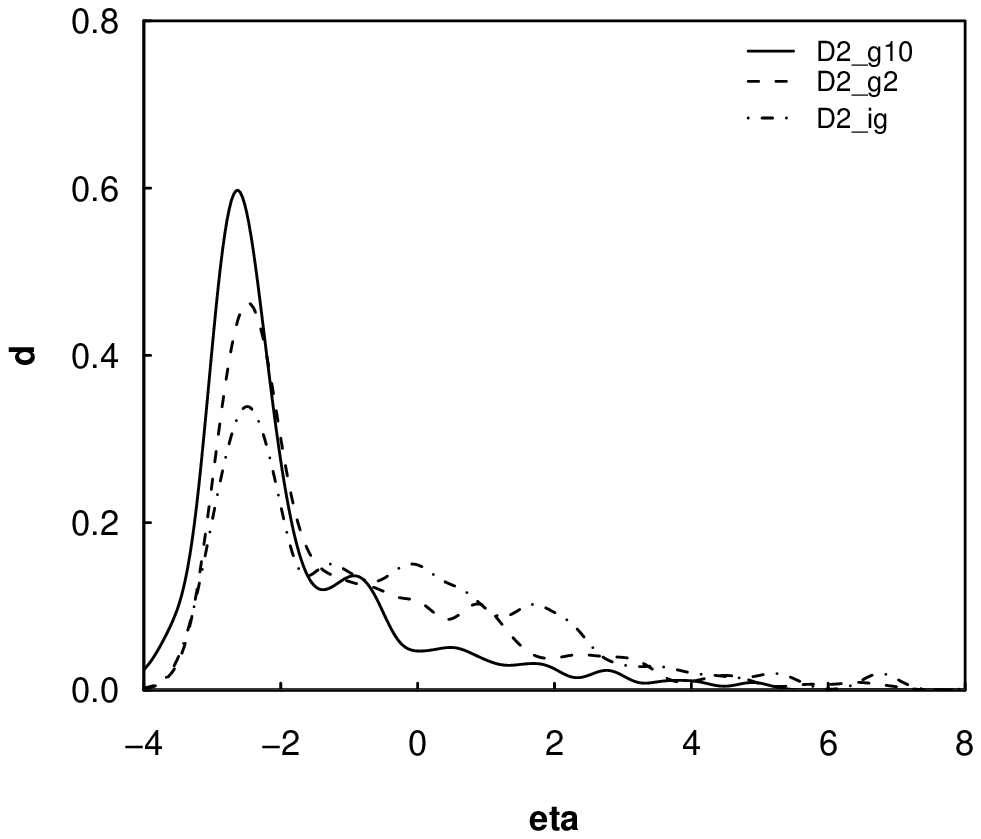}} 
\hspace*{2mm}\\
\resizebox{0.35\textwidth}{!}{\includegraphics*{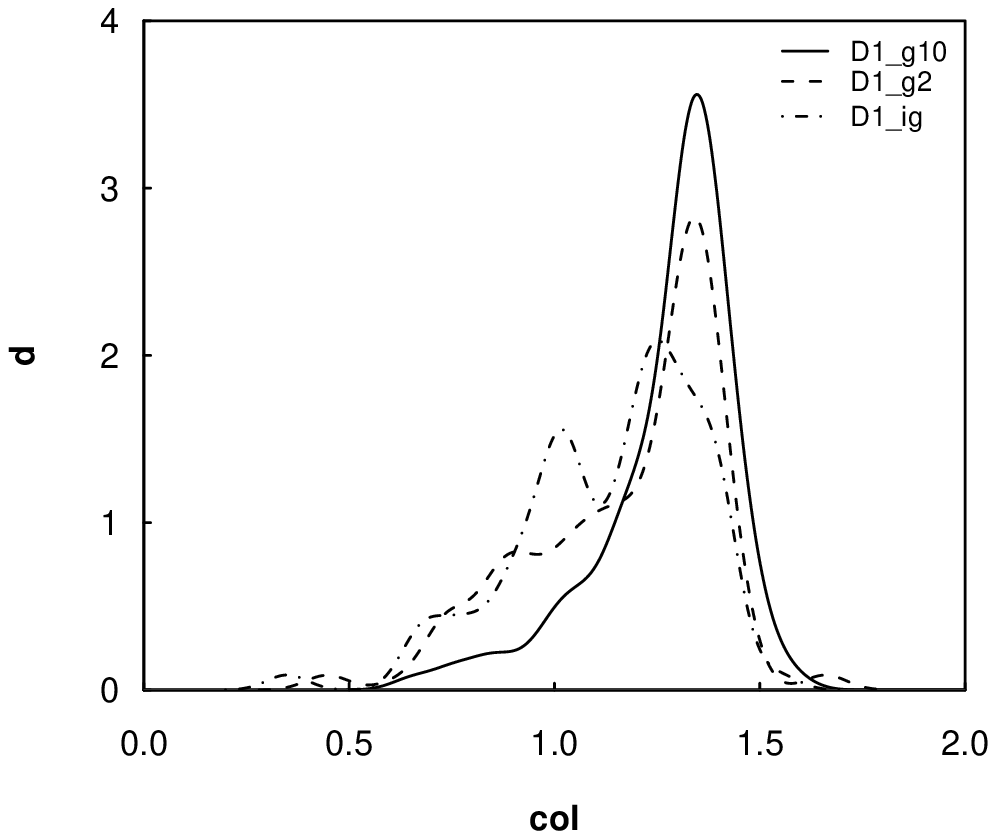}}
\resizebox{0.35\textwidth}{!}{\includegraphics*{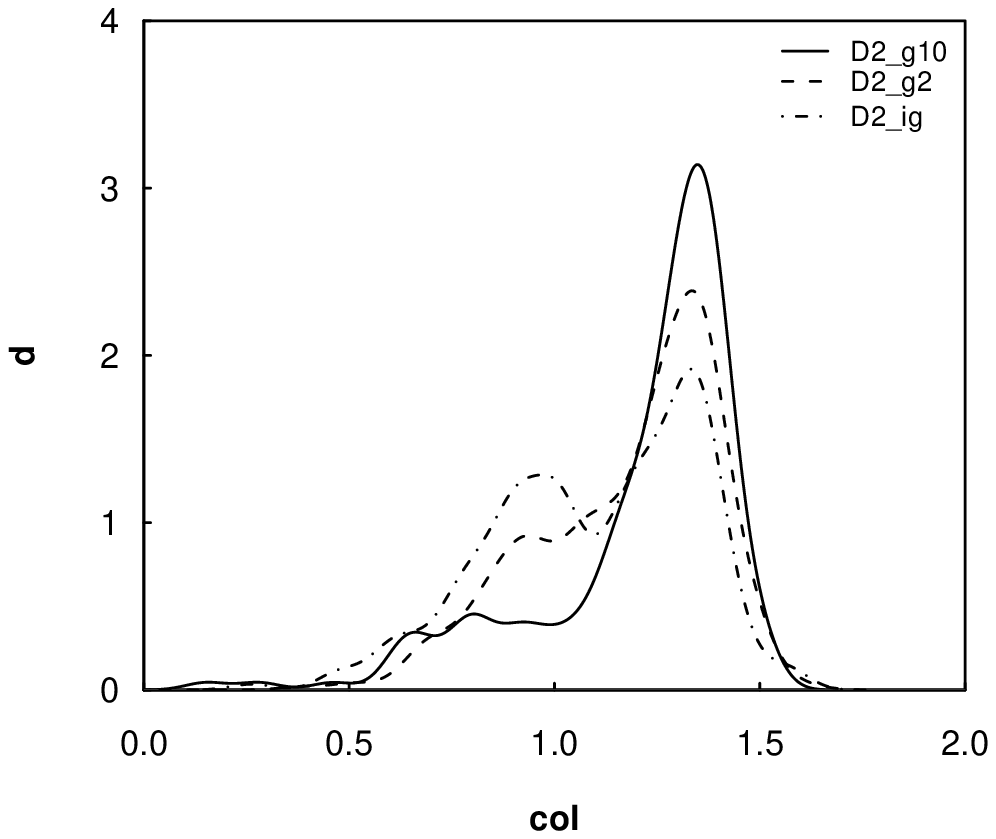}} 
\caption{The distributions of the absolute magnitude (upper panels),
the spectral parameter $\eta$ (middle panels) and
 the color index $col$ (lower panels) 
of galaxies in groups of various richness
in  the supercluster SCL126. Left panels: the core (region $D1$), 
right panels: the outskirts (region $D2$). 
The populations are:
$g10$ -- rich groups ($N_{gal} \geq 10$),
$g2$ -- poor groups ($N_{gal} < 10$), and $ig$ -- isolated
galaxies (those galaxies which do not belong to
any group). The Kolmogorov-Smirnov test results for the difference of the
distributions are in Table~\ref{tab:s1269d1d2}.
}
\label{fig:scl126d1d2gr}
\end{figure*}

\begin{figure*}[ht]
\centering
\resizebox{0.35\textwidth}{!}{\includegraphics*{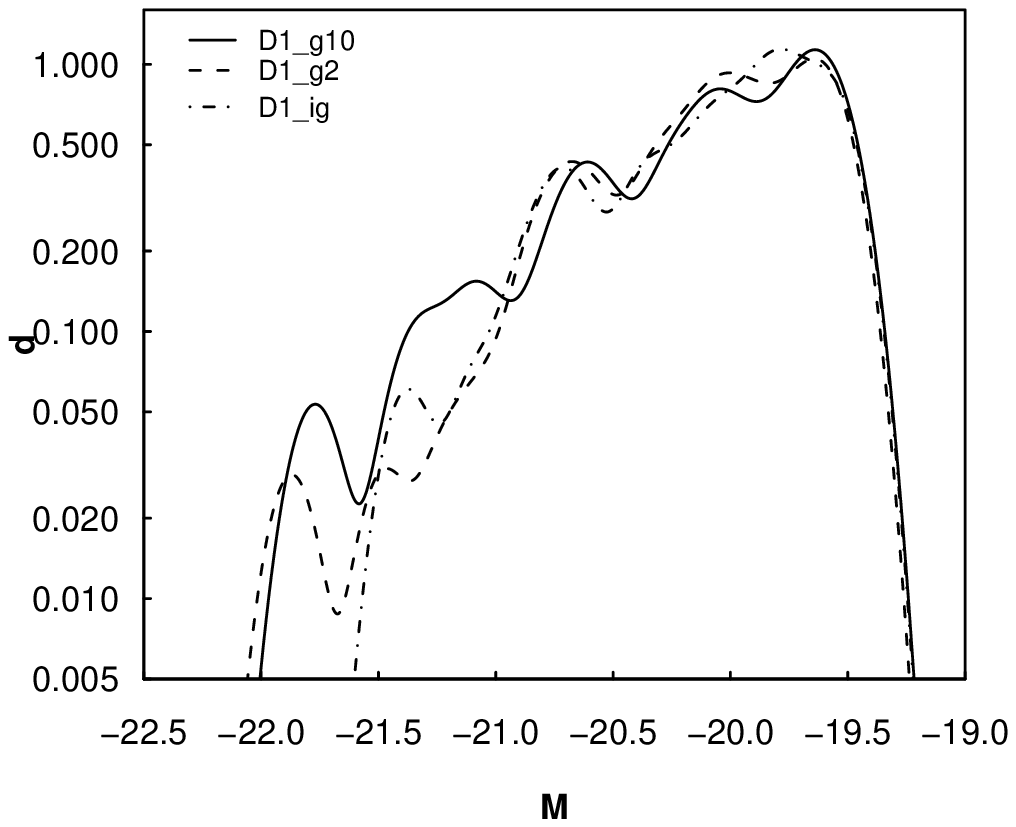}}
\resizebox{0.35\textwidth}{!}{\includegraphics*{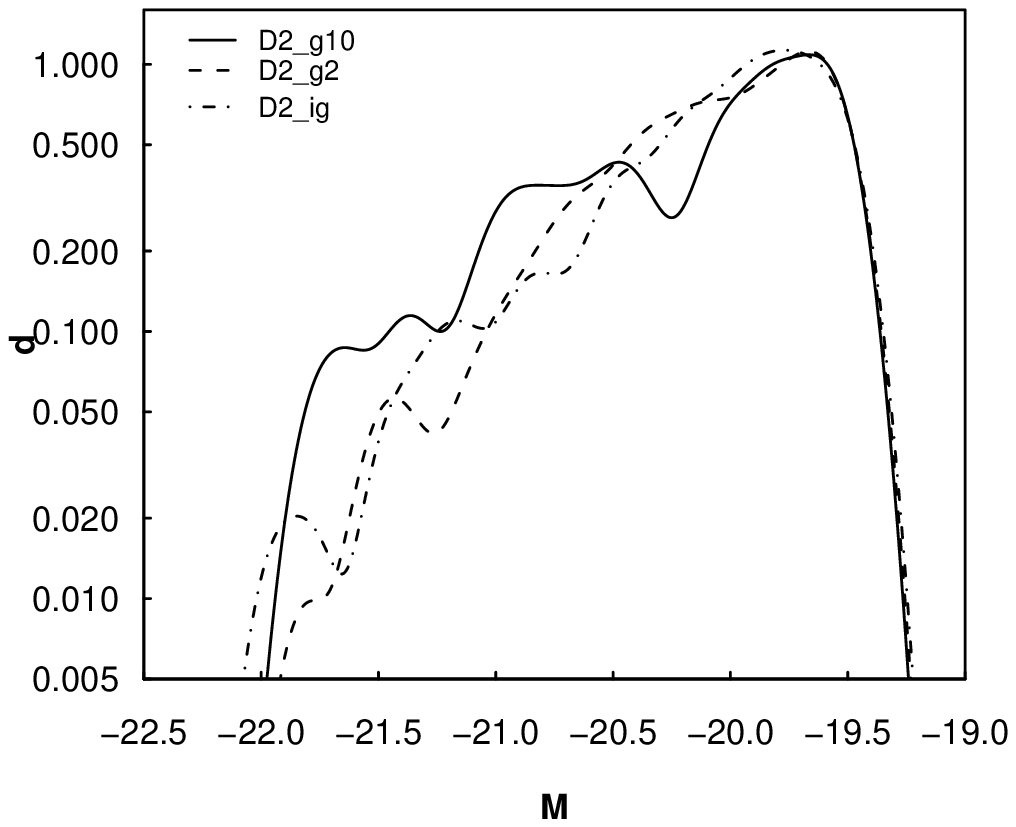}}
\hspace*{2mm}\\
\resizebox{0.35\textwidth}{!}{\includegraphics*{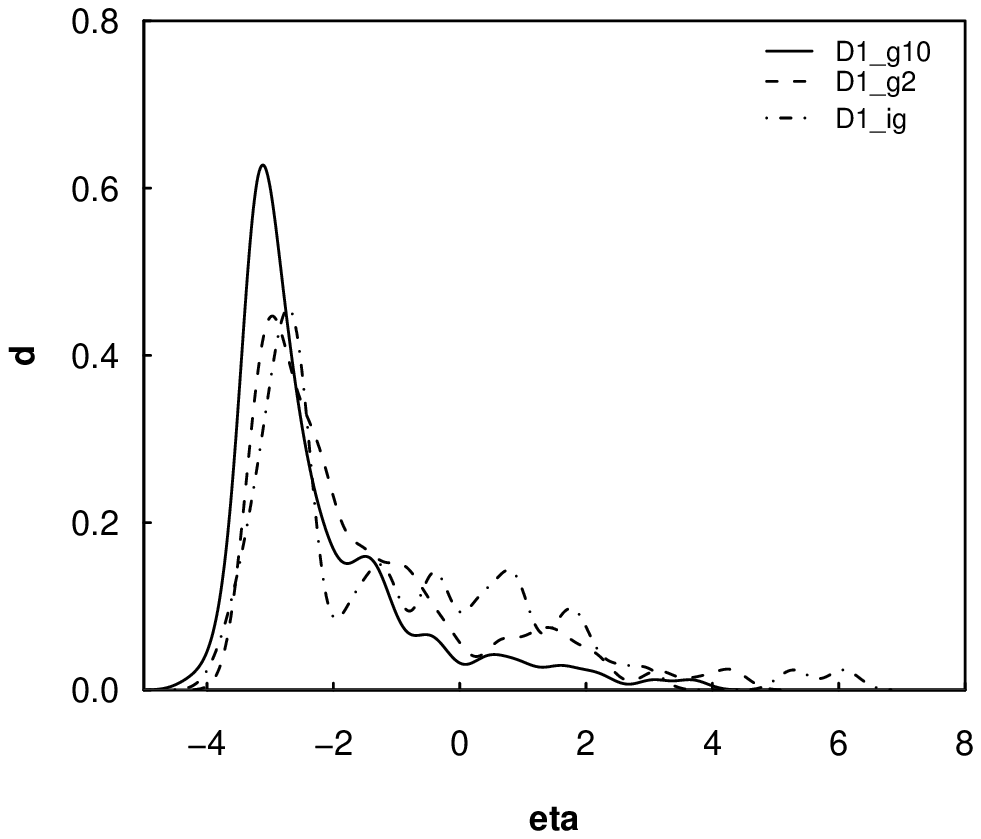}}
\resizebox{0.35\textwidth}{!}{\includegraphics*{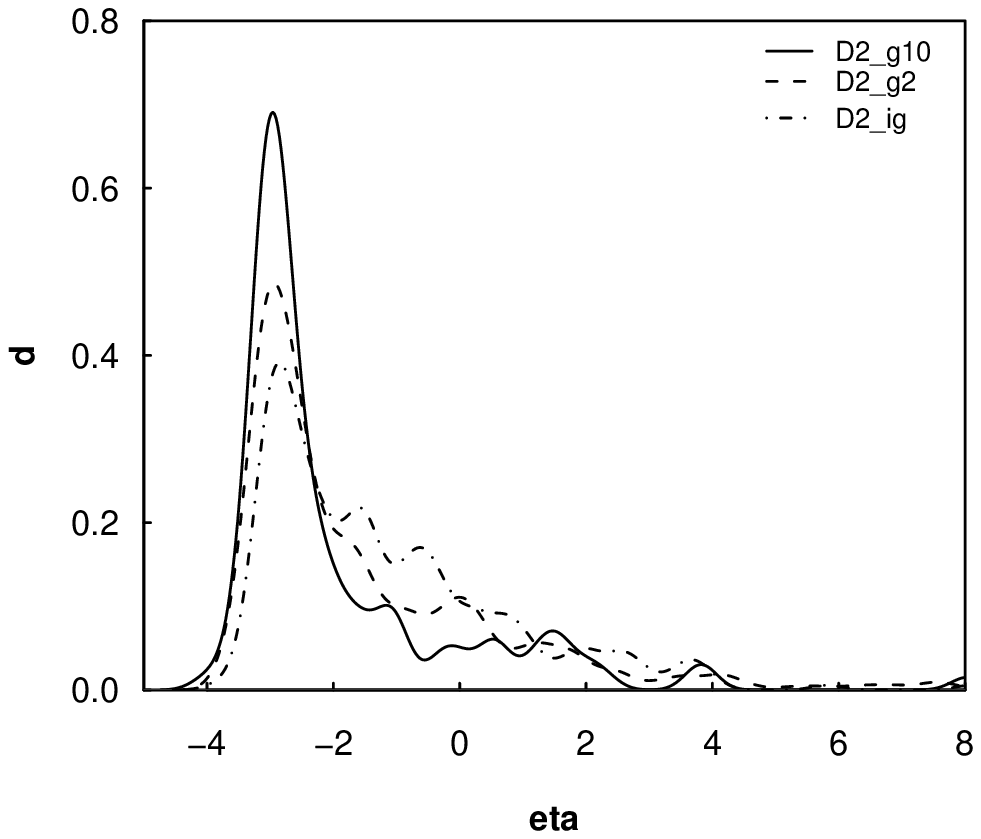}} 
\hspace*{2mm}\\
\resizebox{0.35\textwidth}{!}{\includegraphics*{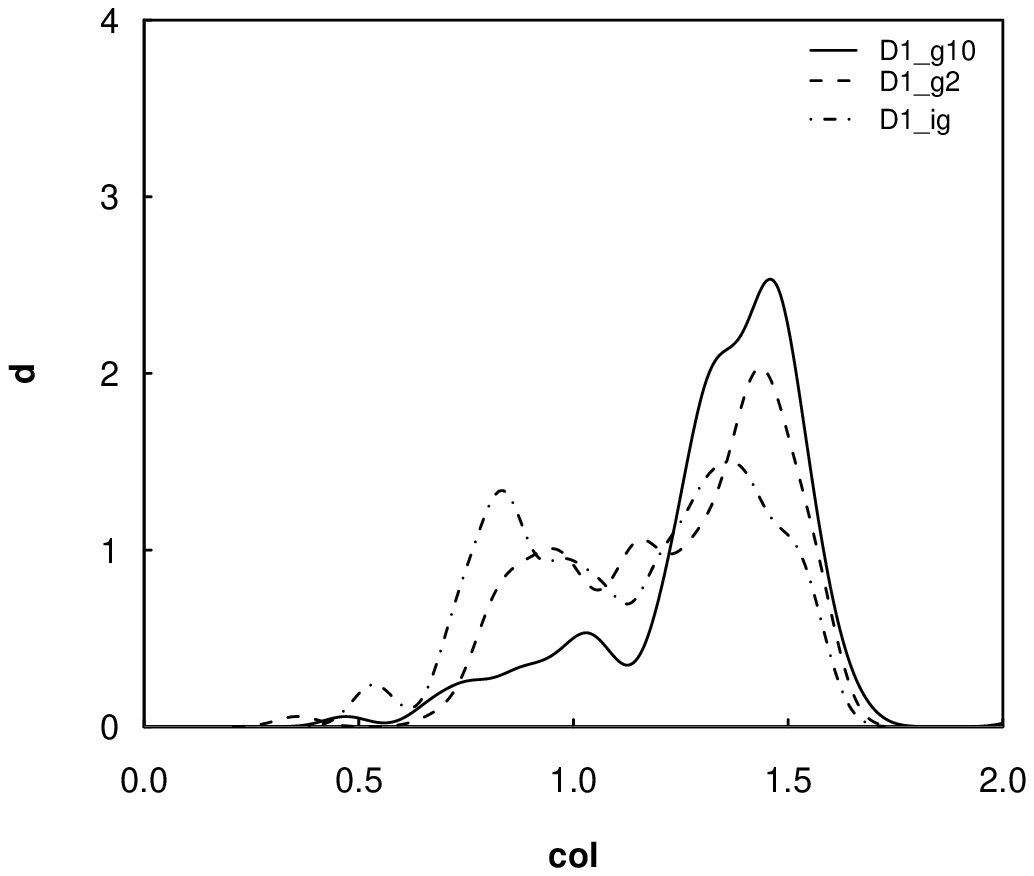}}
\resizebox{0.35\textwidth}{!}{\includegraphics*{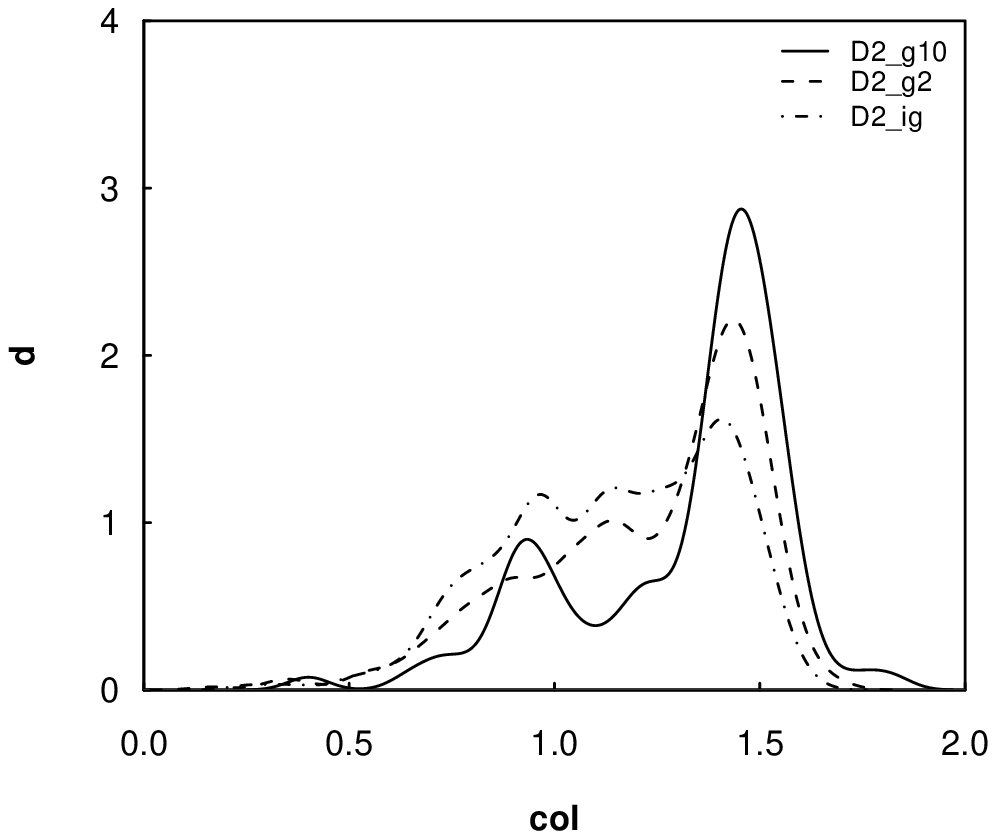}} 
\caption{The distributions of the absolute magnitude (upper panels),
the spectral parameter $\eta$ (middle panels) and
 the color index $col$ (lower panels) 
of galaxies in groups of various richness
in  the supercluster SCL9. Left panels: the core (region $D1$), 
right panels: the outskirts (region $D2$). 
The populations are:
$g10$ -- rich groups ($N_{gal} \geq 10$),
$g2$ -- poor groups ($N_{gal} < 10$), and $ig$ -- isolated
galaxies (those galaxies which do not belong to
any group).The Kolmogorov-Smirnov test results for the difference of the
distributions are in Table~\ref{tab:s1269d1d2}.
}
\label{fig:scl9d1d2gr}
\end{figure*}

Let us study the galaxy content of groups in the cores and in the outskirts 
of our superclusters. 
We plot the differential luminosity functions and the distributions of the 
spectral parameter $\eta$ and the color index $col$ for galaxies in groups of 
different richness in Figs.~\ref{fig:scl126d1d2gr}--\ref{fig:scl9d1d2gr}, for 
the core regions and outskirts of the superclusters SCL126 and SCL9, 
respectively. In Table~\ref{tab:s1269d1d2}  we present the ratios of the numbers 
of bright and faint galaxies $B/F$ in superclusters, and the ratios of early and 
late type galaxies $E/S$, as classified by the spectral parameter $\eta$. We 
also calculate the ratio of the numbers of red and blue galaxies, $r/b$, 
classified by the color index $col$. We give here the 
statistical significance of the differences between the galaxy content in core 
regions and in outskirts  in these two superclusters according to the 
Kolmogorov-Smirnov test.

{\scriptsize
\begin{table}[ht]
\caption{The superclusters SCL126 and SCL9:
the galaxy content in the core region D1 and in the outskirts (region D2), 
in rich and poor groups, and of
those galaxies which do not belong to groups.}

\begin{tabular}{rllllllll} 
\hline 
(1)&(2)&(3)&(4)& (5)&(6)&(7)&(8)& (9) \\      
\hline 
           &        &  \multispan2  SCL126   &            & \multispan2  SCL9 & \\        
Region     &   $D1$ &  $D2$ & \multispan2  Kolmogorov-Smirnov &   $D1$ &  $D2$ & \multispan2  Kolmogorov-Smirnov\\        
           &        &       & \multispan2  \mbox{test results}       &        &       & \multispan2  \mbox{test results} \\        
           &         &          &        $D$     &  $P$       &         &          &       $D$     &  $P$   \\      
N$_{gal}$  &     488     &  820  & &                          &     342    &  834  & &  \\                          
F$gr_{10}$ &     0.46    &  0.22 & &                          &     0.40    &  0.13 & &  \\                         
F$gr_2$    &     0.35    &  0.50 & &                          &     0.40    &  0.53 & &  \\                         
F$ig$      &     0.18    &  0.28 & &                          &     0.20    &  0.34 & &  \\                         
$E/S$ &&&&&&&&\\                                                                                                        
~~$All$    &   2.34 & 1.36 & 0.13 & 5.39e-05                 &   2.31 & 1.84 & 0.10 & 0.02 \\                      
~~$Gr_{10}$&   4.35 & 2.81 & 0.13 & 0.088                    &   4.07 & 3.20 & 0.17 & 0.05 \\                      
~~$Gr_2$   &   1.68 & 1.40 & 0.08 & 0.452                    &   2.05 & 2.07 & 0.06 & 0.77 \\                      
~~$IG$     &   1.26 & 0.78 & 0.19 & 0.240                    &   1.16 & 1.33 & 0.12  & 0.38 \\

$r/b$ &&&&&&&&\\                                                                                                        
~~$All$    &   3.40 & 2.15 & 0.11 & 0.001                    &   2.76 & 2.34 & 0.08 & 0.07  \\                   
~~$Gr_{10}$&   7.37 & 4.01 & 0.11 & 0.16                     &   5.00 & 3.77 & 0.13 & 0.29  \\                     
~~$Gr_2$   &   2.46 & 2.36 & 0.06  & 0.79                     &   2.34 & 2.61 & 0.06 & 0.84  \\                     
~~$IG$     &   1.62 & 1.26 & 0.12 & 0.28                     &   1.48 & 1.71 & 0.11 & 0.48  \\                     
$B/F$ &&&&&&&&\\                                                                                                        
~~$All$    &   0.42 & 0.45 & 0.07 & 0.165                    &   0.96 & 0.87 & 0.04 & 0.72  \\                  
~~$Gr_{10}$&   0.47 & 0.59 & 0.09 & 0.34                      &   1.00 & 0.91 & 0.10 & 0.61  \\                     
~~$Gr_2$   &   0.44 & 0.46 & 0.14 & 0.01                      &   1.00 & 0.95 & 0.05 & 0.97  \\                     
~~$IG$     &   0.27 & 0.35 & 0.09 & 0.66                      &   0.81 & 0.75 & 0.06 & 0.99  \\                     
\label{tab:s1269d1d2}                        
\end{tabular}
\tablecomments{
The columns in the Table are as follows:
\noindent column 1: Population ID and number ratio type. 
N$_{gal}$ -- the number of galaxies in a given
density region, 
F$gr_{10}$ -- the fraction of galaxies
in rich groups with at least 10 member galaxies, 
F$gr_2$ -- the fraction of galaxies in poor groups
with 2--9 member galaxies, 
F$ig$ -- the fraction of galaxies which do not belong to groups. 
\noindent Columns 2--3 and 6--7: the ratio of the numbers of galaxies
in the populations from regions $D1$ and $D2$ (sect. 2.2). 
Columns 4--5 and 8--9: the maximum difference ($D$) and the probability ($P$) that the
distributions of population parameters are taken
from the same parent distribution.
}
\end{table}
}

\subsection{SCL126} 

Fig.~\ref{fig:scl126d1d2gr} (upper row) shows the luminosities of 
galaxies in groups of different richness in the supercluster SCL126. 
This figure shows that the brightest galaxies reside in rich groups 
($N_{gal} \geq 10$), both in the core region  and in the outskirts of the 
supercluster. Luminosities of galaxies in poor groups are fainter, and 
galaxies which do not belong to groups are the faintest, although in the 
core region of the supercluster there are also some rather bright 
galaxies among them. 

According to the spectral parameter $\eta$ (Fig.~\ref{fig:scl126d1d2gr}, middle 
row), rich groups in the core region of the supercluster SCL126 are populated 
mainly with early type, quiescent galaxies (type 1 galaxies, $\eta < -1.4$). The 
fraction of late type galaxies in them is very small. The fraction of late type 
galaxies increases, as we move to poor groups. In the core region of the 
supercluster the fraction of late type galaxies is largest among those galaxies 
which do not belong to groups. 

In the outskirts region, $D2$, rich groups are still mainly populated with early 
type galaxies, but the fraction of late type galaxies in them is larger 
($E/S = 2.81$) than in  rich groups  in the core region, where $E/S = 4.35$. The 
fraction of late type galaxies in poor groups is larger; the fraction of these 
galaxies is the largest among those galaxies which do not belong to any group. 
In the supercluster outskirts, the fraction of late type galaxies  among those 
galaxies which do not belong to groups, is larger than among this population in 
the supercluster core.

Table~\ref{tab:s1269d1d2} shows that according to the Kolmogorov-Smirnov
test,  the differences between the distributions of the
spectral parameter $\eta$ for galaxies in 
the core region and in the outskirts have very high statistical significance.
The differences in the galaxy content of  groups, and of those galaxies
which do not belong to groups from
the core region $D1$ and in the outskirts $D2$ are smaller, and
their statistical significance is marginal.

The distribution of the color index $col$ for galaxies in the supercluster
SCL126 (Fig.~\ref{fig:scl126d1d2gr}, lower row) shows
that members of rich groups from the core region of the
supercluster are mainly red galaxies, $col > 1.07$. The fraction of blue 
 galaxies is larger in poor groups, and this fraction is the
largest among those galaxies which do not belong to groups. In the outskirts
region, even in rich groups there is a larger fraction of galaxies having $col
< 1.07$, in comparison with the core region. Mostly, blue galaxies can be
found among those galaxies which do not belong to groups. The differences
between the galaxy content of poor groups from the core region and the outskirts
are smaller. 

The statistical significances of these differences are given in
Table~\ref{tab:s1269d1d2}.
This table shows that according to the Kolmogorov-Smirnov
test,  the differences between the distributions of
the color index  $col$ for galaxies in 
the core region and in the outskirts have very high statistical significance.
Among other populations, the statistical significance of differences is
marginal.

In other words, galaxies of different type in the supercluster SCL126 are
segregated: red galaxies are preferentially located in rich groups, and blue 
galaxies in poor groups and among those galaxies not in any
group.  This is in a good accordance with the results about the fine structure
obtained with the fourth Minkowski functional $V_3$ where we saw that red
galaxies in the supercluster SCL126 are more clumpy, while blue,
galaxies formed a less clumpy populations around them.

\subsection{SCL9} 

The distribution of galaxies from different populations 
in the supercluster SCL9 is shown  in Fig.~\ref{fig:scl9d1d2gr} and 
Table~\ref{tab:s1269d1d2}.  

We see, in accordance with the data about 
SCL126, that galaxies in rich groups are, in average, brighter than 
galaxies in poor groups; galaxies which do not belong to groups are 
fainter than group galaxies. The largest difference between the distribution
of galaxies by luminosity in the superclusters SCL126 and SCL9 is that
in the supercluster SCL9 the luminosities of the brightest galaxies  
in rich and poor groups are of the same order. 

In the supercluster SCL9 the fraction of early type, red galaxies is again the
largest in rich groups and the smallest among those galaxies which do not
belong to groups. However, in this supercluster rich groups contain also a
large fraction of late type, blue  galaxies, especially in the outskirts of
the supercluster. The fraction of  blue galaxies in poor groups
of the core region is also relatively large. Thus, in contrast to the
supercluster SCL126, in the supercluster SCL9 early and late type,
and red and blue galaxies reside
together in groups of different richness. This is the same structure that was
shown by the fourth Minkowski functional - blue galaxies form numerous small
clumps.

Table~\ref{tab:s1269d1d2} shows that according to the Kolmogorov-Smirnov test, 
the differences between the distributions of the spectral parameter $\eta$ 
for galaxies 
in the core region and in the outskirts in the supercluster SCL9 are 
statistically 
significant. The differences between the distributions of the color index  
$col$ for 
galaxies in the core region and in the outskirts have a lower statistical 
significance. 
The differences between the distributions of luminosities for galaxies in the 
core region and in the outskirts have a marginal statistical significance only.

\subsection{Summary} 

The Table~\ref{tab:s1269d1d2}  shows that the overall galaxy
content of supercluster cores and outskirts is different. In the core regions,
the fraction of galaxies located in rich groups is about 0.45, while the
fraction of galaxies in rich groups in the outskirt region is about 0.20. 
The fraction of those
galaxies which do not belong to any group in the core region is smaller than in
the outskirt region.  

The core regions contain about 1.5 times more of early type,
red galaxies than the outskirts. The statistical significance of these
differences is high. What we see here is an evidence for a large scale 
morphology-density relation in superclusters -- high density cores of 
superclusters
contain relatively more early type,  red galaxies than the lower density
outer regions.

Interestingly, the galaxy content of groups in the 
superclusters SCL126 and SCL9 is 
not similar. In the supercluster SCL126 galaxies of different type  are 
segregated: red galaxies are preferentially located in rich groups, and blue 
galaxies in poor groups and among those galaxies not in any group. In the 
supercluster SCL9 early and late type, and red and blue galaxies reside together 
in groups of different richness. 

This is in a good accordance with the results about the fine structure
of superclusters delineated by galaxies from different populations
obtained using the fourth Minkowski functional $V_3$.

In both superclusters, those galaxies classified as late type but
having red colors,
and blue galaxies with early type spectra are mostly located in poor groups 
or they do not belong to any group; only about 10\% of them are
located in rich groups.

Another  important difference between these superclusters is that
in the supercluster SCL126 the brightest galaxies reside in rich groups
while in the supercluster SCL9 the luminosity of the brightest galaxies
in rich and poor groups is the same.

In summary, the high-density cores of the superclusters contain relatively more 
early-type, quiescent, red  galaxies than the lower-density regions, where there 
are more blue, star-forming galaxies. In the core regions the fraction of galaxies 
in rich groups is larger than this fraction in the outskirts. This shows that both 
the local (group/cluster) and the global (supercluster) environments are 
important in influencing types, colors and star formation rates of galaxies. 

Earlier studies (\citep[among others]{dg76,d80,e87,phill98,n01,n02,
zehavi02,go03,hogg03,hogg04,balogh04,depr03,depr04,ma03b,cr05,blant04,
blant06} have shown the difference between the galaxy populations in
clusters and in the field.  Our results show that this difference exists also
between the core regions and the outer regions of rich superclusters (see also
Paper III).

\section{Discussion}

\subsection{Selection effects}

Both rich superclusters, SCl126 and SCL9, are not fully covered by 
the survey volume of the 2dFGRS. However, in the case of the 
supercluster SCL126 only its small, outer part remains outside the survey 
(see references in sect. 2). All rich clusters in this supercluster are 
included. Thus, most probably, including the full supercluster may add a 
relatively small number of mainly blue galaxies to this supercluster; 
this would not change our results about the galaxy content and fine 
structure in the supercluster core region and about rich groups in the
outskirts of the supercluster.

With the supercluster SCL9, the situation is more complicated. The part which
remains outside contains about half of the Abell clusters in this supercluster
(sect. 2). Therefore, including them would increase the maximum values of the
fourth Minkowski functional, $V_3$, since this characteristic counts the
number of isolated clumps in an object under study. Thus, the differences
between the values of $V_3$ for superclusters SCL126 and SCL9 would even
increase. Even in the present analysis we saw that the values of $V_3$ for
blue galaxies for SCL9 are larger than those values for
SCL126. Thus, even if in the remaining part of the supercluster blue
galaxies were distributed homogeneously such as not to increase the present
values of $V_3$ (not a very probable case!) the differences between the $V_3$
values for galaxies of different type would not disappear.

Similarly, the differences in the galaxy content of groups from the core region 
and the outskirts would not disappear unless the distribution and clumpiness of 
galaxies in the (excluded) part of the supercluster SCL9 were completely 
different from what we see in its 2dFGRS part. Probably we should not expect to 
see large variations  in the galaxy content of different parts of one 
supercluster. Moreover, in paper III we showed that the scatter in galaxy 
populations of superclusters is small.

For a more detailed study we plan to analyse a larger number of rich
superclusters. For that, we started to generate supercluster catalogues using
the SDSS data. In particular, the region of the supercluster SCL126 is covered
by both surveys, therefore giving us a good chance to compare the properties
of this supercluster, enclosed by both data sets.

\subsection{Comparison with other very rich superclusters}

The richest relatively nearby supercluster is the Shapley Supercluster
\citep[and references therein]{proust06} \citep{bar00,qui00}.  The main core
of this supercluster contains at least two Abell clusters and two X-ray
groups.  
\citet{hai06} 
studied the galaxy populations in the core region of this supercluster
using data about galaxies with fainter absolute magnitude limit
than used in our study, thus we can compare their results with ours
qualitatively only. They
demonstrated that the colors of galaxies in the core
region of the Shapley supercluster depend on their environment, with redder
galaxies being located in clusters.  They also found a large amount of faint
blue galaxies between the clusters. We found the same trends with
colors, especially
in the core region of the supercluster SCL126. Haines et al. found also
that in the core of the Shapley supercluster there where the fraction of blue
galaxies is the lowest, the X-ray emission is the strongest. As seen above, we
also find that in rich groups, which have X-ray sources, the fraction of blue
galaxies is very small (the core region of SCL126).

Porter and Raychaudhury studied recently another rich supercluster partly
covered by the 2dF surevy -- the Pisces-Cetus supercluster (SCL10 in E01)
\citep{pr05,pr06}. They used data about Abell clusters, galaxy groups
\citep[by][]{eke04} and galaxies in this supercluster.  
Porter and Raychaudhury used the spectral parameter  $\eta$ to determine
the star formation rates for galaxies in groups in
this supercluster.
They concluded that
galaxies in rich clusters have lower star formation rates than galaxies in
poor groups in agreement with our results. Porter and
Raychaudhury also demonstrated that in the filament between the clusters in this
supercluster the fraction of star forming galaxies is higher at larger
distance from clusters than close to clusters.

\citet{gray04} studied the environment of galaxies of different 
colors in the supercluster A901/902. They divided galaxies into 
red (quiescent) and blue (star-forming) populations, using  $(U - V)$ colors
of galaxies brighter than $M_v \approx -14$. 
\citet{gray04} obtained
strong evidence that the highest density regions in clusters are populated
mostly with red, quiescent galaxies, while blue, star-forming galaxies dominate in
outer/lower density regions of clusters.
Our samples do not contain very faint galaxies;
for brighter galaxies, 
we find qualitatively the same trends in our superclusters.

\citet{wolf05} showed that in the supercluster A901/902 there exists a 
population of dusty red galaxies which show red colors and also signatures of 
star formation. In this supercluster these galaxies are located at intermediate 
densities. This agrees well with our finding about galaxies with mixed
classification (galaxies with 
red colors and spectra typical to late type galaxies): 
these galaxies populate mostly intermediate density
regions of superclusters, being members of poor groups 
or they do not belong to any group.

\citet{hil05} find that the fraction of early spectral type galaxies is
significantly higher in clusters with a high X-ray flux. Several of the X-ray
clusters studied by Hilton et al. belong to the core region of SCL126 (A1650,
A1651, A1663 and A1750), where we also found a high fraction of early type
galaxies. One of the clusters under study by Hilton et al. is located in the
supercluster SCL 9 (A2811).

\citet{balogh04} compared the populations of star-forming and quiescent
galaxies in groups from the 2dFGRS and SDSS surveys using the strength of the
$H_{\alpha}$ emission line
as star formation indicator. The spectral parameter $\eta$ is correlated with the 
equivalent width 
of the $H_{\alpha}$ emission line:  approximately, early type galaxies
correspond to quiescent galaxies and late type galaxies
to star-forming galaxies. 
\citet{balogh04} found that the
relative numbers of star-forming galaxies in groups with high velocity
dispersion is lower than in groups with low velocity dispersion. We found that
rich groups contain a smaller fraction of late type galaxies than poor
groups, qualitatively in accordance with Balogh et al.

\citet{pl04} showed that the dynamical
status of groups and clusters of galaxies depends on their large-scale
environment. Here we show that also the richness of a group and its galaxy
content depend on the large-scale environment. The first of these effects
was described as an environmental enhancement 
of group richness in \citet{e03c} and \citet{e05}.

Thus, 
our present data reveals the dependency of the properties of galaxies in 
superclusters on both the local density, in groups, (as shown also by the 2DF 
team, e.q.  \citep{balogh04} and references therein)
and on the global density (see also Paper III) in 
the supercluster environment. Moreover, 
we showed that the fine stucture of superclusters as 
determined by galaxies from different populations, 
differs. 

The differences in the distribution of galaxies from different populations in 
individual superclusters are related to a different overall morphology of these 
superclusters.

\subsection{Morphology of superclusters and their formation and 
evolution} 

We showed that there exist several differences between the morphology of
galaxy populations of individual rich superclusters, which cannot be explained
by selection effects only.

The supercluster SCL126 has a very high density core with several Abell and X-
ray clusters in a region with dimensions less than 10 \Mpc. Such very high
densities of galaxies have been observed so far only in very few
superclusters. Among them are the Shapley supercluster \citep{bar00}, the
Aquarius supercluster \citep{car02}, and the Corona Borealis supercluster
\citep{small98}. A very small number of such a high density cores of
superclusters is consistent with the results of N-body simulations which show
that such high density regions (the cores of superclusters that may have
started the collapse very early) are rare \citep{gra02}. The fraction of early
type, red galaxies in the core region of the
supercluster SCL126 is very high (higher than in SCL9), both in groups and
among those galaxies which do not belong to groups.

This, together with the presence of a high density core and overall more
homogeneous structure than in the supercluster SCL9 may be an indication that
the supercluster SCL126 has formed earlier (is more evolved dynamically by the
present epoch) than the supercluster SCL9.

However, in another part of the core region of the supercluster SCL126 the X-ray 
cluster A1750 shows signs of merging, both according to X-ray and optical data 
\citep{don,bel}.  

\citet{don} studied this cluster using observational data from ROSAT and ASCA, 
and also the velocities of 
galaxies in this cluster. Their analysis suggests that we observe
an ongoing merger in  this binary cluster.
 
Additional evidence about complex merger events in this cluster was provided 
by \citet{bel}, who used XMM-Newton data to study in detail the surface 
brightness, temperature and entropy distribution in this cluster. Their data 
indicate that two components of this binary cluster have 
just started to interact. In 
addition, their data suggest that there occured another merging event   in one 
of the clusters, perhaps in the past 1-2 Gy. They relate these merging 
events also to the large scale environment of this cluster 
(it is a member of a very 
rich supercluster, SCL126). In summary, these studies suggest that the 
region around the rich cluster A1750 in the supercluster SCL126 may be 
dynamically younger than the main core region.

This all shows that the formation and evolution history of superclusters
is a complex subject. We can try to model it, but the best way is to
follow their real evolution in time, looking for superclusters in
deep surveys. An especially promising project is the 
ALHAMBRA Deep survey by \cite{moles05}
that will provide us with data about (possible) galaxy
systems at very high redshifts which can be searched for and analysed using
morphological methods. Comparing superclusters at different redshifts
will clarify many questions about their evolution.

In order to have a larger sample of local supercluster templates,
we plan to continue our studies of rich superclusters. We
mentioned above that we plan to use for that
data from the Sloan Digital Sky Survey to
extend the sample of relatively nearby superclusters. 

In Paper RI we showed that the richest superclusters in the Millenium 
simulation do not describe the large morphological variety of the 
observed superclusters. Moreover, Minkowski Functionals show that the 
fine structure as delineated by bright and faint galaxies in Millenium 
Simulations do not follow the fine structure delineated by galaxies of 
different luminosities in observed superclusters. This suggest that  the 
model does not yet explain all the features of observed superclusters.
 
Earlier studies of the Minkowski functional of the whole SDSS survey region
\citep[and references therein]{gott06} also conclude that N-body simulations
with very large volume and more power at large scales are needed to model
structures like the supercluster SCL126 more accurately than present
simulations.  Similar conclusions were reached by \citet{e06}.  Also the
galaxy formation and its environmental dependence is not yet well understood.
Our present results indicate that the properties of galaxies and their
evolution history have been affected by both local and global densities in
superclusters.  The details of these processes have to be modelled in 
future simulations.

\section{Conclusions}

We have presented a morphological study of the two richest superclusters from 
the 2dF Galaxy Redshift Survey. We studied the internal structure and 
galaxy populations of  these superclusters.  Our main conclusions are 
the following.

\begin{itemize}

\item{}
The values of the fourth Minkowski functional $V_3$, which contain information 
about both the local and global morphology, show   the fine structure of 
superclusters as determined by galaxies from different populations. The fourth 
Minkowski functional $V_3$ and the morphological signature $K_1$-$K_2$ show a 
crossover from low-density morphology (outskirts of supercluster) to high-
density morphology (core of supercluster) at a mass fraction $m_f \approx 0.7$.

\item{} In the supercluster SCL126, the functional
  $V_3$ shows that the number of clumps in
  the distribution of red   galaxies is larger than the number of
  clumps determined by blue  galaxies, especially in the outer
  regions of the supercluster, where
  the maximum values of $V_3$ are, correspondingly, 8 for red galaxies and 5 
  for blue galaxies.
  Thus in the outskirts of the supercluster, blue galaxies form a more
  homogeneous population than  red galaxies. In the core region, 
  the clumpiness of galaxy populations is of the same order.

\item{} In the supercluster SCL9, the values of $V_3$ are large for both early
  type and red galaxies, and late type and blue galaxies. 
  In the outskirts of the supercluster the differences between
  the clumpiness of galaxy populations are small
  (the values of $V_3$ are the same, 7--8 for both early and late type
   galaxies), while in the core
  region of this supercluster the clumpiness of the late type, blue galaxy
  population is larger (the maximum value of $V_3$ is 17)
  than the clumpiness of the early type, red galaxy population,
  where the maximum $V_3 = 12$.

\item{}
Our superclusters contain galaxies with mixed classifications: 
these galaxies show spectra typical to late type galaxies and red colors,
or they are early type galaxies with blue colors. 
These galaxies are mostly located at intermediate densities 
in the outskirts of our superclusters, being members of poor groups or
do not belonging to any group. Due to these galaxies, 
at intermediate mass fractions 
the curves of the fourth Minkowski functional $V_3$ for early type
galaxies and for red galaxies, as well as for late type galaxies and 
blue galaxies, differ in some details.

\item{} Groups in high-density cores of superclusters are richer than in
  lower-density (outer) regions of superclusters. In high-density cores,
  groups contain relatively more  early-type, red galaxies than the
  groups in lower-density regions in superclusters, where there are more 
  late type, blue galaxies.  Therefore, both the richness of a group and its
  galaxy content depend on the large scale environment where it resides.

\item{} 
  In cores of superclusters, the fractions of early type, red galaxies 
  is larger than these fractions in outskirts.
 In the supercluster SCL126 the
  morphological segregation of red and blue galaxies is stronger
  than in the supercluster SCL9.  In SCL126, the most luminous galaxies in
  rich groups have larger luminosities than most luminous galaxies in poor
  groups, while in SCL9 the luminosity of the brightest galaxies in rich and
  poor groups is comparable.

\item{} The differences in overall morphology, fine structure and galaxy
  content of the supercluster suggest that there are differences in their
  evolutional history which affect their present-day properties.

\end{itemize}
 
Our study shows the importance of the role of superclusters as a high density
environment that affects the evolution and the present-day properties of their
member galaxies and the groups/clusters of galaxies that constitute the
supercluster.

The forthcoming Planck satellite observations will determine the anisotropy of
the cosmic background radiation with unprecedented accuracy and angular
resolution. As a by-product, Planck measurements will provide an all-sky
survey of massive clusters via the Sunyaev-Zeldovich (SZ) effect. For the
Planck project, detailed information of supercluster properties is important,
helping to correlate the SZ-signals with the imprints of local superclusters.
As a continuation of the present work, we are
preparing supercluster catalogues for the Planck community.

\acknowledgments
We thank the anonymous referee for careful reading of the manuscript and for
many useful comments which helped to improve the paper.
We are pleased to thank the 2dFGRS Team for the publicly available data
releases.  The present study was supported by the Estonian Science Foundation
grants No. 6104 and 7146, and by the Estonian Ministry for Education and
Science research project TO 0060058S98. This work has also been supported by
the University of Valencia through a visiting professorship for Enn Saar and
by the Spanish MEC project AYA2006-14056 (including FEDER).  J.E.  thanks
Astrophysikalisches Institut Potsdam (using DFG-grant 436 EST 17/4/06), and
the Aspen Center for Physics for hospitality, where part of this study was
performed.  PH and PN were supported by Planck science in Mets\"ahovi, Academy
of Finland. In this paper we have used R, a language for data analysis and
graphics \citep{ig96}.

\appendix

\section{Estimating probability densities}
\label{sec:densities}

As said in the main text, all the distributions (probability
densities) shown in this paper have been obtained using the R environment
\citep{ig96}, \texttt{http://www.r-project.org} (the 'stats' package).
We do not show the customary error limits in our figures; we explain here why.

\begin{figure}[ht]
\centering
\resizebox{0.45\textwidth}{!}{\includegraphics*{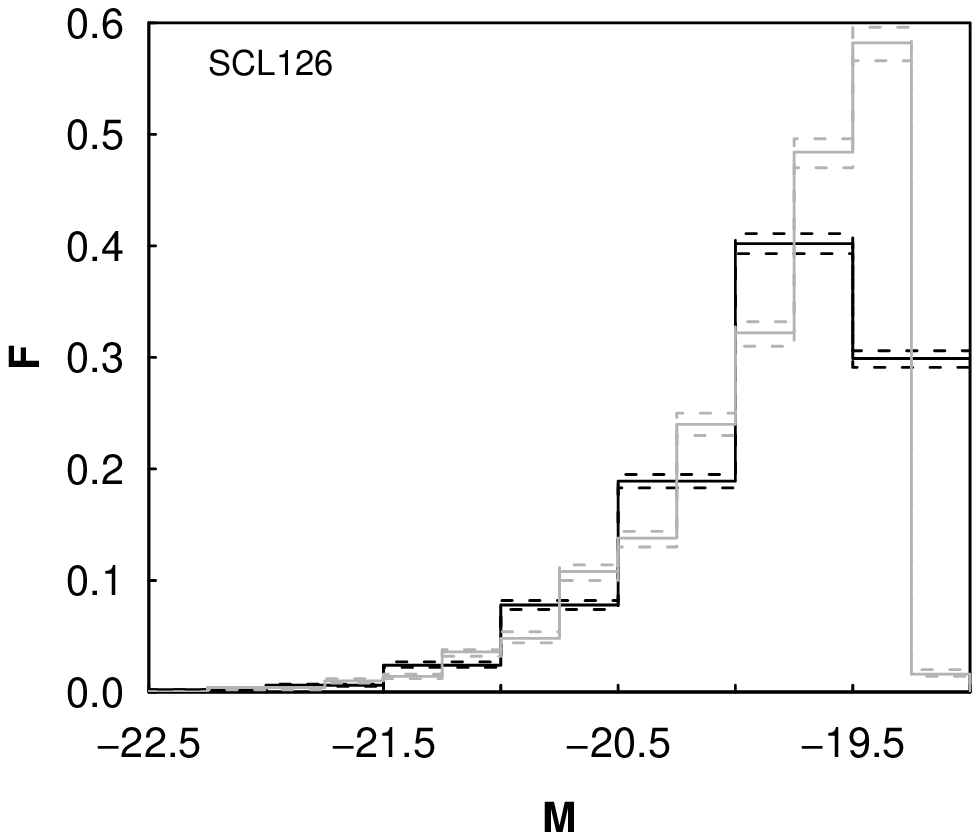}}
\caption{The differential luminosity function histograms
$F=dN/dM$, where $M$ is the absolute magnitude of a galaxy,
for galaxies in the supercluster SCL126; here the
solid line shows the luminosity function and the dashed lines indicate
the Poisson errors. Black lines show the histogram for the 0.5 magnitude
bins, gray lines -- for the 0.25 magnitude bins.
}
\label{fig:r2bmagerr}
\end{figure}

Fig.~\ref{fig:r2bmagerr} shows the differential luminosity function histogram
with Poisson error limits.  This Figure shows, first, that Poisson errors are
small, and, secondly, that these errors are not very useful since they are
defined mainly by the bin widths -- the histograms for the different bin
widths clearly do not coincide within the formal error limits, and, therefore,
they cannot represent the true density distribution.
As known for long in statistics 
\citep[see, e.g., a good pedagogical presentation by][]{wandjones}, 
the most important part in probability
density estimation is the right choice of the bin (kernel) width, and
the 'density' command in the 'stats' package does that, minimizing the MISE
(mean integral standard error) of the estimate. 
Also, it is long known that kernel estimates
\[
f(x)=\frac1{N}\sum_i^N K(x-x_i;h),
\]
where the data are $\{x_i\},\, i\in[1,N]$, and $K(x;h)$ are suitable kernels
of a width $h$, 
are preferred to binning. Both these estimates depend on the bin (kernel)
widths that can be found by optimizing the MISE; but histograms depend,
additionally, on the placing of the bins. Also, the minimum MISE in
case of binning is larger than for kernel estimates.

The MISE is defined as
\begin{eqnarray}
\mbox{MISE}&=&E\int\left(\hat{f}(x)-f(x)\right)^2dx\\
	&=&\int\mbox{Var}\left(\hat{f}(x)\right)\,dx+
	\int\left[\mbox{Bias}\left(\hat{f}(x)\right)\right]^2dx.
\label{eq:mise}
\end{eqnarray}
The two terms in the last equality depend in a different way on the
kernel width (bandwidth). When we increase this width, the
variance decreases, but the bias increases -- this is what happens when
we take wider bins for Fig.~\ref{fig:r2bmagerr}. The 'stats' package
chooses the optimal bandwidth as that which minimizes the MISE.
We could also try to minimise the local mean standard error MSE($x$),
but this asks for adaptive density estimation (using bandwiths that
depend on $x$), and the procedure is much more complex.

For a kernel estimate, when binning is replaced by convolution with
a suitable kernel, the minimum MISE is of
the order of $n^{-4/5}$, where $n$ is the size of the sample, and the
optimal kernel width is of the order of $\hat{\sigma}n^{-1/5}$,
where $\hat{\sigma}$ is the estimate of the rms deviation of the data 
\citep[see, e.g.][]{silverman}. The expression 'of the order to' can be
mostly read as 'equal to', as the proportionality coefficients are
usually close to unity. This kernel width means that there are about
$n\cdot n^{-1/5}=n^{4/5}$ points per kernel -- we get our accustomed
Poissonian error, but only for the optimal kernel width. However, this
is only a rough estimate, as the local error, the mean standard error 
MSE$(x)$ depends on the true density, being larger in the regions where 
this density has minima or maxima, and is difficult to estimate.
Thus the 'stats' package does not provide the MSE, and we do not show
error bars in our figures.

\begin{figure}[ht]

\centering
\resizebox{0.45\textwidth}{!}{\includegraphics*{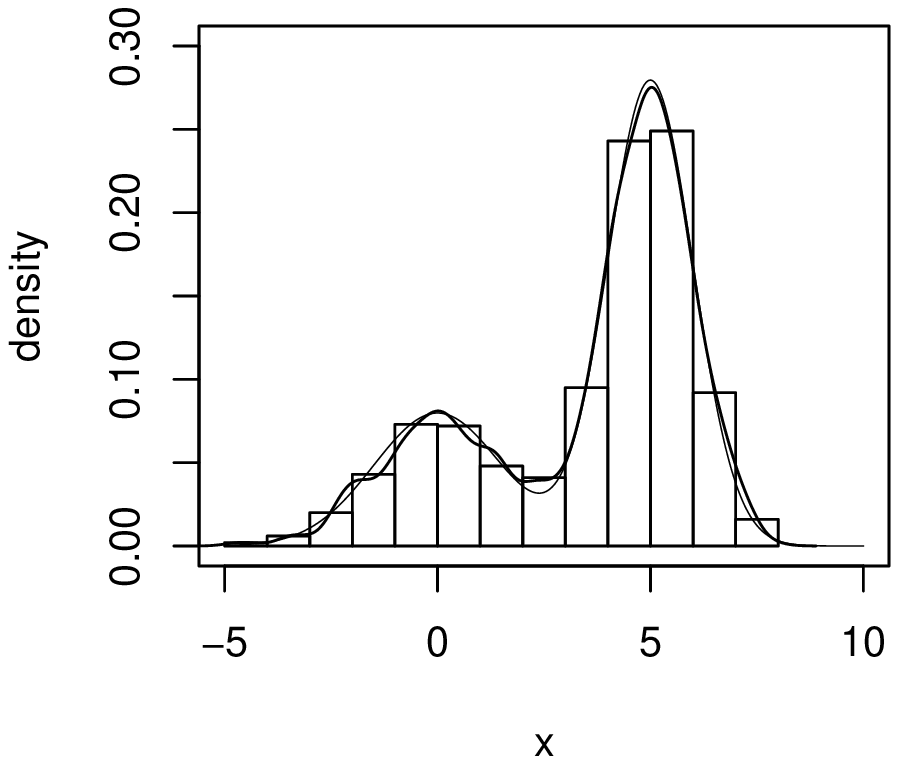}}
\caption{Comparison of the histogram and kernel density estimates. 
The true density is shown by a thin line, and the kernel estimate
obtained by the R 'stats' package for a sample of the size 1000 
-- by a thick line. For comparison, we show a histogram (obtained also
by 'R'). It is clearly seen that the histogram is inferior to the
kernel estimate.
}
\label{fig:testfig}
\end{figure}

You can compare histograms and kernel densities 
in Fig~\ref{fig:testfig}. Here we generated a random sample of 1000
values for a weighted sum of two normal distributions, calculated its
histogram, and applied the R procedure to obtain the optimized kernel
density. The optimal kernel width is 0.36, three times smaller than the bin
width; however, the kernel density follows well the smooth true
density, and recovers the details much better than the histogram does.

As the total number of
galaxies in our superclusters is large, the differences
between the estimated and true densities are small. 
These differences can be easily estimated. Let us take SCL9 as an example
(SCL126 has more galaxies, and the errors are slightly smaller).
The total number of galaxies in SCL9 is 1176, so the MISE for the distributions
involving the whole supercluster is $\approx0.0035$, and about 0.006 for
separate populations (the corresponding rms errors are 0.06 and 0.08). 
The local errors (MSE) are, in average, a few times smaller (by the
ratio of the total range of a random variable to $\hat{\sigma}$).
Our use of a constant width kernel
does not allow us to better estimate the local errors,
but we can always strictly
compare densities, using the full data -- the integral distributions (e.g.,
the Kolmogorov-Smirnov test).  

\section{Spatial densities}
\label{sec:spatial} 

When studying the morphology of superclusters of galaxies, a necessary
step is to convert the spatial positions of galaxies into spatial densities.
The standard approach is to use kernel densities 
\citep[see, e.g.][]{silverman}:
\[
\varrho(\mathbf{x})=\sum_i K(\mathbf{x-x}_i;h) m_i,
\]
where the sum is over all galaxies, $\mathbf{x}_i$ are the coordinates
of the $i$-th galaxy, and $m_i$ is its mass (or luminosity, if we are
estimating luminosity densities; 
for number densities we set 
$m_i\equiv1$).  The function $K(\mathbf{x};h)$ is the kernel of the
width $h$, and a suitable choice of the kernel determines the quality
of the density estimate. If the kernel widths depend either on the
coordinate $\mathbf{x}$ or on the position of the $i$-th galaxy, the
densities are adaptive. 

Kernels have to be normalized and symmetrical:
\[
\int K(\mathbf{x};h)dV=1,\qquad \int \mathbf{x}K(\mathbf{x};h)dV=0.
\]

For the usual case, when densities are calculated for a spatial grid,
good kernels are generated by box splines $B_J$ (usually
used in $N$-body mass assignment).  Box splines have compact support
(they are local), and they are interpolating on a grid:
\[
\sum_i B_J(x-i)=1,
\]
for any $x$, and a small number of indices that give non-zero values for
$B_J(x)$.  In this paper we restrict us to
the popular $B_3$ splines:
\[
B_3(x)=\frac1{12}\left[|x-2|^3-4|x-1|^3+6|x|^3-4|x+1|^3+|x+2|^3\right]
\]
(this function is different from zero only in the interval $x\in[-2,2]$).
We define the (one-dimensional) $B_3$ box spline kernel 
of width $h=N$ as
\[
K_B^{(1)}(x;N)=B_3(x/N)/N.
\]
This kernel preserves the interpolation property (mass conservation)
for all kernel widths that are integer multiples of the grid step, $h=N$.
The 3-D $K_{B}^{(3)}$ box spline kernel we use is given by the direct 
product of three one-dimensional kernels:
\[
K_B(\mathbf{x};N)\equiv K_B^{(3)}(\mathbf{x};N)=K_B^{(1)})(x;N)K_B^{(1)}(y;N)
        K_B^{(1)}(z;N),
\]
where $\mathbf{x}\equiv\{x,y,z\}$.
Although it is a direct product, it is isotropic to a good degree
(to a few per cent in the outer regions).

As seen before, the best (optimal) kernel width is usually determined 
by minimizing the MISE; in our case this is about 6--8\hmpc.
We have defined (see RI) superclusters as density enhancements of a common
scale of 8\hmpc, so we will use this value. 

\section{Minkowski functionals and shapefinders} 
\label{sec:define} 

Consider an
excursion set $F_{\phi_0}$ of a field $\phi(\mathbf{x})$ (the set
of all points where density is larger than a given limit,
$\phi(\mathbf{x}\ge\phi_0$)). Then, the first
Minkowski functional (the volume functional) is the volume of 
this region (the excursion set):
\begin{equation}
\label{mf0}
V_0(\phi_0)=\int_{F_{\phi_0}}d^3x\;.
\end{equation}
The second MF is proportional to the surface area
of the boundary $\delta F_\phi$ of the excursion set:
\begin{equation}
\label{mf1}
V_1(\phi_0)=\frac16\int_{\delta F_{\phi_0}}dS(\mathbf{x})\;,
\end{equation}
(but not the area itself, notice the constant).
The third MF is proportional to the \index{integrated mean curvature}
integrated mean curvature
of the boundary:
\begin{equation}
\label{mf2}
V_2(\phi_0)=\frac1{6\pi}\int_{\delta F_{\phi_0}}
    \left(\frac1{R_1(\mathbf{x})}+\frac1{R_2(\mathbf{x})}\right)dS(\mathbf{x})\;,
\end{equation}
where $R_1(\mathbf{x})$ and $R_2(\mathbf{x})$ 
are the principal radii of curvature of the boundary.
The fourth Minkowski functional is proportional to the integrated
Gaussian curvature (the Euler characteristic) \index{Euler!characteristic}
of the boundary:
\begin{equation}
\label{mf3}
V_3(\phi_0)=\frac1{4\pi}\int_{\delta F_{\phi_0}}
    \frac1{R_1(\mathbf{x})R_2(\mathbf{x})}dS(\mathbf{x})\;.
\end{equation}
At high (low) densities this functional gives us the number of isolated 
clumps (voids) in the sample 
\citep{mart05,saar06}.

As the argument labeling the isodensity surfaces, we chose the (excluded) mass
fraction $m_f$ -- the ratio of the mass in regions with density {\em lower}
than the density at the surface, to the total mass of the supercluster. When
this ratio runs from 0 to 1, the iso-surfaces move from the outer limiting
boundary into the center of the supercluster, i.e. the fraction $m_f=0$
corresponds to the whole supercluster, and $m_f=1$ to its highest density
peak.

We use directly only the fourth Minkowski functional in this paper;
the other functionals are used to calculate the shapefinders
\citep{sah98,sss04}. 
The shapefinders are defined as a
set of combinations of Minkowski functionals: $H_1=3V/S$ (thickness),
$H_2=S/C$ (width), and $H_3=C/4\pi$ (length).  The
shapefinders have dimensions of length and are normalized to give $H_i=R$
for a sphere of radius $R$.  For a convex surface, the shapefinders $H_i$
follow the inequalities $H_1\leq H_2\leq H_3$.  Oblate ellipsoids (pancakes)
are characterized by $H_1 << H_2 \approx H_3$, while prolate ellipsoids
(filaments) are described by $H_1 \approx H_2 << H_3$.

\citet{sah98} also defined  two dimensionless
shapefinders $K_1$ (planarity) and $K_2$ (filamentarity): 
$K_1 = (H_2 - H_1)/(H_2 + H_1)$ and $K_2 = (H_3 -
H_2)/(H_3 + H_2)$.

In the $(K_1,K_2)$-plane filaments are located near the $K_2$-axis,
pancakes near the $K_1$-axis, and ribbons along the diagonal, connecting 
the spheres at the origin with the ideal ribbon at $(1,1)$. 

\subsection{Algorithms}
\label{subsec:algorithms} 

Different algorithms are used to calculate the Minkowski
functionals; here we use a simple grid-based algorithm, based
on integral geometry (Crofton's intersection formula), proposed
by Schmalzing and Buchert (\cite{jens97}). 

To start with, we find the density thresholds for given filling
fractions $f$ by sorting the grid densities. This is a common step
for all grid-based algorithms. Vertexes with higher
densities than the threshold form the excursion set. This set is
characterized by its basic sets of different dimensions -- points
(vertexes), edges formed by two neighbouring points, squares (faces)
formed by four edges, and cubes formed by six faces. The algorithm
counts the numbers of elements of all basic sets, and finds the values of the
Minkowski functionals as
\begin{eqnarray}
\label{crofton}
V_0&=&a^3N_3\;,\nonumber\\
V_1&=&a^2\left(\frac29N_2-\frac23N_3\right)\;,\nonumber\\
V_2&=&a\left(\frac29N_1-\frac49N_2+\frac23N_3\right)\;,\nonumber\\
V_3&=&N_0-N_1+N_2-N_3\;,
\end{eqnarray}
where $a$ is the grid step, $N_0$ is the
number of vertexes, $N_1$ is the number of edges, $N_2$ is the
number of squares (faces), and $N_3$ is the number of basic cubes in
the excursion set.

\section{Clumpiness: morphological templates} 
\label{sec:proto} 

In Paper RI we generated a series of empirical models which served us as 
morphological templates to understand the behaviour of shapefinders. 
These models showed
that the morphological signature of rich superclusters corresponds to
a multibranched filament; the simplest model for that was a long filament with
short filaments across it. In RI we did not study the fine structure of
superclusters as expressed by their clumpiness. In these models the locations
of points which mimic positions of galaxies were generated randomly. Thus our
empirical models in that paper do not recover completely the inner structure of
superclusters.

For this appendix, we generated a series of empirical models to
understand better the substructure of superclusters.
These models will also help us to determine how well our methods
distinguish  different types of substructure.

In these models, the overall distribution 
of points (which mimic individual galaxies) resembles a thin filament 
with a size $10\times 20\times 100$ (in grid units), as in RI. We place 
inside this filament a series of clusters, located randomly. Their 
richness, number and size varies so that the total number of objects is 
always 1000 or 500 (approximately the number of galaxies in superclusters
under study, and in individual galaxy populations). 
We used the  richness values 5, 10 and 20; the number of clusters is, 
correspondingly,
100, 50 and 25. The size of clusters was 1, 5, and 10 in grid units.
Small clusters mimic real groups and clusters, large clusters --
overdensity regions.
Clusters are located randomly and may overlap, forming additional 
overdensity regions, as, for example, in cores of superclusters,
where also galaxy populations are mixed.

\begin{figure*}

\centering
\resizebox{0.28\textwidth}{!}{\includegraphics*{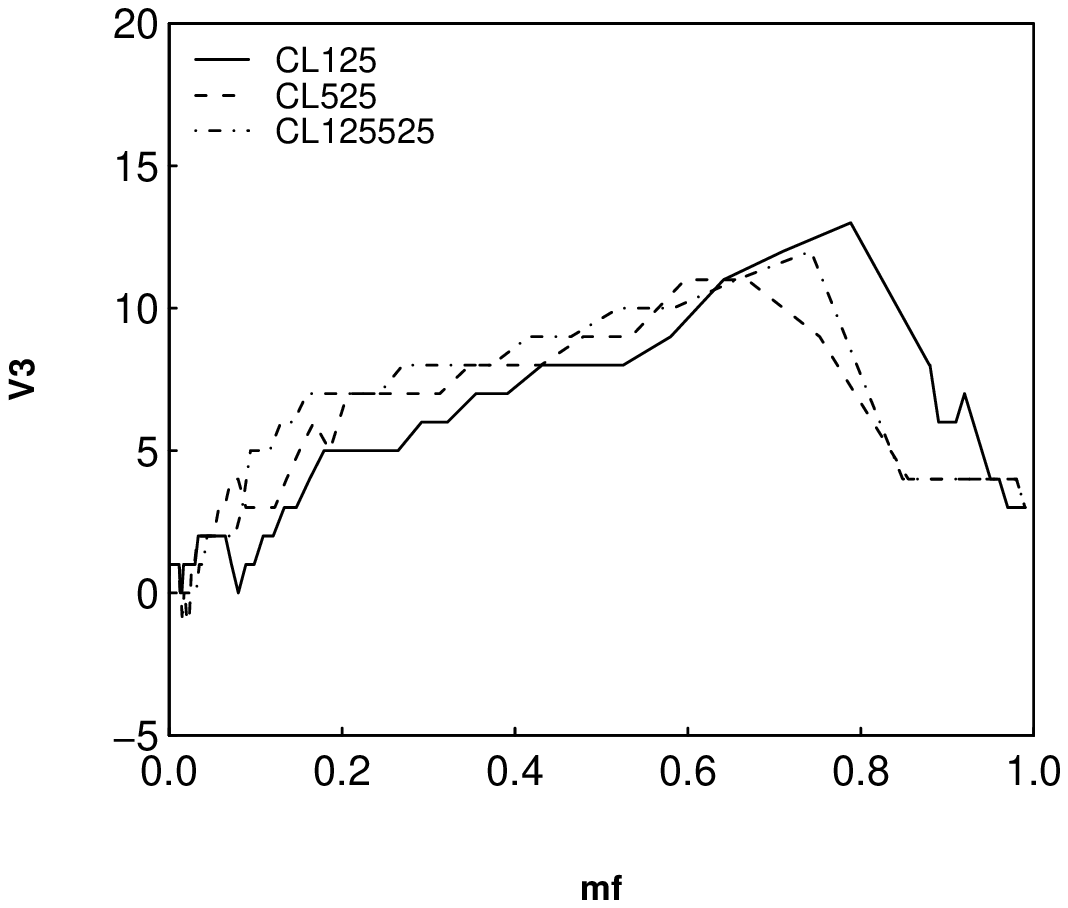}}
\resizebox{0.28\textwidth}{!}{\includegraphics*{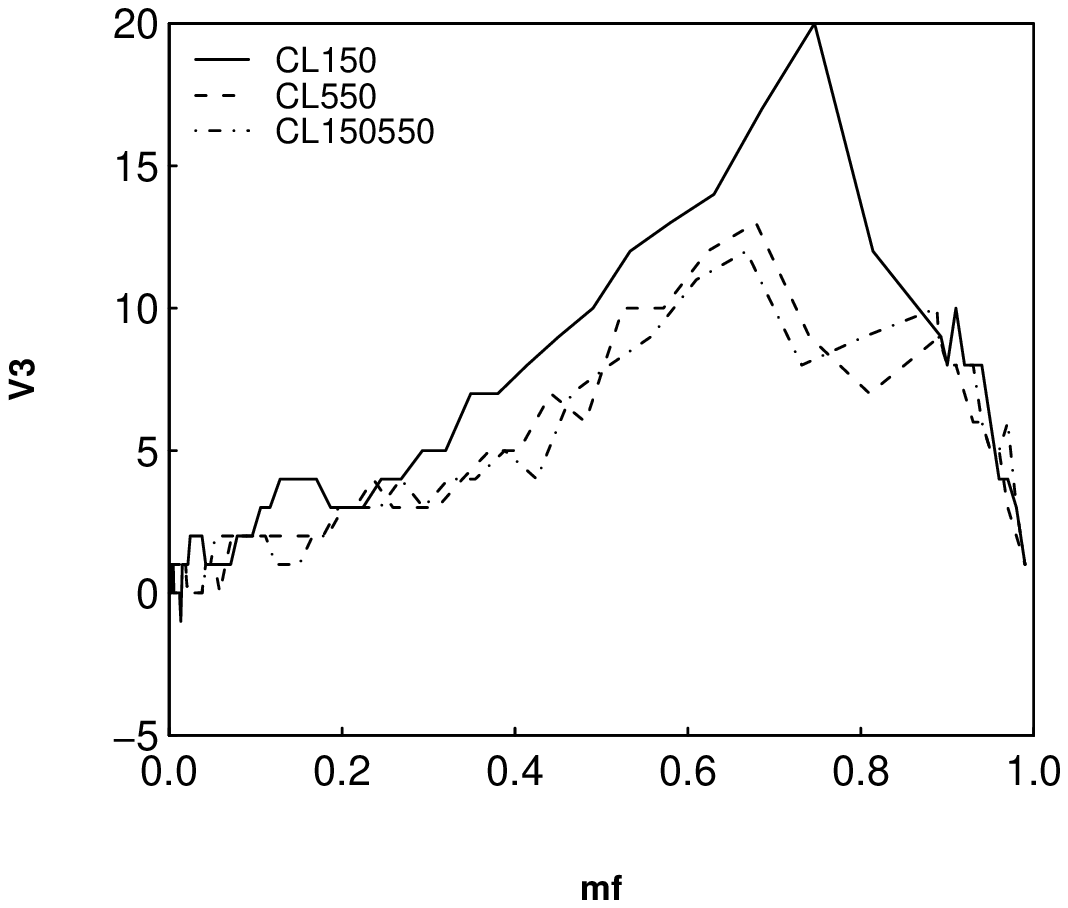}}
\resizebox{0.28\textwidth}{!}{\includegraphics*{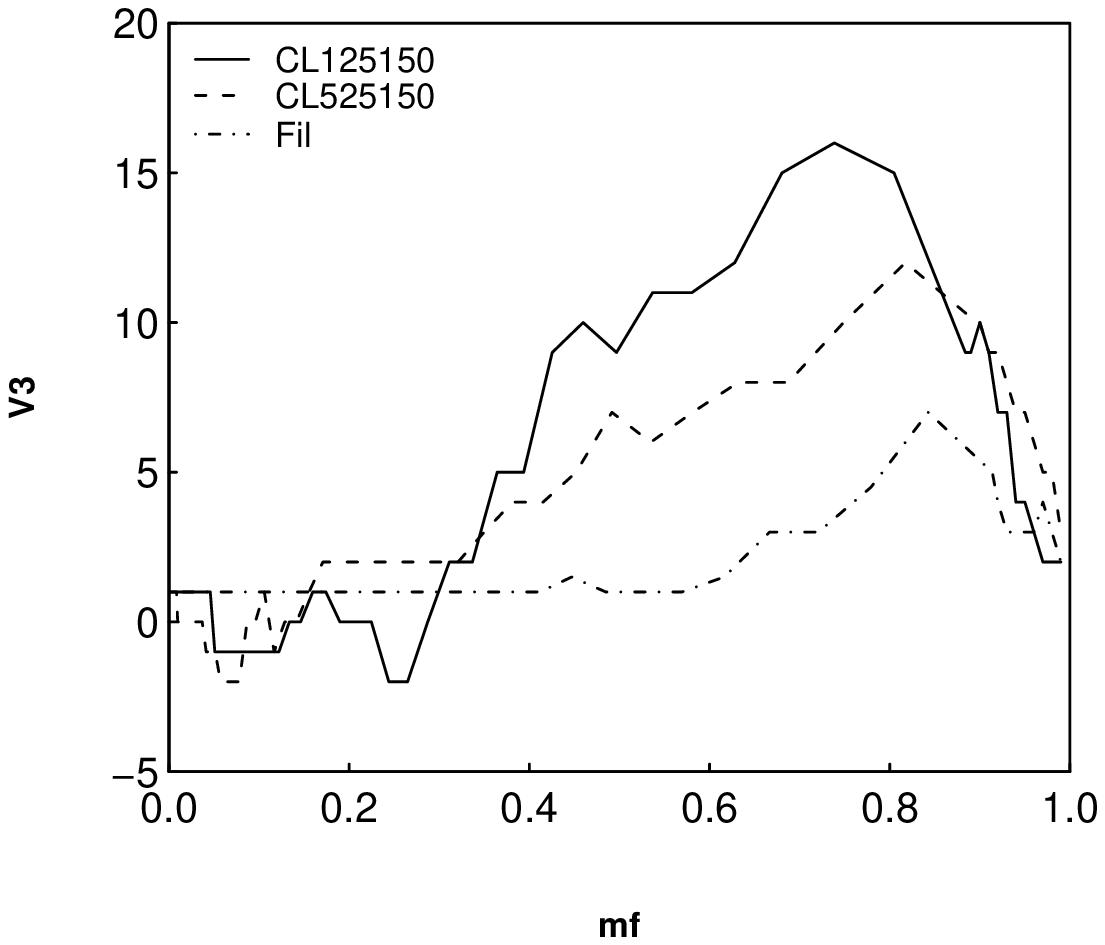}}\\
\resizebox{0.28\textwidth}{!}{\includegraphics*{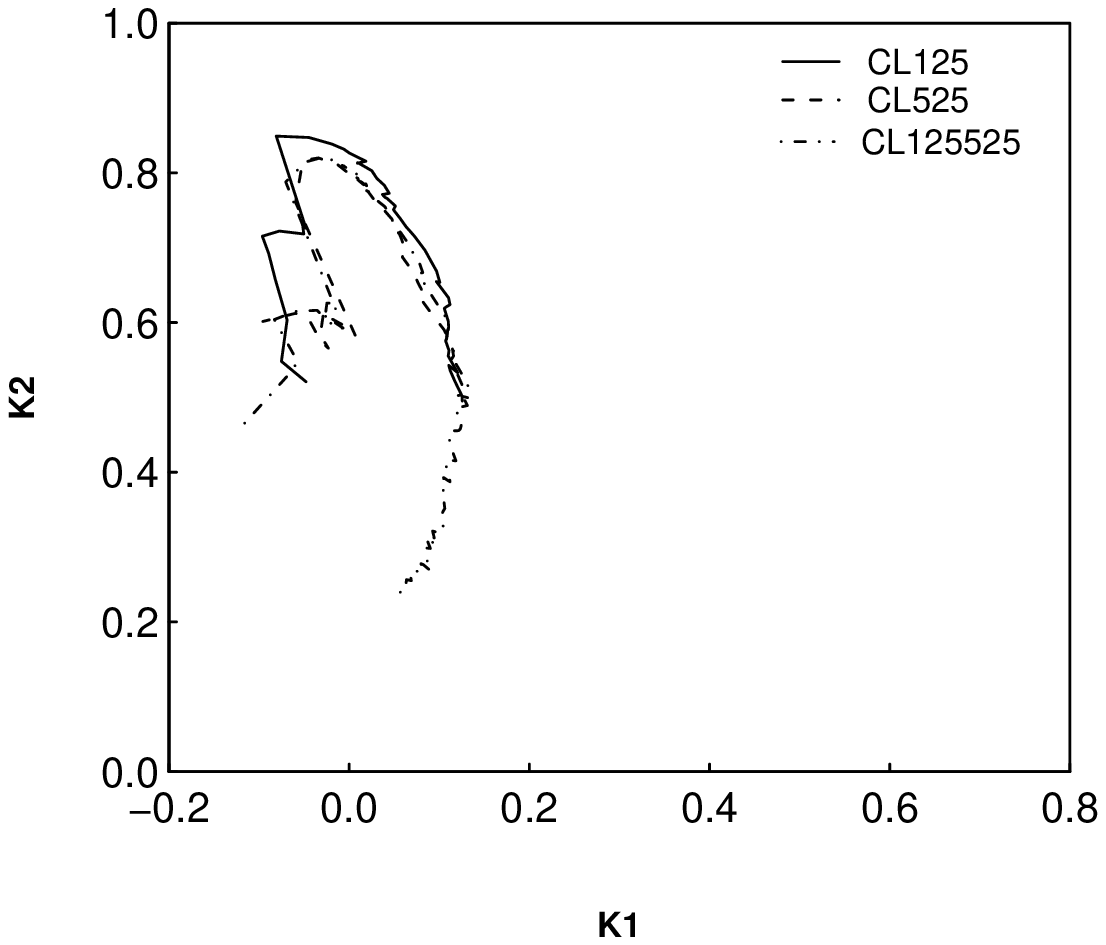}}
\resizebox{0.28\textwidth}{!}{\includegraphics*{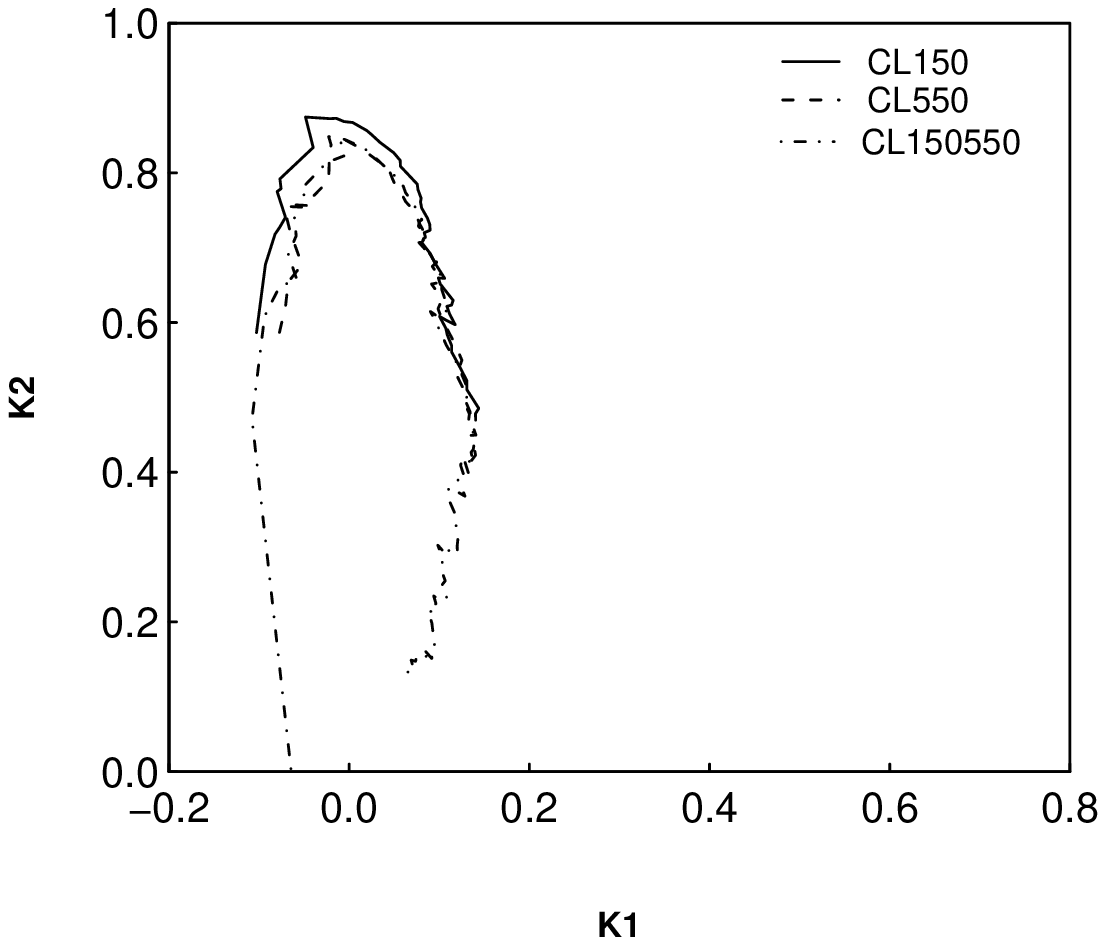}}
\resizebox{0.28\textwidth}{!}{\includegraphics*{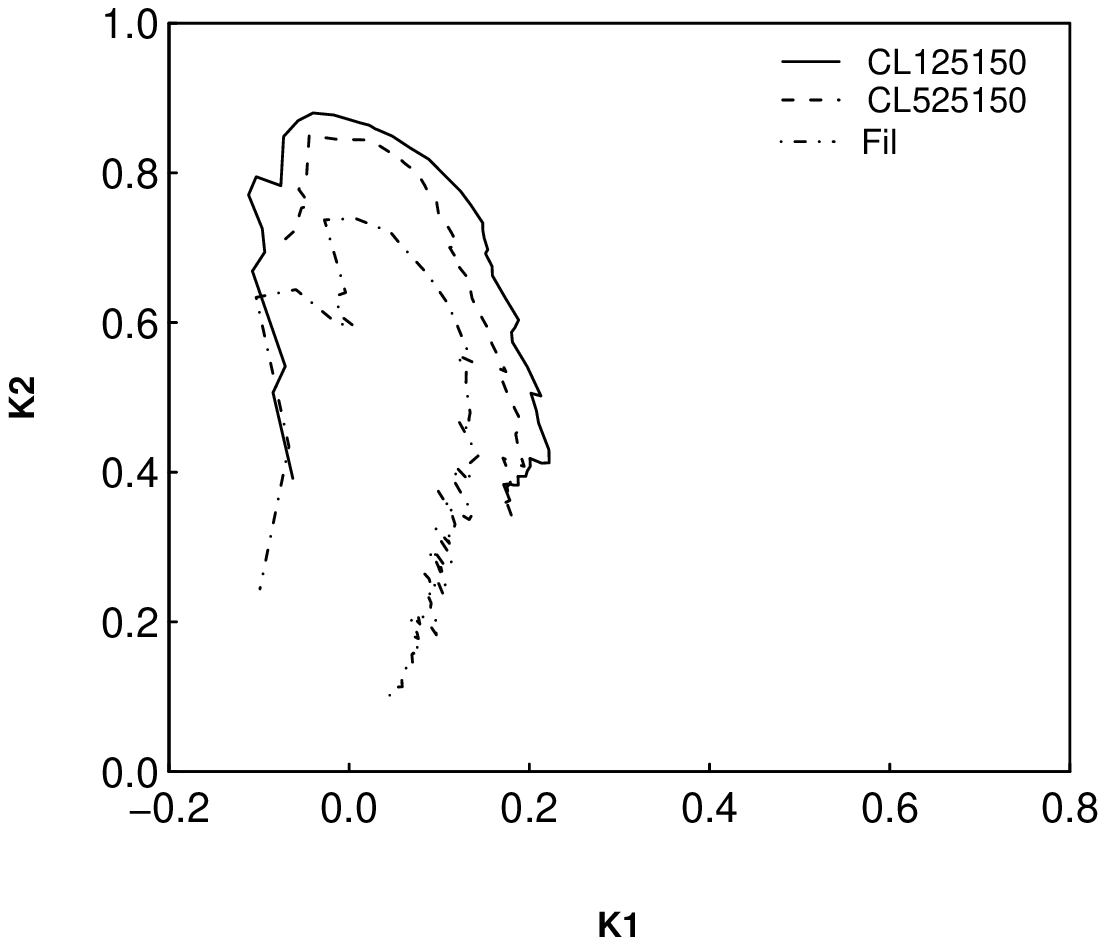}}
\caption{
The Euler characteristic (upper row)
and the morphological signature $K_1$-$K_2$ (lower row) for
the empirical models.
Left panels:
$N_{cl} = 25$, $N_{gal} = 20$ and the size of the clusters is 1 and 5
(in grid units), and the combined model where the centres of clusters
coincide. 
Solid line - model CL125, dashed line - model CL525, and dot-dashed line - 
model CL125525.
Middle panels:
$N_{cl} = 50$, $N_{gal} = 10$ and the size of clusters is 1 and 5
(in grid units), and the combined model where the centres of clusters
coincide. 
Solid line - model CL150, dashed line - model CL550, and dot-dashed line - 
model CL150550.
Right panels:
the combined models where the centres do not coincide.
'Fil' denotes a simple filament with randomly distributed points.
Solid line - model CL125150, dashed line - model CL525150, and dot-dashed line - 
model 'Fil'.
}
\label{fig:clus1}
\end{figure*}

We plot the Euler characteristics and morphological signatures for
these models in Figs.~\ref{fig:clus1}. 

The models used in this figure are as follows. In the first set
the number of clusters is 25, each cluster
has 20 galaxies in it. The sizes of clusters are 1 and 5. In the
third model in this set, these
clusters are combined together. In the latter model clusters have the same centre
coordinates. Thus the model CL125525 (the combined models CL125 and CL525)
mimics the distribution where groups are surrounded by lower density galaxies
(e.g., ellipticals by spirals).

In the second set the number of clusters is 50, with 10 member galaxies, and
in the combined model, cluster centres again coincide (the models CL150, CL550, and
CL150550). In the third set (CL125150 and CL525150) we added to the models with 25
clusters (sizes 1 and 5) the model CL150, to have a smaller number of rich groups
and a large number of poor groups, but in this case the
centres of groups do not coincide. These models mimic the situation where
different populations of galaxies are located in clumps of different richness
and size, but in separate systems (partly, since clumps may overlap randomly). 
We plot here also the curves for a simple filament  for comparison.

We do not show in this figure models with 100 small clusters and those with
cluster size 10; in all these models the values of $V_3$ were smaller than in
real superclusters. In the case of large clusters the value of $K_2$ becomes too 
large.

Fig.~\ref{fig:clus1} shows that, first, the morphological signature
is recovered correctly by our empirical models.
Second, the $V_3(m_f)$ curves for different models are different,
showing that they discriminate well different substructure.

These models show that the supercluster SCL126 is better modelled by a small
number of richer/bigger clumps, and the supercluster SCL9 -- by a large number of
smaller clumps. Clusters overlap, thus the maximum values of $V_3$ are smaller
than the original numbers of clusters in the models. The case when any
population of galaxies is located randomly within the supercluster volume does not
correspond to the real distribution of galaxies. Of course, our very simple models
in which clusters of one galaxy population are surrounded by low density "clouds" 
of galaxies from another population do not recover well the details of
clumpiness of different galaxy populations of observed superclusters. In real
superclusters, populations of galaxies are more strongly mixed.

\end{document}